\documentclass[12pt]{iopart} 

\usepackage{iopams} 
\usepackage{graphicx} 

\usepackage{setspace}
\usepackage{lineno}
\usepackage{booktabs}
\usepackage{adjustbox}
\usepackage{subcaption}
\usepackage{xcolor}
\usepackage{xr}
\usepackage{hyperref}

\usepackage[finalnew]{trackchanges}

\addeditor{Jacob Landsberg}  

\usepackage[style=authoryear, backend=bibtex]{biblatex}
\DeclareCiteCommand{\cite}
  {\usebibmacro{prenote}} 
  {\printnames{labelname}~(\printdate)} 
  {\multicitedelim} 
  {\usebibmacro{postnote}} 

\makeatletter
\long\def\@makecaption#1#2{\vskip \abovecaptionskip
  {\footnotesize {\bf #1.} #2\par} 
  \vskip \belowcaptionskip}
\makeatother

\begin{document}

\title[]{\change{AI-Informed Model Analogs for Subseasonal-to-Seasonal Prediction}{AI-Informed Model-Analogs for Understanding Subseasonal-to-Seasonal Jet Stream and North American Temperature Predictability}}
\author{Jacob B. Landsberg$^{1,2}$, Matthew Newman$^3$, Elizabeth A. Barnes$^{1,2}$}
\address{$^1$ Department of Atmospheric Science, Colorado State University, Fort Collins, CO, USA}
\address{$^2$ Faculty of Computing and Data Sciences, Boston University, Boston, MA, USA}
\address{$^3$ NOAA Physical Sciences Laboratory, Boulder, CO, USA}
\ead{jlandsbe@colostate.edu}
\doublespacing
\begin{abstract}
Subseasonal-to-seasonal (S2S) forecasting is crucial for public health, disaster preparedness, and agriculture, yet both forecasting and diagnosing sources of potential forecast skill on this timescale remains particularly challenging. We adapt an interpretable AI-informed analog forecasting approach, previously used for longer timescales, to improve S2S model-analog prediction and understanding of its climate drivers. Using an artificial neural network, we learn a mask of weights to optimize analog selection and showcase its versatility across two prediction tasks: 1) regional continuous prediction of Month 1 midwestern U.S. summer temperatures and 2) classification of Month 1-2 North Atlantic wintertime upper atmospheric winds. The AI-informed analogs outperform traditional model-analog forecasting approaches, as well as climatology and persistence baselines, for deterministic and probabilistic skill metrics on both climate model and reanalysis data; moreover, this skill gap grows for extreme predictions. Moreover, our interpretable-AI framework allows analysis of learned masks of weights, yielding improved understanding of the role of underlying physical processes upon predictability. We find skin temperature and the Northern Hemisphere to be more important predictors of North Atlantic wintertime upper atmospheric winds than upper atmospheric winds 1-2 months prior and the Southern Hemisphere, respectively.
\end{abstract}
\maketitle

\section{Introduction}

Forecasting on S2S timescales, typically defined as 2 weeks to $\sim$2 months, is vital for public health, disaster preparedness, agriculture, and energy/water management \parencite{White:2017aa}. Despite the clear benefits of skillful predictions on these timescales, S2S forecasting remains especially difficult. Often referred to as a `predictability desert' \parencite{Robertson:2018ab,Chen:2024aa}, S2S forecasts cannot solely rely on the initial atmospheric conditions, as is often done in short-term numerical weather prediction, or on the slow-varying boundary conditions that underpin climate outlooks \parencite{Robertson:2018ab,Vitart:2018aa}. Instead, forecasters must integrate information from initial conditions, boundary conditions, and S2S modes of variability, like the Madden Julian Oscillation (MJO) \parencite{Zhang:2013aa}, to produce skillful predictions \parencite{Vitart:2018aa}. Still, on S2S timescales, the strength of these sources of predictability and their teleconnections remain unclear \parencite{Merryfield:2020aa,Vitart:2018aa} and skill, e.g. accuracy of summertime surface temperature prediction in North America, remains relatively low \parencite{Breeden:2022aa,Pegion:2019aa}. 

A variety of tools have been used to approach the S2S forecasting challenge. Dynamical models have slowly but steadily improved S2S forecast skill \parencite{Peng:2023aa} and data-driven approaches, like fully-AI models, can now forecast phenomena such as the North Atlantic Oscillation (NAO) and MJO at S2S lead times \parencite{ling2024fengwuw2sdeeplearningmodel,Chen:2024aa} with similar skill to dynamical models. To further improve forecasts, there has recently been a renewed focus on pinpointing climate states that represent times of enhanced predictability \parencite[e.g.,][]{Mariotti:2020aa,Mayer:2021aa, Albers:2019aa}. Identifying these ‘windows of opportunity’ is a potential approach to improve skill on S2S timescales by allowing forecasters to know when forecast uncertainty is high or when they can leverage these times of enhanced predictability for more accurate forecasts \parencite{Mariotti:2020aa}.

Here, we tackle S2S prediction by combining a variety of these methodologies and employing an AI-informed model-analog forecasting approach. Analog forecasting rests on the premise that climate states with similar initial conditions tend to evolve in a consistent manner \parencite[e.g.,][]{Lorenz:1969aa,Zhao:2016aa}. By identifying past states resembling current conditions, their subsequent evolution can offer plausible trajectories for future conditions. For a variety of forecasts, from the tropics to the northern high latitudes, analog forecasting has been shown to rival the skill of global climate models \parencite{Lou:2023aa,Ding:2019aa,Walsh:2021aa} all while offering several key advantages. Unlike fully-AI models, analog forecasting is intuitive, interpretable, and can uphold physical laws \parencite{Rader:2023aa,Ding:2018aa}; moreover, compared to global dynamic climate models, analog forecasting is highly computationally efficient \parencite{Ding:2019aa}. 

Analogs offer an interpretable, physical model that is helpful for diagnosing errors and probing physical drivers, while their fast computational speed allows for the quick generation of ensembles of forecasts. Creating proficient ensembles is a key way to improve skill on S2S timescales \parencite[e.g.,][]{Han:2023aa,Palmer:2004aa,Krishnamurti:1999aa}, provide probabilistic forecasts \parencite[e.g.,][]{Mullan:2006aa,Leutbecher:2008aa,Weisheimer:2014aa}, and even help explore windows of opportunity \parencite[e.g.,][]{Leutbecher:2008aa,Weisheimer:2014aa}—essential on S2S timescales. For instance, with a calibrated ensemble of forecasts, one can use ensemble member agreement as a sign of a lower forecast uncertainty to identify windows of opportunity \parencite[e.g.,][]{Ferranti:2018aa}. However, despite these advantages in computation and interpretability, successful analog forecasting hinges on having both a robust library of analogs and a reliable method to identify sufficiently similar past states. 

To address this need for a large analog library, we turn to climate models, which have orders of magnitude more climate realizations than we have observational data \parencite{Ding:2018aa,Wu:2023aa}. Yet, even with climate models, finding perfect analogs is impractical—estimates suggest over $10^{30}$ years of data would be needed to match two atmospheric flow stream patterns in just the Northern Hemisphere within observational error \parencite{VAN-DEN-DOOL:1994aa}. Hence, determining the conditions that make a climate state an adequately close analog, rather than a perfect one, is crucial. For example, \textcite{Ding:2018aa} use regional matching to identify close analogs for seasonal tropical Indo-Pacific Ocean prediction; \textcite{esd-13-1437-2022} use global matching for multi-decadal global predictions; and \textcite{Wu:2023aa} use area-specific matching for annual-to-multi-year Pacific Decadal Oscillation prediction. These methods for selecting analogs have been shown to work for certain problems, although they demand either a huge library of analogs (as in global matching) or depend on prior knowledge of physical drivers and teleconnections (as in regional or area-specific matching). 

Here, we explore an alternative, AI-based spatial weighting approach originally introduced by \cite{Rader:2023aa}. We train a neural network to output a mask of weights that highlights where it is most important for initial conditions to match, such that two states will evolve similarly. Using a learned set of weights to find optimal analogs reduces reliance on prior knowledge and enables investigation of which regions and variables are most essential for two climate states to follow similar future trajectories. This method of optimized analog forecasting was first successfully applied to annual-to-decadal sea surface temperature prediction \parencite{Rader:2023aa} and has since been extended to multi-year-to-decadal 2-meter temperatures  \parencite{fernandez2025multiyeartodecadaltemperaturepredictionusing} and seasonal-annual El Niño-Southern Oscillation (ENSO) predictions \parencite{toride2024usingdeeplearningidentify}. 

Here, we show that this AI-based model-analog forecasting approach can achieve skill beyond traditional model-analog methods on S2S timescales while maintaining interpretability and computational efficiency. We highlight the benefits of using AI-based analogs across two prediction tasks: 1) regional continuous prediction of Month 1 midwestern U.S. summer temperatures and 2) classification of Month 1-2 North Atlantic wintertime upper atmospheric winds. Through these predictions tasks we show the AI-based model-analog approach outperforms traditional model-analog forecasting approaches, climatology, and persistence on reanalysis data on S2S timescales, exhibiting especially strong performance for extreme temperature prediction. Moreover, via this interpretable AI-forecasting framework we analyze the learned masks of weights to better understand the S2S sources of predictability that underpin this improvement.

\section{Methods}
\subsection{Prediction Tasks}
We demonstrate the skill of our model-analog forecasting approach for the prediction tasks described in Table \ref{tab:climate_data}, opting for a varied pair of examples to test the generalizability of the method across different S2S prediction problems. We apply the AI-informed model-analog approach to both classification and continuous prediction tasks, to different regions, seasons, variables, and lead times. These prediction tasks each have a unique learned mask of weights to optimize the choice of analogs. While we focus on monthly prediction, we also find skill over traditional model-analog methods for week 3-4 prediction in Task \#3, included in the Supplemental Section \ref{sec:CA}.

\begin{table}[h!]
    \centering
    \renewcommand{\arraystretch}{1.5} 
    \resizebox{\textwidth}{!}{
    \begin{tabular}{lccc}
    \toprule
    \textbf{Prediction Task} & \textbf{1} & \textbf{2} & \textbf{3  (Supplemental)} \\
    \midrule
    \textbf{Region} & Midwestern U.S. & North Atlantic & Southern California \\
    \textbf{Data Frequency} & Monthly & Monthly & Smoothed Daily \\
    \textbf{Prediction Time} & Month 1 & Month 1-2 & Week 3-4 \\
    \textbf{Prediction Season} & July - Sept. & Dec. - Feb. & June Week 3 - Sept. Week 3 \\
    \textbf{Input $\to$ Target} & Skin Temp $\to$ Skin Temp & Skin Temp + U250 $\to$ U250 & Skin Temp $\to$ Skin Temp \\
    \textbf{Target Type} & Single Value & Field & Single Value \\
    \textbf{Classification/Value} & Continuous & Tercile Classification & Tercile Classification \\
    \bottomrule
    \end{tabular}
    }
    \caption{The three prediction tasks.}
    \label{tab:climate_data}
\end{table}

\subsection{AI-Informed Analog Approach}
\label{forecasting_approach}
Most traditional analog forecasting methods follow a similar approach: to predict how a certain climate state (referred to as a state of interest or SOI) will evolve, one finds the closest $k$ matches (for $k \ge 1$, in the analog) library. ``Closeness" is often measured by minimizing a distance measure, like mean-squared error (MSE), between the SOI and potential analogs either across the entire globe, or across a region of interest. One can then use the trajectories of these closest matches as a prediction for how the SOI will evolve into the future. Rather than predicting the evolution of the whole globe, one often evaluates the analog skill in a specific region of interest, which we refer to as the target region. 

\begin{figure}[ht]
    \centering 
    \includegraphics[width=\textwidth]{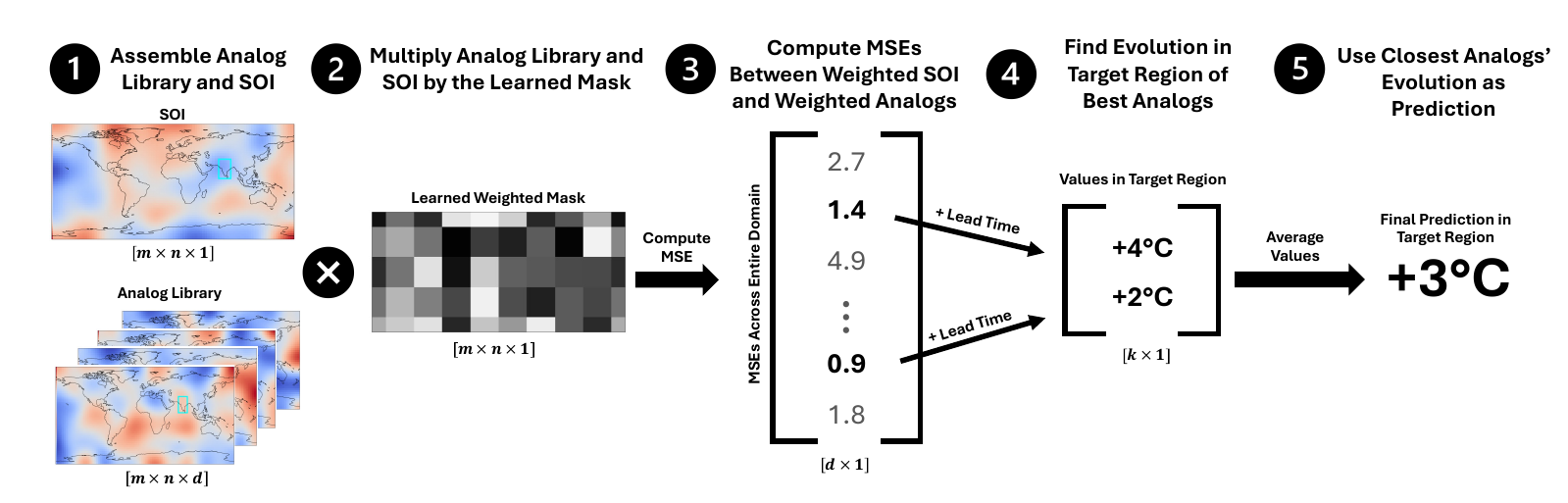}
    \caption{Schematic of the steps in the AI-informed analog approach: 1) Assemble the SOI and the library of $d$ potential analogs. 2) Multiply all potential analogs and the SOI by the learned mask of weights. 3) Compute the MSE between the weighted SOI and the $d$ weighted potential analogs, and select the $k$ closest analogs (in this example $k=2$). 4) Find the values of the analogs in the target region after the desired lead time. 5) Use the target field of the $k$ closest analogs' evolution in the target region as the prediction for the SOI (this example is a continuous prediction problem, so the mean value is taken).}
    \label{fig:analog_forecasting_schematic2}
\end{figure}

We take a similar approach, except here we utilize a soft mask of weights to measure closeness between the SOI and potential analogs. Prior to computing the distance measure between the SOI and each potential analog, we multiply the entire library (after seasonal subsetting) and the SOI by a learned mask of weights (Steps 1 and 2 in Figure \ref{fig:analog_forecasting_schematic2}). This mask, therefore, highlights or dampens the importance of conditions matching in certain areas of the globe for a potential analog to be considered close to the SOI. We then use MSE to compute the closest analogs after weighting, selecting the $k$ closest analogs (Step 3 in Figure \ref {fig:analog_forecasting_schematic2}). Lastly, we use the $k$ closest analogs' mean evolution (for continuous prediction problems) or majority vote (for classification) in the target region as our final prediction (Steps 4 and 5 in Figure \ref{fig:analog_forecasting_schematic2}).

\subsection{Data}
\label{sec:data}
We predict both U250 and skin temperature, relying on output from the Community Earth System Model 2–Large Ensemble (CESM2-LE) \parencite{Danabasoglu:2020aa} in order to have a sufficiently large analog library and to learn the weighted mask. As we will show, the AI-informed model-analog approach produces skillful predictions when evaluated on both CESM2-LE data, in a perfect-model framework, and on ECMWF Reanalysis v5 (ERA5) data \parencite{Hersbach:2020aa}. We make use of monthly-mean fields that are approximately resolved at 1.25° x .9° (natively for CESM2-LE data, and bilinearly interpolated for ERA5 data).  For all data sources, we convert the data to anomalies about the climatological seasonal cycle and then \change{to standard deviations across the subsetted season}{normalize} at each grid point\add{ by dividing by the standard deviation of the anomalies}. However, between data sources, we handle the anthropogenic effects of climate change slightly differently, as will be discussed next. 
\subsubsection{CESM2-LE Data}
\hfill\\
We use monthly CESM2-LE data from 1850-2100 that employs CMIP6 historical and SSP3-7.0 future radiative forcing scenarios \parencite{Simpson:2023aa}. We take all 100 members to calculate the ensemble mean, which we subtract from each individual member to both remove the effects of anthropogenic climate change and to convert the data to anomalies from the seasonal cycle. To increase speed and reduce memory load, we then use only a third of the members for training and the analog library. These members are divided between the analog library and SOIs, with fields from 19 members composing the library and fields from 14 members serving as the SOIs (see Table \ref{tab:monthly_ensemble_members} for member details). We partition the SOIs with a 10/2/2 member split for training, validation, and testing respectively.

\subsubsection{ERA5 Data}
\hfill\\
We use ERA5 data from January 1940 to September 2024. We fit and subtract a third-order polynomial at each grid point and each calendar month to define detrended anomalies from the seasonal cycle. The ERA5 data acts as a second test set to evaluate skill on observations. 

\subsection{Artificial Neural Network to Learn Mask of Weights}
\label{neural_network}
\begin{figure}[ht]
    \centering
    \includegraphics[width=\textwidth]{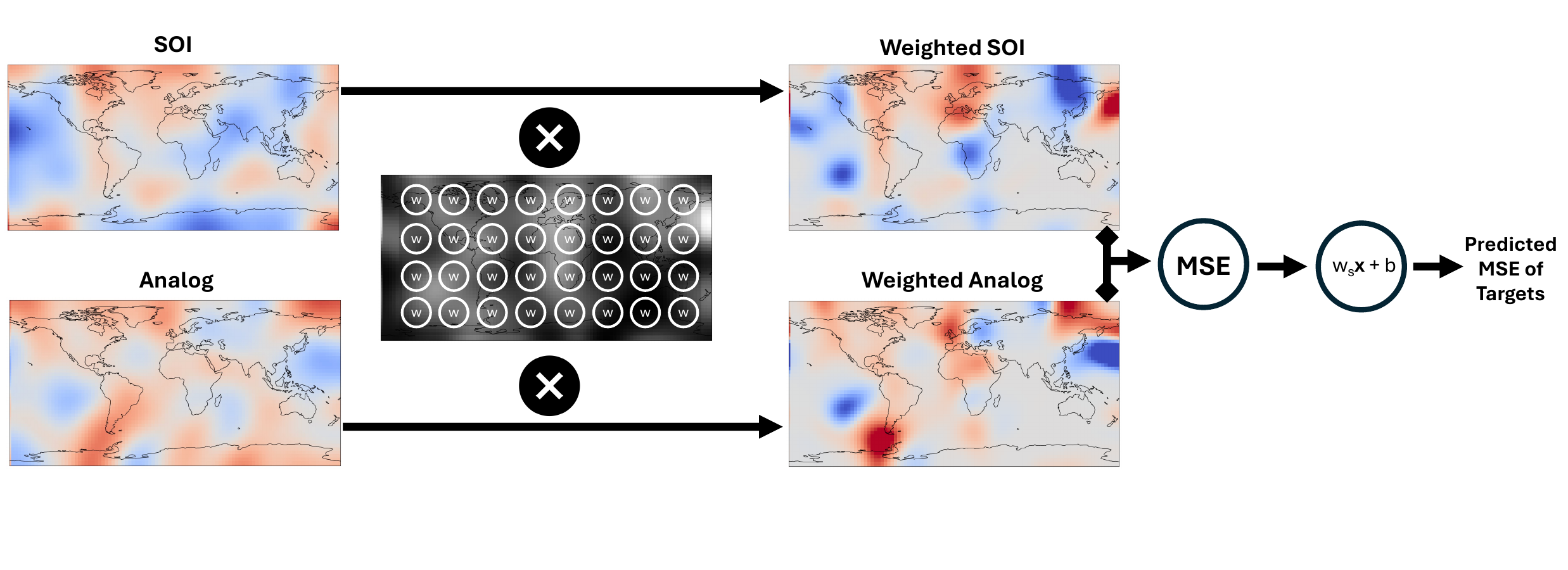} 
    \caption{Schematic of the neural network setup to learn the weighted mask. One SOI and one analog are multiplied by a layer of learnable weights. The MSE between the two weighted inputs is computed and passed through a linear scaling layer. This output represents the predicted difference in the two maps' targets. Loss is computed as the MSE between the predicted difference of the targets and the true difference of the targets.}
    \label{ann_schematic}
\end{figure}

The goal of the artificial neural network is to optimize the premise of analog forecasting; namely, that smaller differences in initial conditions lead to smaller differences in future conditions. To optimize this relationship between initial and future conditions, we task the network with predicting how similarly two states will evolve in the target region given their initial conditions. We employ a network that is similar to that of \textcite{Rader:2023aa}, with minor modifications to its final layers.  During each forward pass, the network, depicted in Figure \ref{ann_schematic}, takes two maps as input and uses these initial conditions to predict the similarity of their future states. The SOI map is from the training set and the analog map is randomly selected from the analog library. These maps are both multiplied by a grid of learnable weights (i.e., the mask) of the same size as the inputs, resulting in two weighted maps. The mask is restricted to have a mean  of 1 across all weights, such that during training weight is moved between different areas of the globe, but conserved. The MSE between these two weighted maps is calculated, representing the effective similarity of the two initial states  for this prediction task, and is passed through a single linear scaling layer. The output of this layer represents the network's prediction of the MSE between the two maps in the target region after they have evolved (i.e., after the desired lead time). Loss is computed as the MSE between the predicted difference of the targets and the true difference of the targets. Hence, the primary objective of the network is to align through this weighting process, the network learns a mask that aligns the MSE between two states' inputs to their MSE after evolution in the target region. This process is repeated for each SOI in the training set. Details of the network setup and hyperparameters can be found in Table \ref{tab:hyperparameters}.

Our network deviates from that of \textcite{Rader:2023aa}, in that we use a single linear layer instead of multiple dense layers at the end of the network.  We restrict the linear layer's weight to be $ \ge 0$ to ensure a monotonically increasing relationship between the MSE of the maps' weighted inputs and the predicted MSE of the maps' target regions. This better matches our process for selecting analogs as described in Section \ref{forecasting_approach}, as we expect that two maps with a smaller weighted MSE will also evolve to have a smaller MSE between their targets. This switch to a single linear layer resulted in a negligible change in skill, but increased network parsimony and training speed.

\subsection{Metrics}
We employ deterministic and probabilistic error metrics for each type of prediction task (i.e., classification and continuous prediction). For continuous prediction (Task \#1) we compute mean absolute error (MAE) and continuous ranked probability score (CRPS). 

\noindent MAE is defined as
\begin{equation}
    MAE = \frac{1}{N} \sum_{i=1}^{N} |f_i - o_i|
    \label{eq:mae}
\end{equation} where \( N \) is the number of samples, \( f_i \) is the predicted value for sample \( i \), and $o_i$ is the true value for sample $i$. 

\noindent CRPS is defined as
\begin{equation}
    CRPS(F, x) = \int_{-\infty}^{\infty} \left( F(y) - H(y - x) \right)^2 dy
    \label{eq:crps}
\end{equation} 
where \( F(y) \) is the cumulative distribution function of the forecast, \( x \) is the true value, and \( H(y - x) \) is the Heaviside step function, which is 0 for \( y < x \) and 1 for \( y \geq x \). CRPS ranges from 0 (for a perfect forecast) to $\infty$.

\noindent For classification (Tasks \#1 and \#3), we compute misclassification rate and the multi-class Brier Score (BS). 

\noindent Misclassification rate is defined as
\begin{equation}
    \mathrm{Error\ Rate} = \frac{\mathrm{Number\ of\ Incorrect\ Classifications}}{\mathrm{Total\ Number\ of\ Predictions}}
    \label{eq:error_rate}
\end{equation}
\noindent BS is defined as
\begin{equation}
BS = \frac{1}{N} \sum_{i=1}^{N} \sum_{k=1}^{K} \left( f_{ik} - o_{ik} \right)^2
\label{eq:brier}
\end{equation} 
where \( N \) is the number of samples, \( K \) is the number of classes, \( f_{ik} \) is the predicted probability for class \( k \) for sample \( i \), and \( o_{ik} \) is the true value (1 if the true class is \( k \), otherwise 0). BS ranges from 0 (for a perfect forecast) to 2.

\noindent We convert all types of error to skill scores by comparing them to the error of a climatological forecast:
\begin{equation}    \label{eq:Skill_Score}
    \mathrm{Skill\ Score} = 1 - \frac{Error}{Error_{climatology}}
\end{equation}
 All skill scores are strictly $\le 1$, with a skill score of 1 indicating perfect skill and a skill score of 0 indicating equal skill to a climatological forecast. A negative skill score indicates worse skill than a climatological forecast.

\subsubsection{Baselines}
We include a regional and a global analog baseline in addition to persistence, climatological, and random baselines to evaluate the relative skill of the learned mask approach. To create a global baseline, we select analogs by matching conditions over the entire globe (equivalent to a weighted mask of 1s everywhere). We create a regional baseline by selecting analogs via matching conditions only in the target region (equivalent to a weighted mask of 1s in the target region and 0s everywhere else). The 90th percentile random baseline is formed by repeating the prediction for all SOIs using random analogs 100 times and selecting the 90th percentile of best predictions. \add{Furthermore, we include a limited comparison of analogs in Task \#1 to operational S2S models in Supplemental Section \protect{\ref{sec:model_baselines}} as a reference point.}

\section{Results}

\subsection{Month 1 Temperature Extremes Over the Midwestern U.S.}
\begin{figure}[ht]
    \centering
    \includegraphics[width=\textwidth]{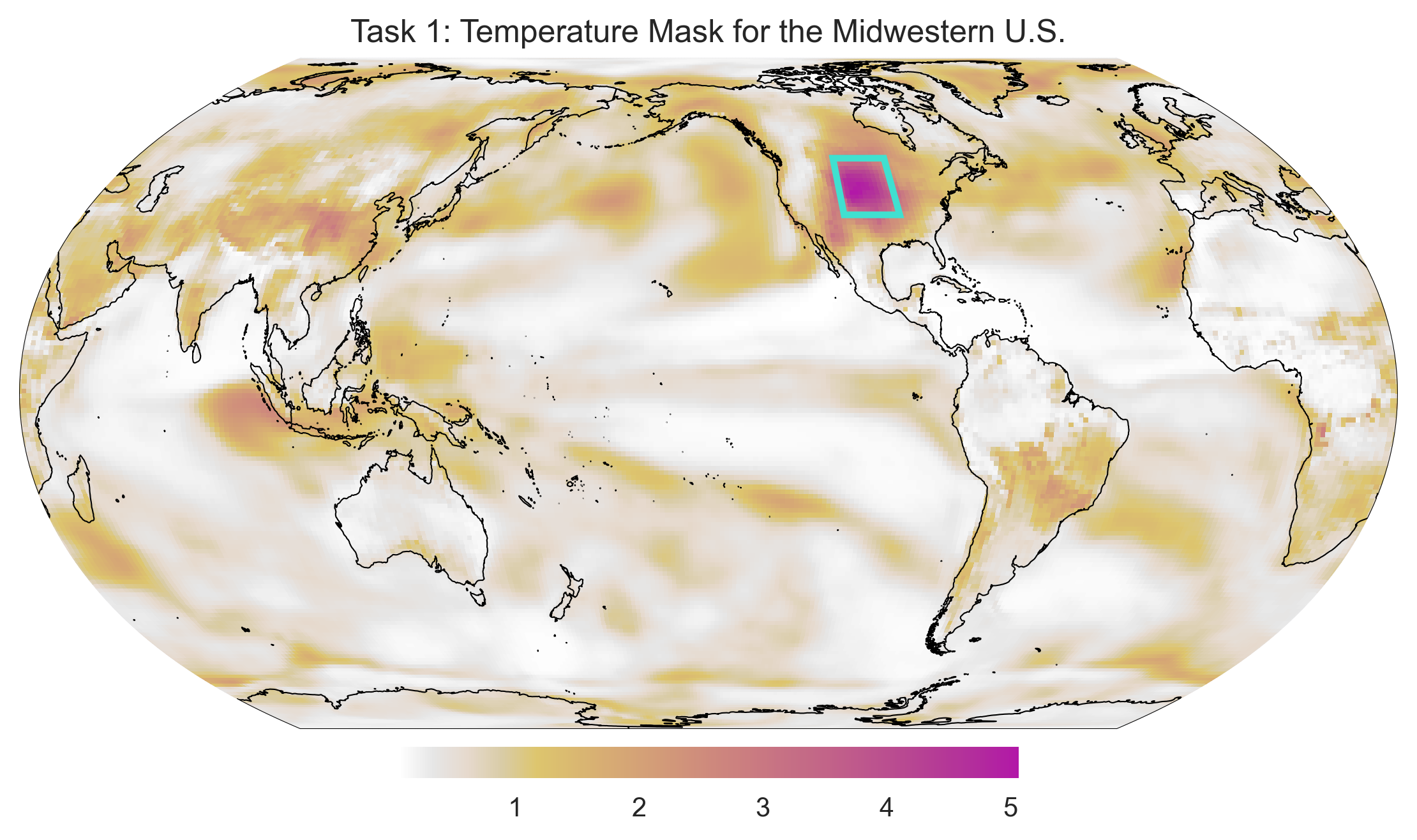}
    \caption{The learned mask for Task \#1, midwestern U.S. summer temperatures. The cyan box outlines the target region.}
    \label{fig:midwest_mask}
\end{figure}

We first explore how well the AI-informed model-analog forecasting approach can perform continuous prediction, by assessing monthly summer midwestern U.S. ($36^\circ-49^\circ \mathrm{N},\ 90^\circ-106^\circ \mathrm{W}$) temperatures (Task \#1) with a focus on extremes. We include this focus on extreme heat prediction, as the midwestern U.S. experiences some of the highest heat index events in the country \parencite[e.g.,][]{Romps_2022}. We predict the \change{temperature (in units of standard deviations---$\sigma$)}{normalized temperature anomalies} each month from July through September, using the learned mask in Figure \ref{fig:midwest_mask}. The mask displays a strong emphasis on the target region, highlighting the regional importance of the central  U.S. for predicting midwestern summer temperatures, and preferential weighting in the mid-latitudes of the Northern Hemisphere as well as the Maritime Continent. The Maritime Continent signal is reminiscent of the MJO, which is influential in Midwest summer climate \parencite{Midwest_MJO}, while the mid-latitude pattern resembles wave trains correlated with summertime North American heatwaves \parencite{Yu2023}. The local signal in the central U.S. itself may be a result of the strong summertime land-atmosphere and soil moisture-temperature coupling in this region \parencite{SummerLandAtmosphereCouplingStrengthintheUnitedStatesComparisonamongObservationsReanalysisDataandNumericalModels}.

\begin{figure}[ht!]
    \centering
    \setlength{\fboxrule}{0.1pt} 
    \setlength{\fboxsep}{1pt}    
    \fbox{
        \begin{minipage}{0.97\textwidth} 
            \centering
            \begin{subfigure}{0.49\textwidth}
                \centering
                \includegraphics[width=\textwidth]{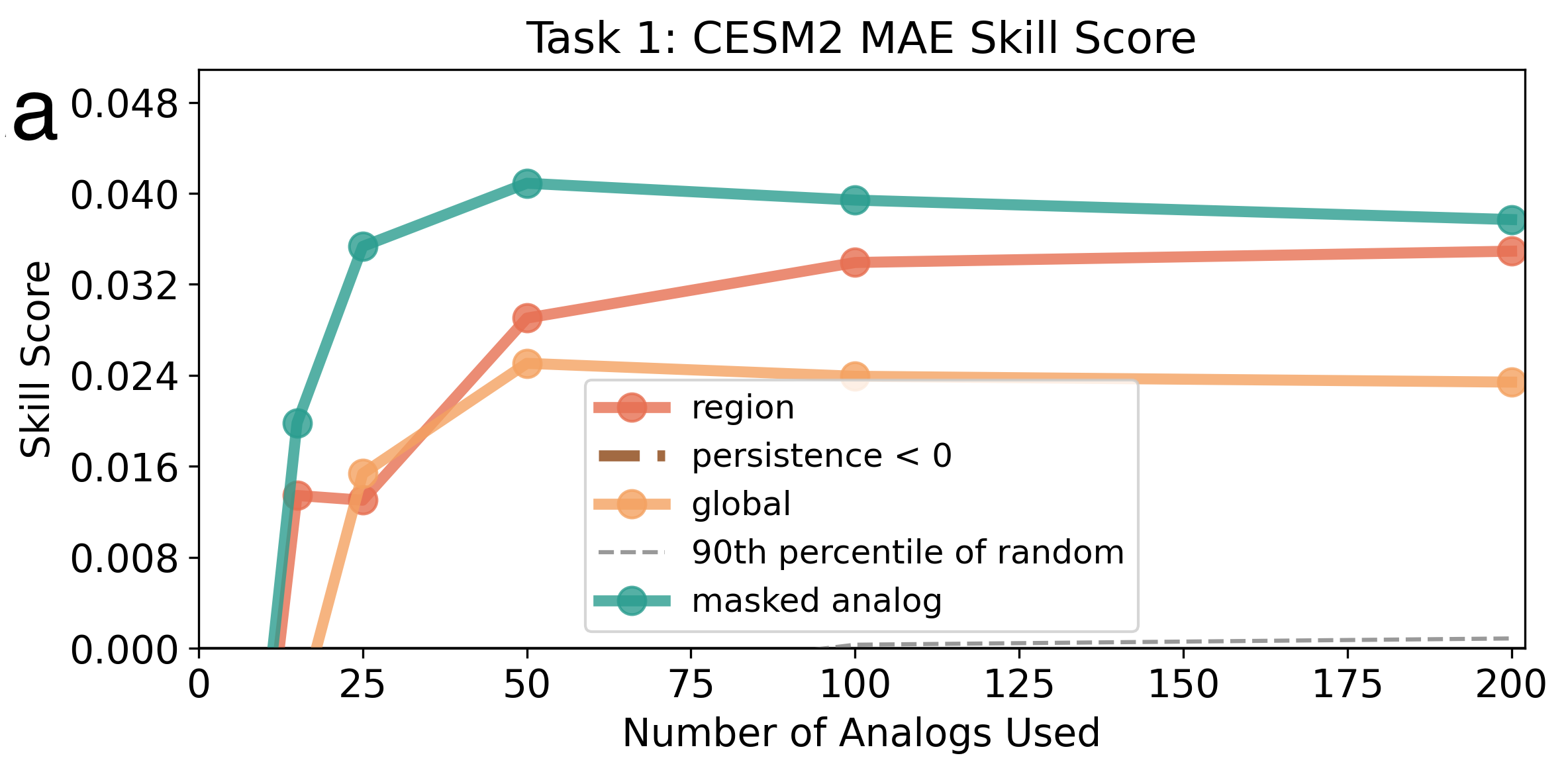}
            \end{subfigure}
            \hfill
            \begin{subfigure}{0.49\textwidth}
                \centering
                \includegraphics[width=\textwidth]{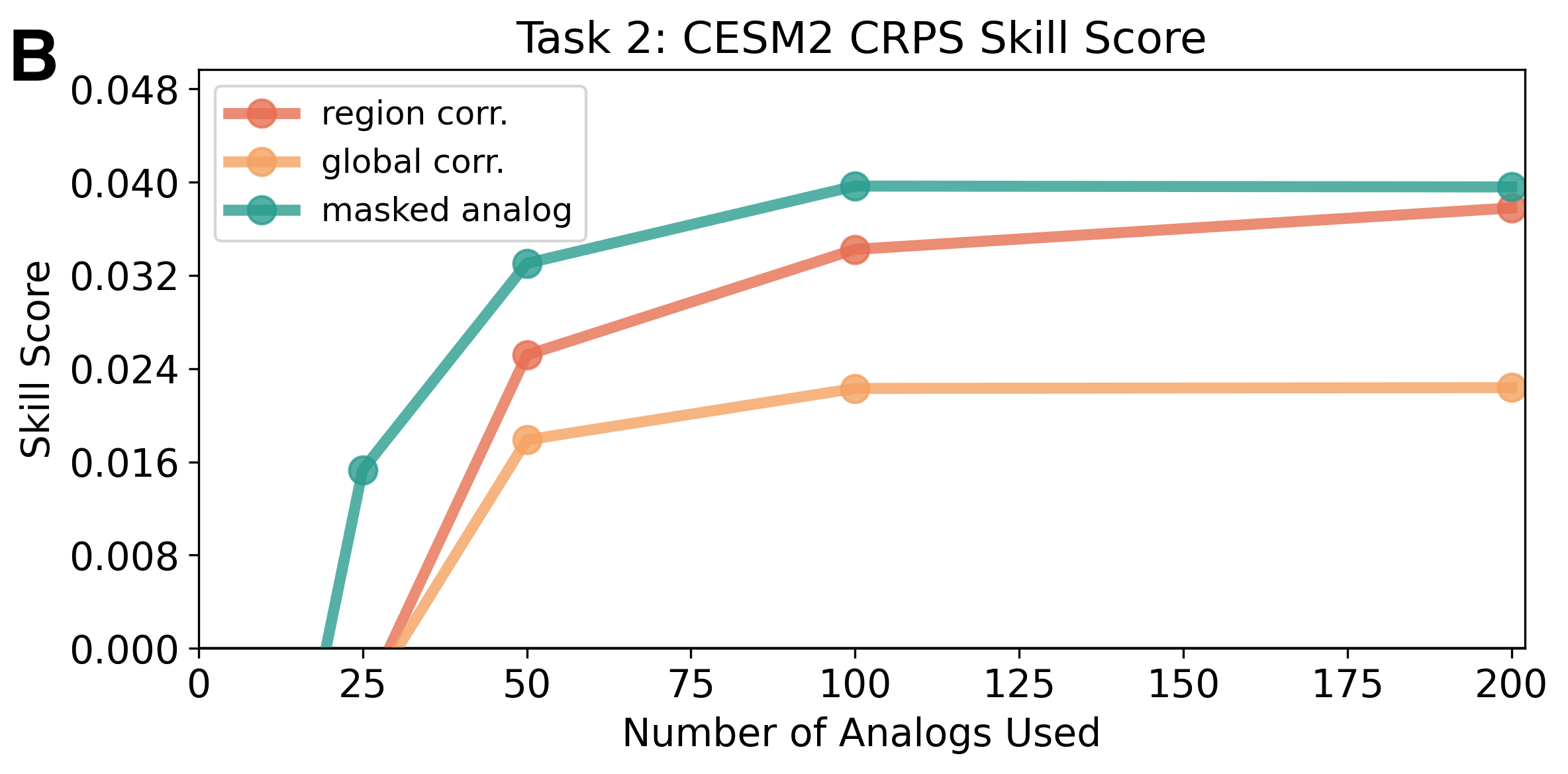}
            \end{subfigure}

            \vspace{2mm} 

            \begin{subfigure}{0.49\textwidth}
                \centering
                \includegraphics[width=\textwidth]{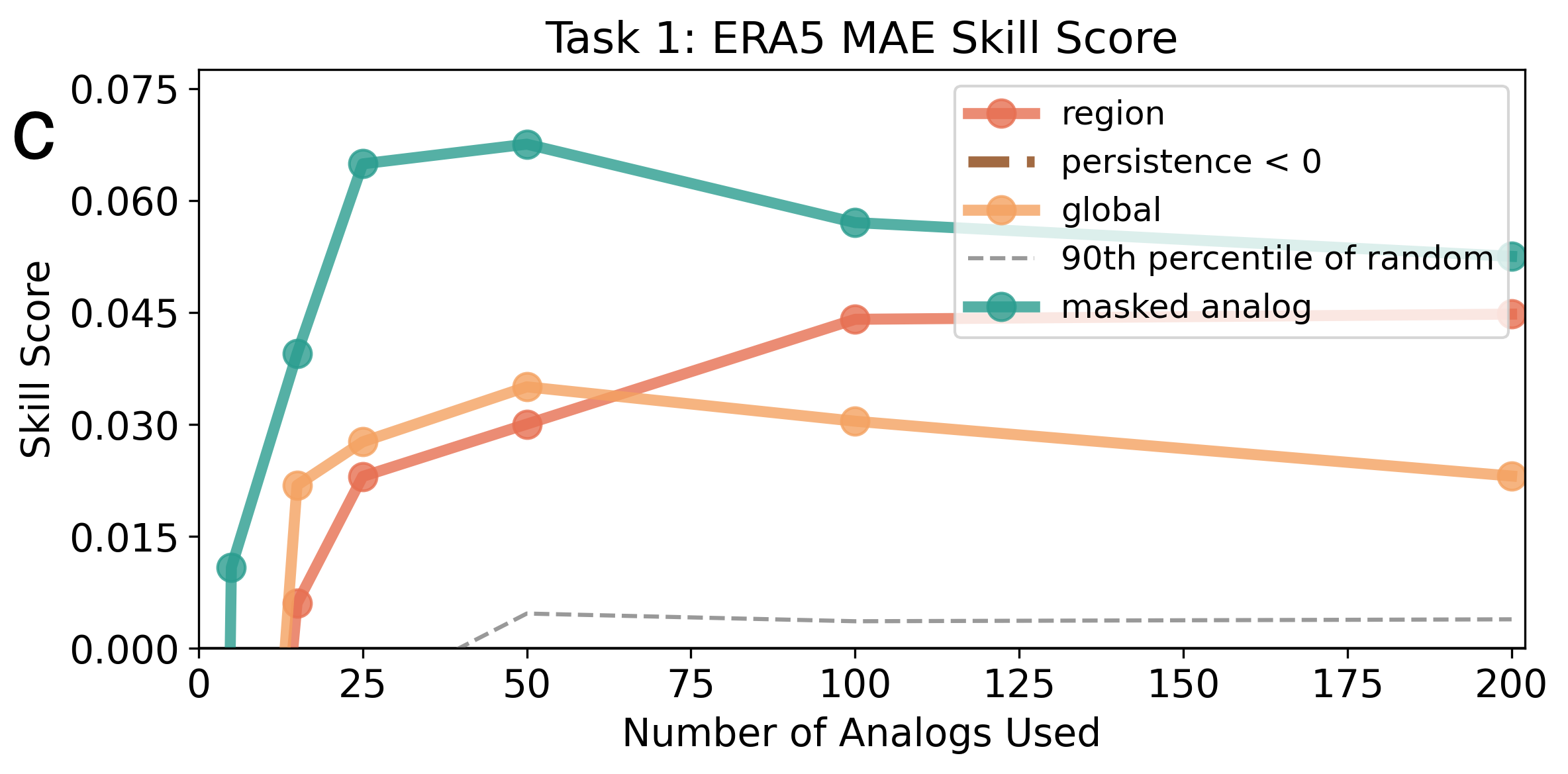}
            \end{subfigure}
            \hfill
            \begin{subfigure}{0.49\textwidth}
                \centering
                \includegraphics[width=\textwidth]{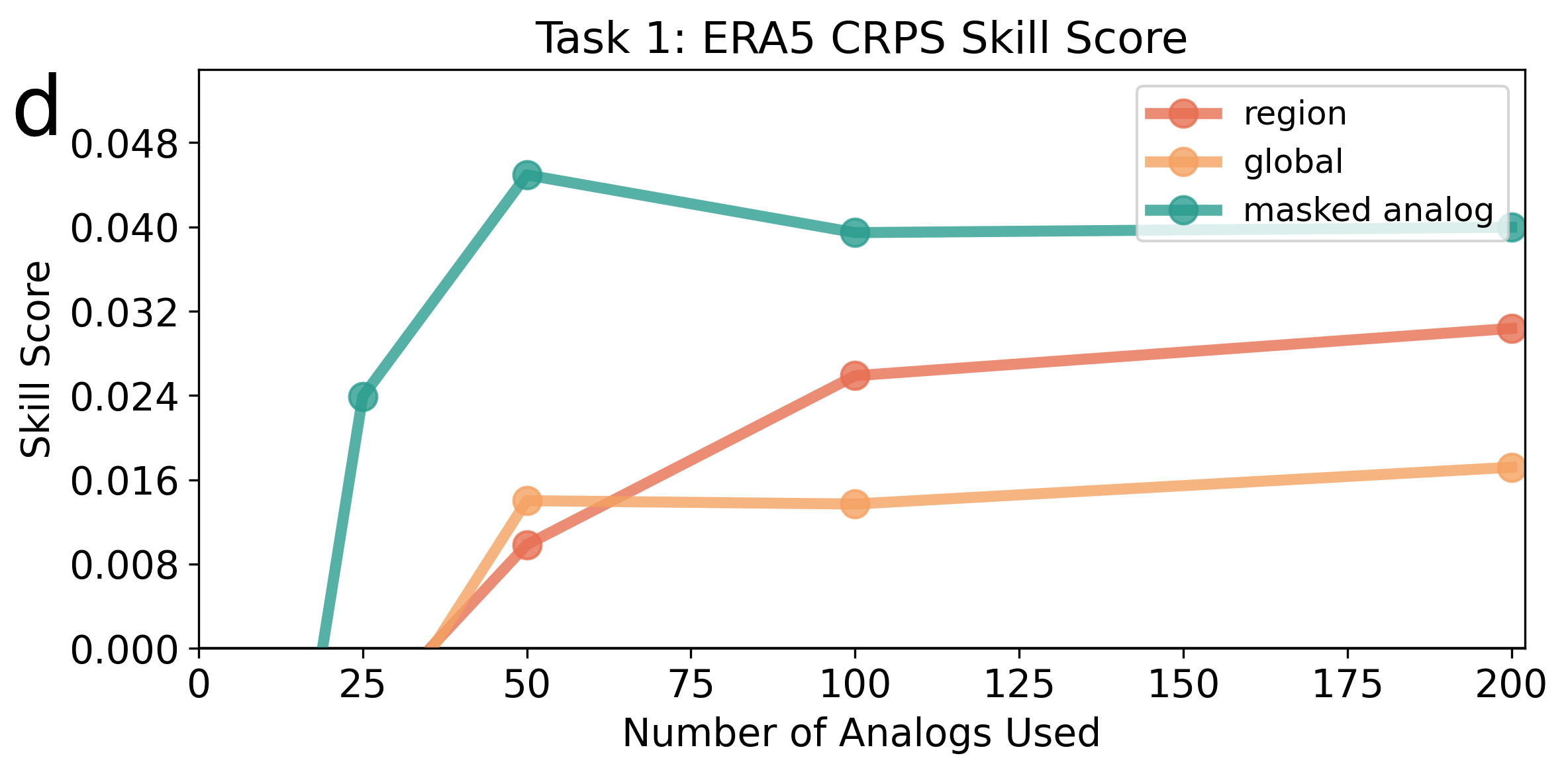}
            \end{subfigure}
        \end{minipage}
    }
    \caption{Skill scores for a) CESM2-LE MAE, b) CESM2-LE CRPS, c) ERA5 MAE, and d) ERA5 CRPS for Month 1 midwestern U.S. temperatures.}
    \label{fig:midwest_skill}
\end{figure}

 The learned mask modestly outperforms all baselines (Figure \ref{fig:midwest_skill}), with MAE skill increases of 17\% and 51\% and CRPS skill increases of 5\% and 48\% tested on CESM2-LE and ERA5, respectively. All skill scores peak at 50 analogs for both CESM2-LE and ERA5 data, except for CESM2-LE CRPS, which peaks at 100 analogs. This overall improvement in temperature forecasting on S2S timescales is important, however, better prediction of extreme temperatures in particular has an outsized impact on enhancing agricultural production, public health, and energy management \parencite{Domeisen:2022aa}. Thus, we focus on assessing the AI-based analog's ability to predict extreme temperatures. Here, we utilize a discard plot, in which we progressively discard samples with lower extremity to visualize how MAE skill changes for more extreme samples. We denote extremity simply as the absolute value of the prediction, i.e. a measure of how far from climatology the prediction is.

\begin{figure}[ht]
    \centering
    \includegraphics[width=\textwidth]{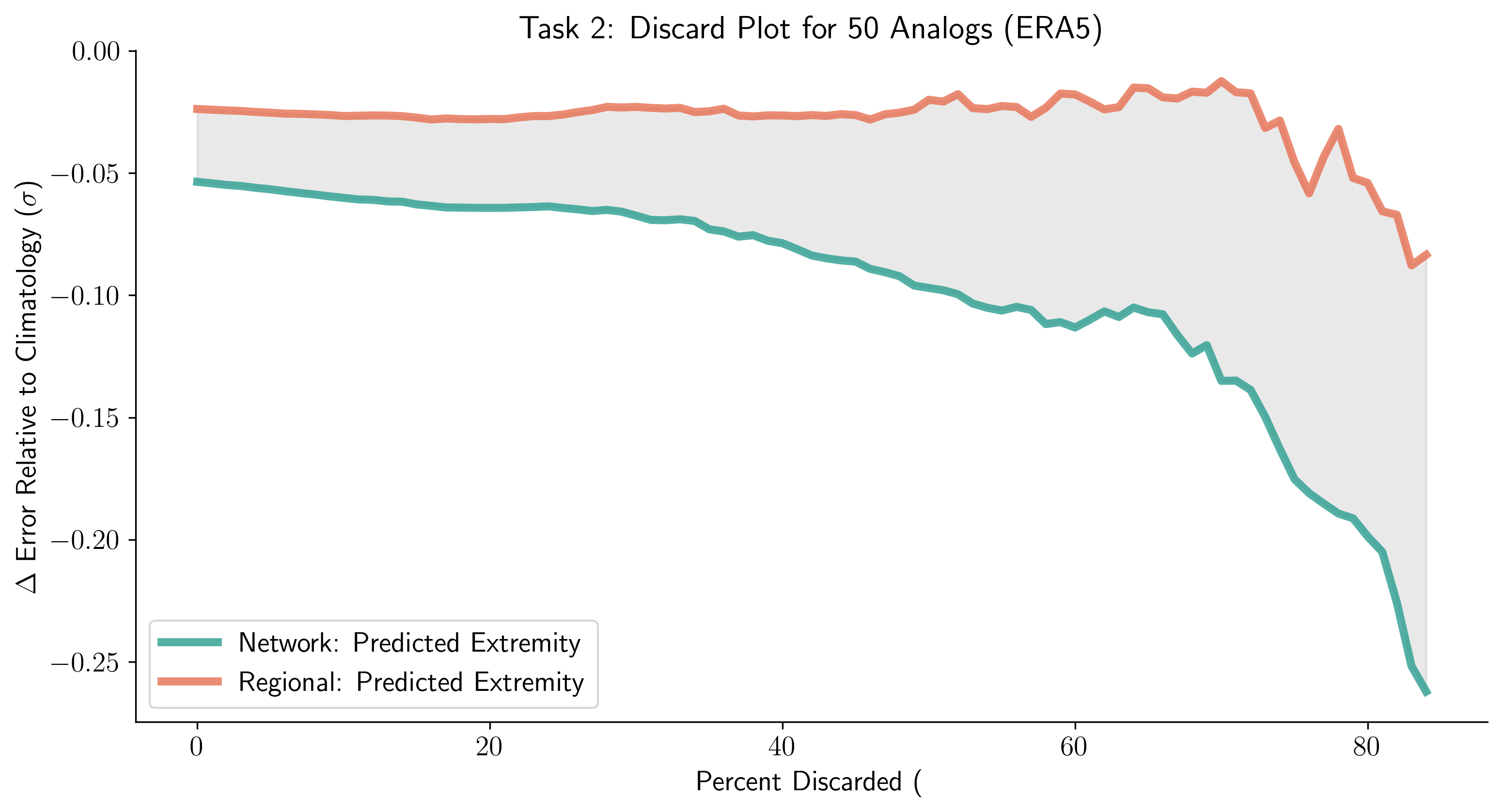}
    \caption{Discard plot based on predicted extremity with an ensemble of 50 analogs for midwestern U.S. summer temperatures, testing on ERA5 data. Data with the lowest extremity is progressively discarded, with the x-axis showing the percentage of data remaining.}
    \label{fig:midwest_discard}
\end{figure}

In Figure \ref{fig:midwest_discard}, we show the discard plot with ERA5 data using an ensemble of 50 analogs. The AI-based model-analogs exhibits a marked increase in skill for samples with more extreme predictions. As our skill score is defined relative to climatology, it may be unsurprising that the AI-informed model-analogs would have lower relative error on more extreme events. However, this is not the case for the regional baseline, where there is only a slight increase in skill for the most extreme samples. This analysis highlights how the skill gap between AI-informed model-analogs and traditionally-selected model-analogs widens for more extreme temperature events. Moreover, as we use \textit{predicted} extremity to discard samples, this information is available \textit{a priori}, allowing forecasters to better understand when the analog ensemble is likely to perform best and building trust in its more extreme predictions. This behavior also holds for CESM2-LE data (Figure \ref{fig:midwest_cesm2_discard}).
\subsection{Month 1-2 Winds in the North Atlantic}
\begin{figure}[ht]
    \centering
    \includegraphics[width=\textwidth]{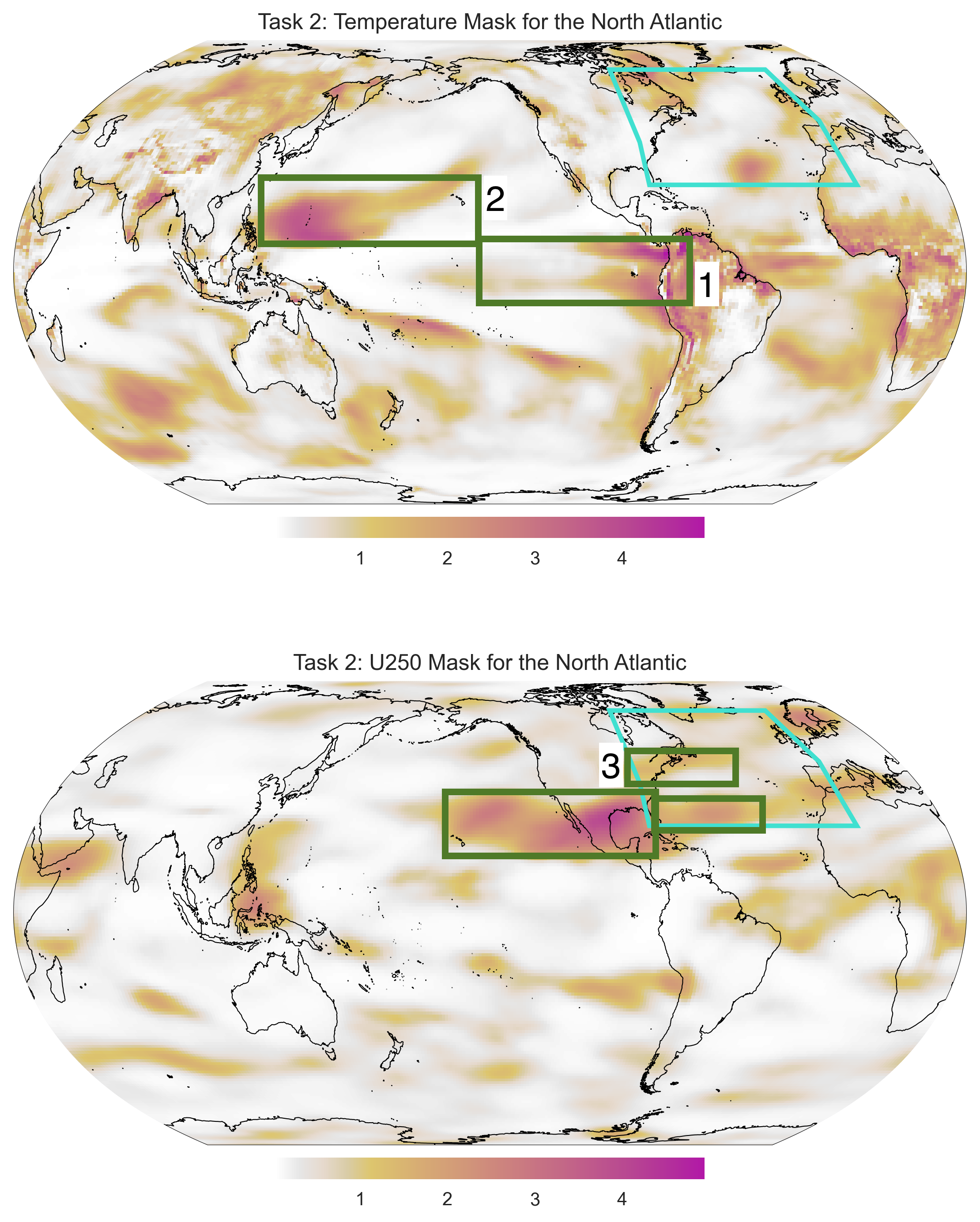}
    \caption{The learned mask for Task \#2, North Atlantic winter U250 classification. The cyan box outlines the target region. The dark green boxes outline areas of high weight in the mask.}
    \label{fig:atlantic_mask}
\end{figure}

Next, we explore the learned mask's ability to perform grid-point classification of upper atmospheric winds in the North Atlantic ($25^\circ-48^\circ \mathrm{N},\ 0^\circ-80^\circ \mathrm{W}$) and probe the mask itself to better understand the relative importance of different areas for successful prediction. At each grid point in the target region (rather than averaging across the target region), we classify the 250 hPa zonal wind (U250). The three target classes are formed by splitting the target temperatures into terciles, ensuring all classes are equally sized. Terciles for classifying the analog library are determined using the data within the analog library and terciles for the test set are defined based on the data in the test set to limit the impact of CESM2-LE biases relative to ERA5. We make predictions for December-January and January-February, using the learned mask in Figure \ref{fig:atlantic_mask}. We chose to examine the winter winds, as the jet stream variability in this region is largest during this time \parencite[e.g.,][]{Hall2017}. In this case, we select analogs using both U250 and surface temperature as inputs. Therefore, we learn a unique mask for each field, although, importantly, these masks are learned together by the network. As with Tasks \#1 and \#3, we include the predictand as an input variable both for predictive power (e.g., to capture jet stream information) and to indirectly allow for the encoding of persistence. Here we also include skin temperature as an additional input variable, as it encodes signal from ENSO and the MJO which can drive North Atlantic atmospheric variability \parencite[e.g.,][]{ENSO_NAO,NAO,MJO_NAO_teleconnections, TheConsistencyofMJOTeleconnectionPatternsAnExplanationUsingLinearRossbyWaveTheory}. Thus, despite the drawback in increased memory load, we find including both U250 and skin temperature are useful inputs, as we discuss further in \ref{sec:physical_interpretation_SST}.

\begin{figure}[ht!]
    \centering
    \setlength{\fboxrule}{0.1pt} 
    \setlength{\fboxsep}{1pt}    
    \fbox{
        \begin{minipage}{0.97\textwidth} 
            \centering
            \begin{subfigure}{0.49\textwidth}
                \centering
                \includegraphics[width=\textwidth]{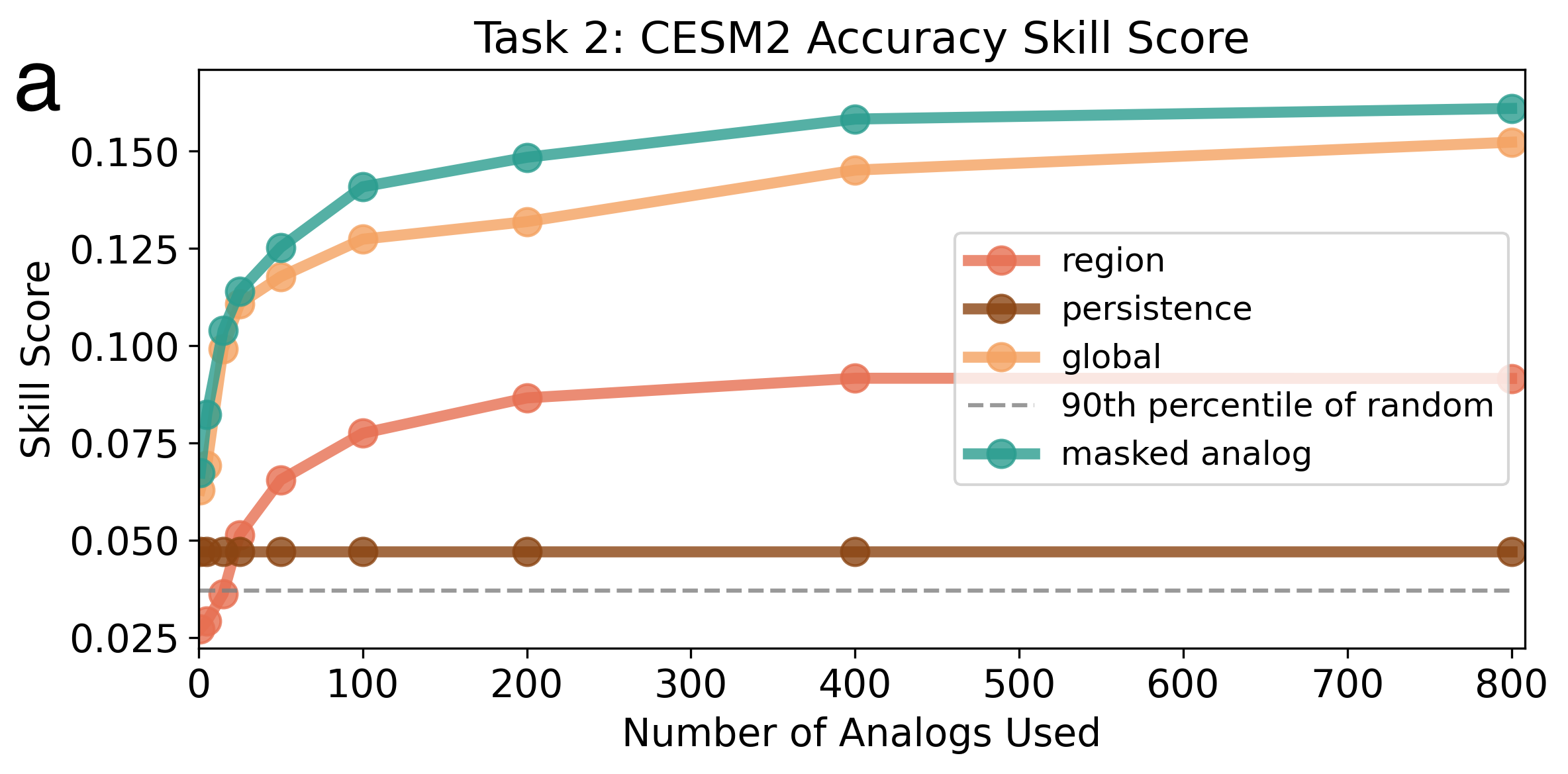}
            \end{subfigure}
            \hfill
            \begin{subfigure}{0.49\textwidth}
                \centering
                \includegraphics[width=\textwidth]{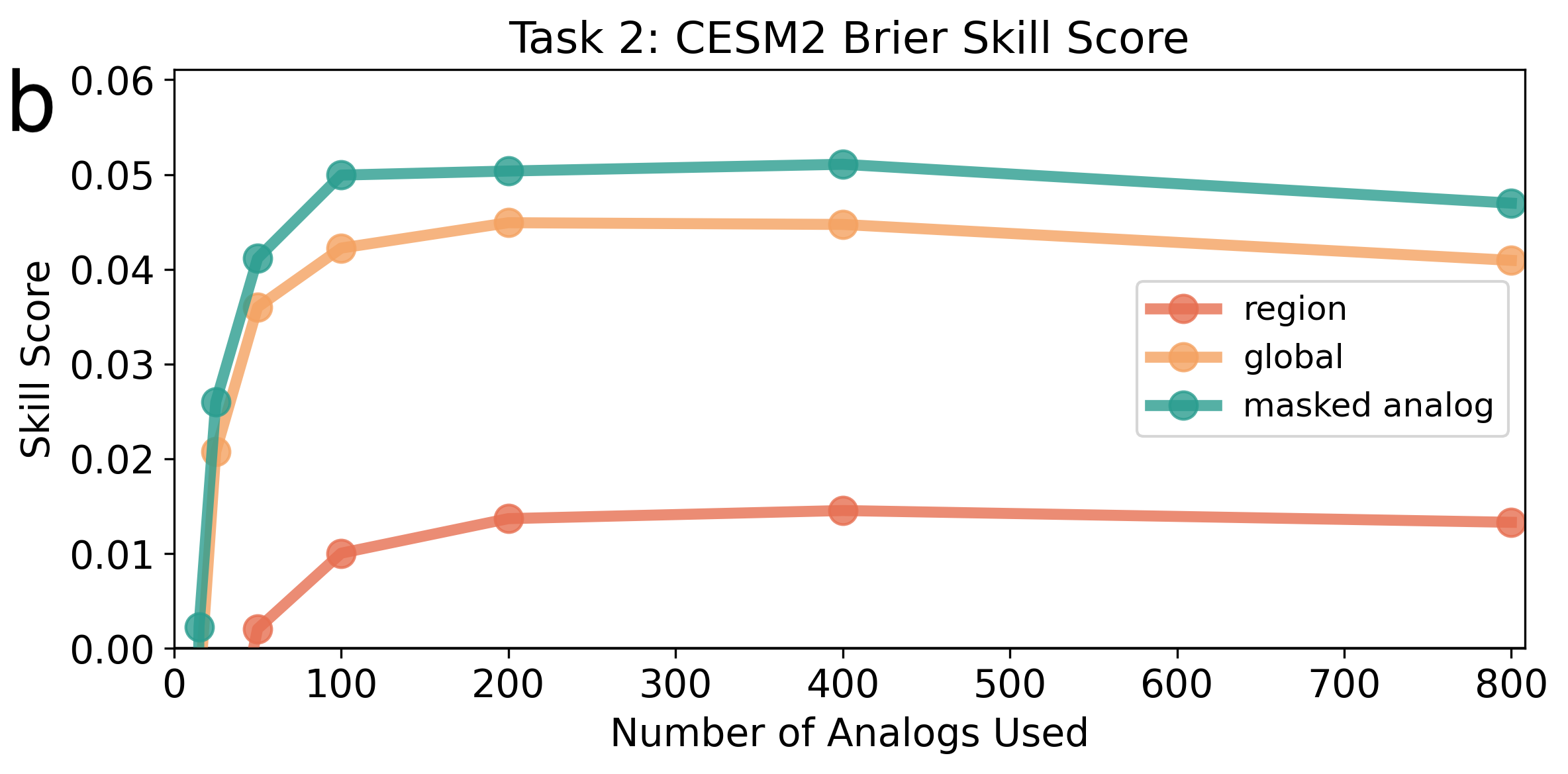}
            \end{subfigure}

            \vspace{2mm} 

            \begin{subfigure}{0.49\textwidth}
                \centering
                \includegraphics[width=\textwidth]{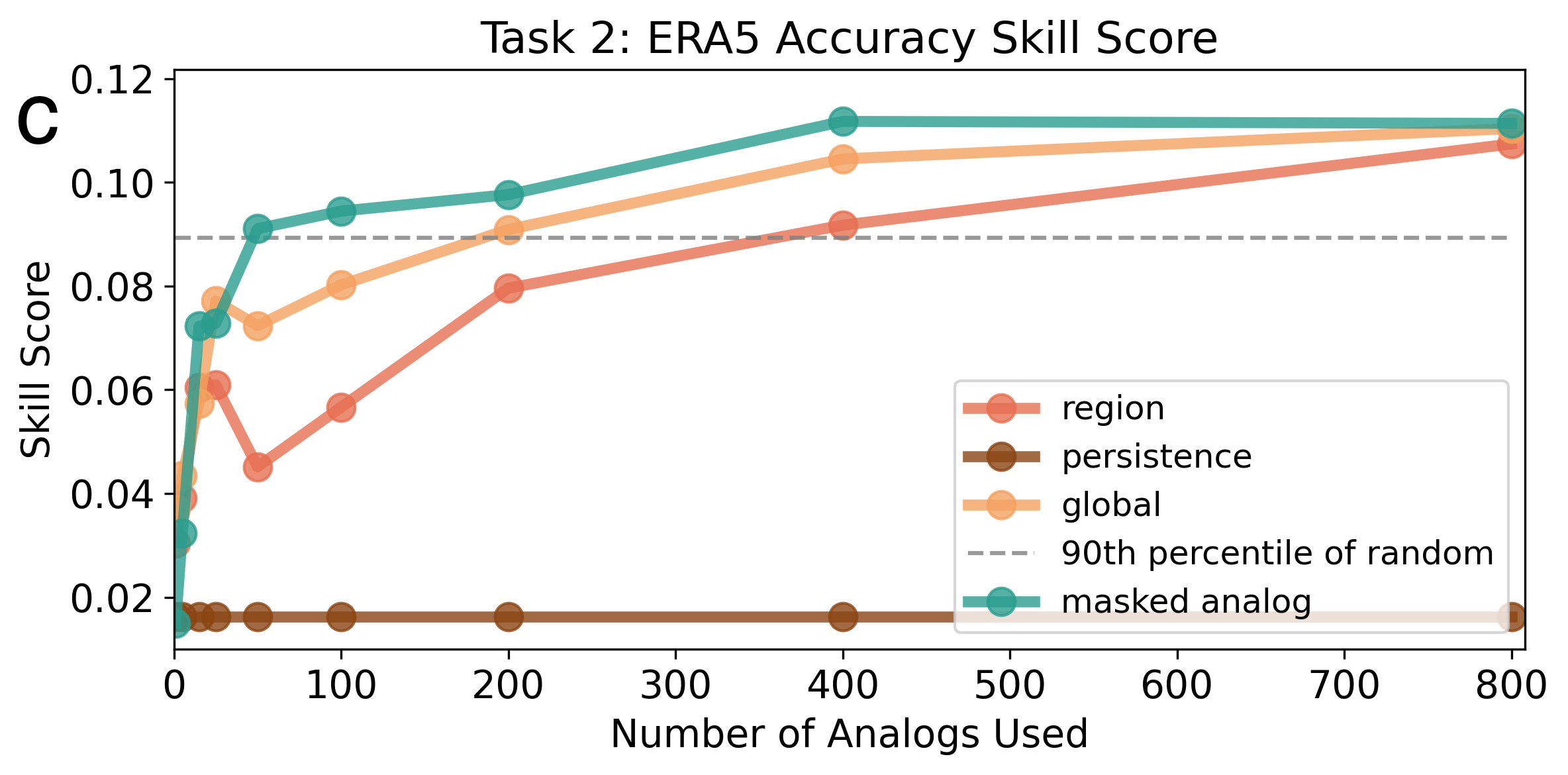}
            \end{subfigure}
            \hfill
            \begin{subfigure}{0.49\textwidth}
                \centering
                \includegraphics[width=\textwidth]{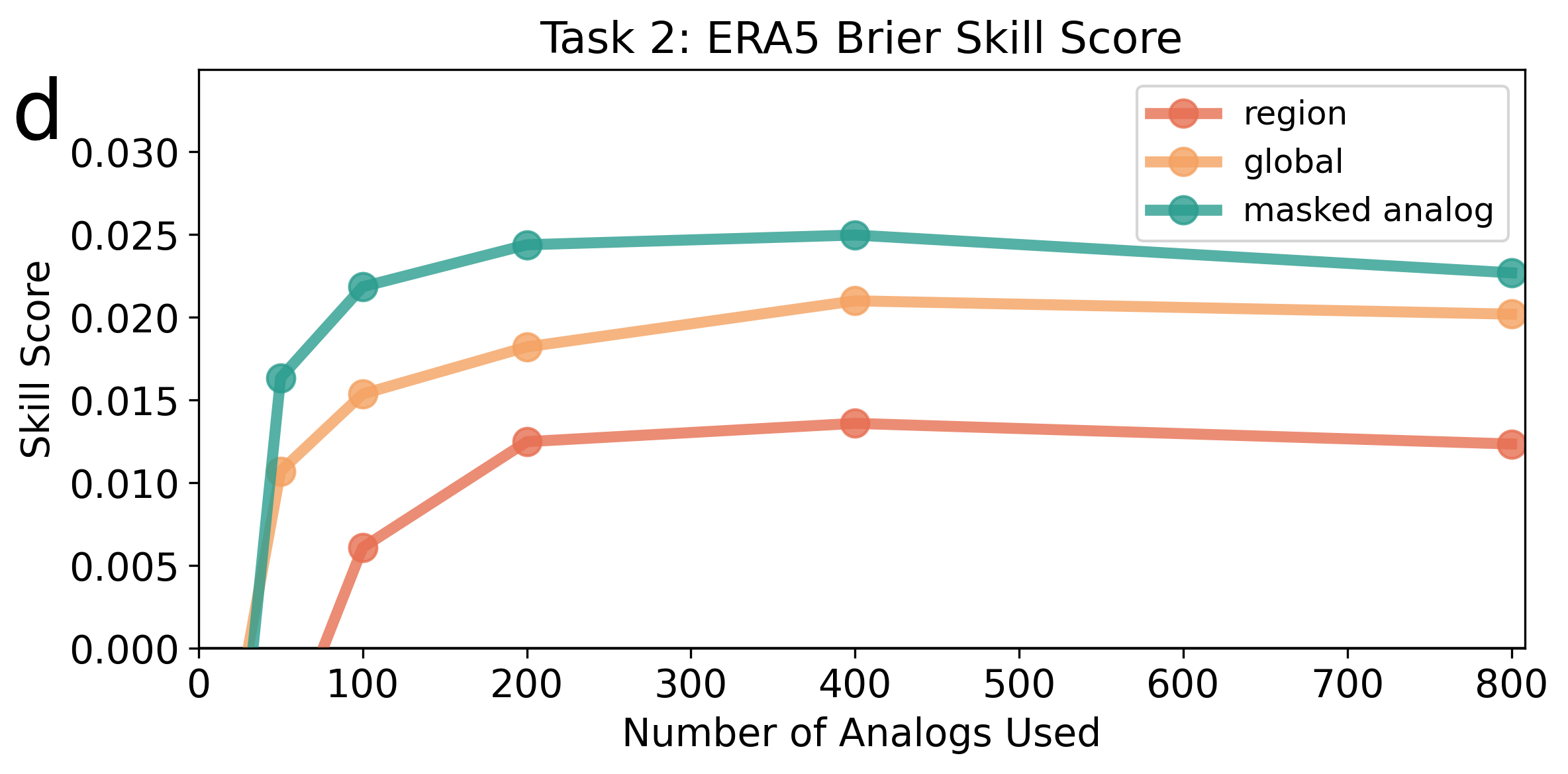}
            \end{subfigure}
        \end{minipage}
    }
    \caption{Skill scores for a) CESM2-LE accuracy, b) CESM2-LE BS, c) ERA5 accuracy, and d) ERA5 BS for Month 1-2 North Atlantic U250 classification.}
    \label{fig:atlantic_skill2}
\end{figure}

We evaluate skill at each grid point in the whole field (e.g., Figure \ref{fig:NA_field_skill}), summarizing these with mean skill scores over the target region (Figure \ref{fig:atlantic_skill2}). The learned mask outperforms all baselines for this grid-point-by-grid-point classification, with accuracy skill increases of 6\% and 1\% and BS skill increases of 19\% and 13\% tested on CESM2-LE and ERA5, respectively. Skill peaks at 400 analogs for both CESM2-LE and ERA5 data, except for CESM2-LE classification, which peaks at 800 analogs. The numbers of analogs in the ensembles are much higher than in Task \#1, since we have moved from a continuous prediction problem to a 3-class classification problem. With a continuous prediction problem, if the number of analogs in the ensemble is too high, the ensemble mean will converge to climatology. An example of this regression to the mean can be found in Figure \ref{fig:regression_to_mean}. With a majority vote classification problem, however, the ensemble can be larger without this issue, as the majority vote will not converge to climatology.

\subsubsection{Physical Interpretation of the Learned Mask}\mbox{}\\
\label{sec:physical_interpretation_SST}
Using this interpretable AI-informed model-analog approach, we can analyze the learned mask to better understand the physical drivers behind analog predictions. Here we discuss three areas of high weight in the mask, shown in Figure \ref{fig:atlantic_mask}: 1) SSTs in the eastern tropical Pacific, 2) SSTs in the Phillipine Sea, and 3) U250 in the Northern Hemisphere. Area 1 resembles the canonical ENSO region. We find this reflected in the analog selection, with selected analogs generally have more similar ENSO states than a random selection (Figure \ref{fig:analog_vs_random_nino34}), as measured by Niño-3.4 indices \parencite{AVerifiedEstimationoftheElNioIndexNio34since1877}. This is in line with work that finds ENSO impacts the North Atlantic region through its influence on the North Atlantic Oscillation (NAO)—the most prominent pattern of atmospheric variability in the region \parencite{ENSO_NAO,NAO}. Area 2 resembles a prominent Rossby wave source region, which is a known source of MJO teleconnections to the North Atlantic \parencite{MJO_NAO_teleconnections, TheConsistencyofMJOTeleconnectionPatternsAnExplanationUsingLinearRossbyWaveTheory}. Lastly, Area 3 highlights the Pacific jet exit region of the subtropical jet stream. The region of high weight over the Eastern Pacific and Mexico aligns with the mean position of the subtropical jet, and both its strength and the two North–South shifted branches, shown in smaller green boxes in Figure \ref{fig:atlantic_mask}, have been found to vary with the NAO state \parencite{JET_NAO}. 

\begin{figure}[htb!]
    \centering
    \setlength{\fboxrule}{0.1pt} 
    \setlength{\fboxsep}{1pt}    
    \fbox{
        \begin{minipage}{0.97\textwidth} 
            \centering
            \includegraphics[width=\textwidth]{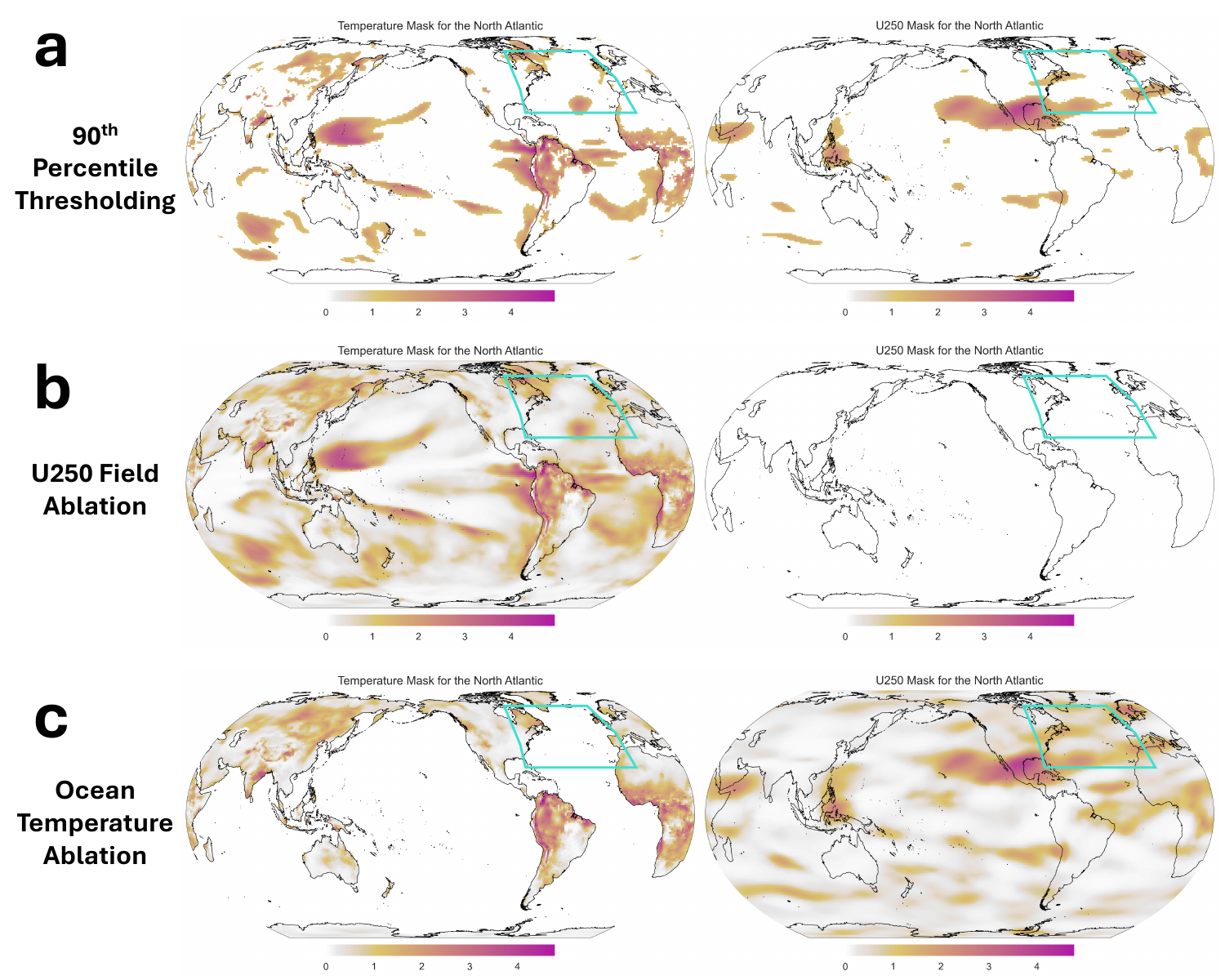}
        \end{minipage}
    }
    \caption{Examples of masks with each of the ablation methods: (a) Threshold ablation, (b) Ablating entire fields, and (c) Ablating specific regions.}
    \label{fig:atlantic_ablation_methods}
\end{figure}

We also directly probe the learned mask to better understand the relative importance of initial conditions in different areas for successful analog prediction. We do so by ablating the mask, i.e. setting the weights to 0, and observing changes in skill. We analyze changes in BS skill with CESM2-LE data and a 400 analog ensemble. We test on CESM2-LE data rather than ERA5 because the impacts on BS are small, and thus, there are too few samples to draw meaningful conclusions from ERA5 data. We employ three ablation methods: 1) threshold ablation, where we increase mask sparsity by setting weights to 0 if they are below either the 40th, 80th, or 90th percentile or incentivizing sparsity during training itself by adding constrained inverse $L_2$ regularization (see \ref{sec:L2} for details), 2) Ablating entire fields (e.g. temperature or U250), and 3) ablating specific regions (e.g. the Northern Hemisphere). Masks for examples of these ablation methods are shown in Figure \ref{fig:atlantic_ablation_methods}. All of these ablation methods, except constrained inverse $L_2$ regularization, are performed after the mask has already been learned.

\begin{figure}[!htb]
    \centering
    \includegraphics[width=.8\textwidth]{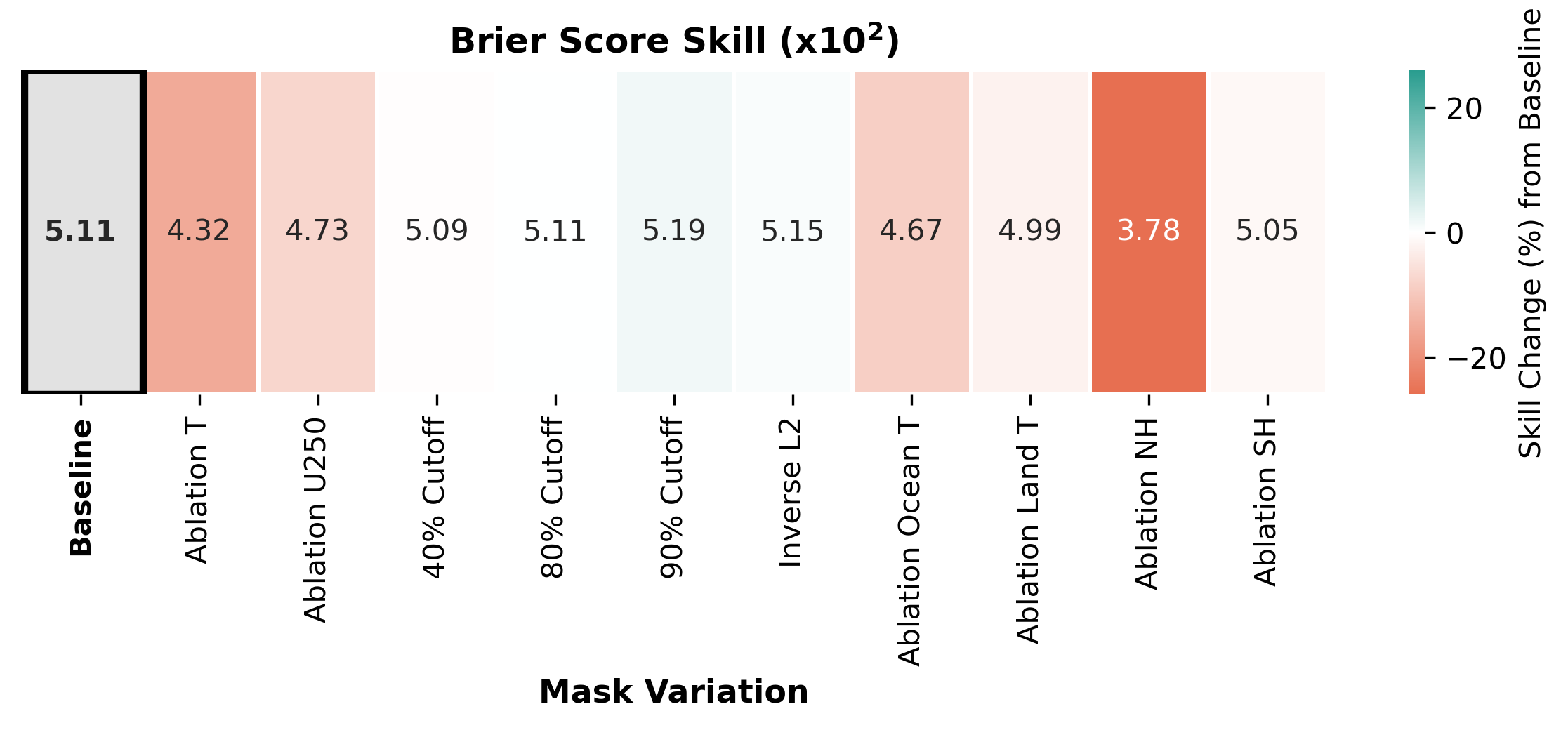}
    \caption{BS skill for different ablation methods evaluated on CESM2-LE data.}
    \label{fig:ablation_bs_skill}
\end{figure}

We focus on a 400 analog ensemble, as this is the number of analogs for which skill peaks (Figure \ref{fig:ablation_bs_skill}), although the general trends remain similar across ensemble size (Figure \ref{fig:bs_expanded_skill}). We find a slight improvement in skill when we increase mask sparsity by thresholding or by introducing constrained inverse $L_2$ regularization. This increase in skill with a sparser map is consistent with \cite{Rader:2023aa}, who found a slight improvement in skill for multi-year predictions using a $\sim95\%$ percentile threshold. However, when we test increasing the sparsity for shorter timescales (e.g. Task \#3), we find minimal change in observation skill and a slight decrease in probabilistic model skill (Figure \ref{fig:CA_90th_percentile_mask}). Considering field ablation, temperature appears to be the more important of the two fields, although ablating either the temperature field or the U250 field results in a significant decrease in skill, highlighting the importance of both for identifying skillful analogs. 
While all ablation methods besides increasing sparsity decreases skill, ablating the Northern Hemisphere, both fields, and ocean temperatures result in the largest drop in skill.

\section{Conclusions}
We demonstrate how an AI-informed model-analog forecasting approach, previously only shown to be skillful on seasonal-to-decadal timescales, can also produce skillful deterministic and probabilistic subseasonal-to-seasonal predictions. \add{While the goal of this work is not to outperform operational S2S forecasting systems, a limited comparison to operational models in Supplemental Section \protect{\ref{sec:model_baselines}}, indicates AI-based model-analogs are in line with operational model performance.} \change{We}{Further, we} showcase this approach's improvement over climatological, persistence, and traditional model-analog forecasting methods in both a perfect model framework and with reanalysis data for classification and continuous prediction tasks. For example, we find 49\% and 19\% increase in probabilistic skill testing on observational data for Tasks \#1 and \#3, respectively. While skill relative to climatology remains modest, especially for shorter timescales (e.g., Task \#1 and Task \#3), this is typical for S2S prediction. For example, the Seasonal-to-Multiyear Large Ensemble (SMYLE) prediction system, which similarly utilizes CESM2, finds zero or minimal skill (ACC $< .3$) when predicting month 1-3 summer Midwestern U.S. temperatures \parencite{smyle}.

Moreover, the interpretability of this AI-based approach enables users to explore the learned mask to gain insight into the relative importance of different areas of the globe for successful prediction. We perform an analysis of the learned mask for North Atlantic U250 classification, assessing which variable fields and areas of the globe are comparatively more important for a successful analog forecast. This type of analysis can help identify key initial conditions that most influence climate state evolution on S2S timescales, guiding both future model development and observational prioritization.

The AI-informed model-analog ensembles additionally provides improved prediction of extremes compared to traditional analog forecasting methods. Since extreme temperature events have a disproportionate impact on human health, agriculture, and energy/water management, improving their prediction is essential for mitigating the most dire consequences \parencite{annurev:/content/journals/10.1146/annurev-earth-071719-055228}.

Like previous work \parencite[e.g.,][]{Rader:2023aa,toride2024usingdeeplearningidentify,fernandez2025multiyeartodecadaltemperaturepredictionusing}, we leverage a model-analog approach rather than rely on observations alone to form our analog library. While we use a single model (CESM2) to form our analog library, \cite{fernandez2025multiyeartodecadaltemperaturepredictionusing} experiment with using multiple models on multi-year-to-decadal timescales to both form the analog library and to train their neural network. Including more climate models or, specifically for S2S timescales, extended-range AI or dynamical weather forecasts \parencite[e.g.,][]{lang2024aifsecmwfsdatadriven,Vitart:2022} may similarly boost skill on these shorter timescales. 

Further, although we have shown that this method produces skillful S2S predictions on near observational data, the approach is still limited by the model's, in this case CESM2's, biases \parencite[e.g.,][]{Pang:2024aa,Wei:2021aa,Woelfle:2019aa}. Future work could also explore how to best incorporate near-observational data into the AI-based model-analog forecasting approach. For example, \cite{fernandez2025multiyeartodecadaltemperaturepredictionusing} found that transfer learning from models to reanalysis improved analog prediction skill; a similar approach could be useful for learning masks on S2S timescales as well. Moreover, reanalysis data could also be incorporated into the analog library itself, perhaps offering even more realistic climate trajectories than those from a library composed solely of model data. These additions to the analog library and/or training set could further advance AI-based model-analog S2S forecasts, especially for regions and seasons where model biases are more pronounced.

\section{Acknowledgments}
J.B.L. acknowledges support from NOAA grants \#NA22OAR4310621 and \#NA19OAR4590151. The authors wish to thank Martin Fernandez for helpful insights for this research \add{and Yan Wang for providing and processing NMME model data}. The authors also wish to thank the editor and two anonymous reviewers for their helpful comments and suggestions.
\section{Data and Code Availability}
CESM2-LE data is available at \url{https://www.earthsystemgrid.org/dataset/ucar.cgd.cesm2le.output.html} \parencite{Danabasoglu:2020aa}. ERA5 data is available at \url{https://cds.climate.copernicus.eu/datasets} \parencite{Hersbach:2020aa}. Code used to perform the analysis and generate the figures is available at \url{https://github.com/jlandsbe/S2S.git}.

\clearpage
\printbibliography

@Article{Yu2023,
author={Yu, Bin
and Lin, Hai
and Mo, Ruping
and Li, Guilong},
title={A physical analysis of summertime North American heatwaves},
journal={Climate Dynamics},
year={2023},
month={Aug},
day={01},
volume={61},
number={3},
pages={1551-1565},
issn={1432-0894},
doi={10.1007/s00382-022-06642-1},
url={https://doi.org/10.1007/s00382-022-06642-1}
}

@article{Midwest_MJO,
author = {Wang, Rui and Huang, Jianping and Lian, Xinbo and Li, Han and Zhao, Yingjie and Zhang, Beidou and Han, Dongliang},
title = {Attribution of Air Temperature Variation to the Incidence of COVID-19},
journal = {Geophysical Research Letters},
volume = {52},
number = {14},
pages = {e2025GL116345},
doi = {https://doi.org/10.1029/2025GL116345},
url = {https://agupubs.onlinelibrary.wiley.com/doi/abs/10.1029/2025GL116345},
eprint = {https://agupubs.onlinelibrary.wiley.com/doi/pdf/10.1029/2025GL116345},
note = {e2025GL116345 2025GL116345},
year = {2025}
}

@article{JET_NAO,
author = {Hunt, Kieran and Nazir Zaz, Sumira},
year = {2022},
month = {08},
pages = {},
title = {Linking the North Atlantic Oscillation to winter precipitation over the Western Himalaya through disturbances of the subtropical jet},
volume = {60},
journal = {Climate Dynamics},
doi = {10.1007/s00382-022-06450-7}
}

@article{MJO_NAO_teleconnections,
author = {Lin, Hai and Brunet, Gilbert and Yu, Bin},
title = {Interannual variability of the Madden-Julian Oscillation and its impact on the North Atlantic Oscillation in the boreal winter},
journal = {Geophysical Research Letters},
volume = {42},
number = {13},
pages = {5571-5576},
doi = {https://doi.org/10.1002/2015GL064547},
url = {https://agupubs.onlinelibrary.wiley.com/doi/abs/10.1002/2015GL064547},
eprint = {https://agupubs.onlinelibrary.wiley.com/doi/pdf/10.1002/2015GL064547},
year = {2015}
}

@article { TheConsistencyofMJOTeleconnectionPatternsAnExplanationUsingLinearRossbyWaveTheory,
      author = "Kai-Chih Tseng and Eric Maloney and Elizabeth Barnes",
      title = "The Consistency of MJO Teleconnection Patterns: An Explanation Using Linear Rossby Wave Theory",
      journal = "Journal of Climate",
      year = "2019",
      publisher = "American Meteorological Society",
      address = "Boston MA, USA",
      volume = "32",
      number = "2",
      doi = "10.1175/JCLI-D-18-0211.1",
      pages=      "531 - 548",
      url = "https://journals.ametsoc.org/view/journals/clim/32/2/jcli-d-18-0211.1.xml"
}

@article{ENSO_NAO,
author = {Sabatani, Davide and Gualdi, Silvio},
year = {2025},
month = {06},
pages = {},
title = {ENSO teleconnections with the NAE sector during December in CMIP5/CMIP6 models: impacts of the atmospheric mean state},
journal = {npj Climate and Atmospheric Science},
doi = {10.1038/s41612-025-01064-2}
}

@inbook{NAO,
author = {Hurrell, James W. and Kushnir, Yochanan and Ottersen, Geir and Visbeck, Martin},
publisher = {American Geophysical Union (AGU)},
isbn = {9781118669037},
title = {An Overview of the North Atlantic Oscillation},
booktitle = {The North Atlantic Oscillation: Climatic Significance and Environmental Impact},
chapter = {},
pages = {1-35},
doi = {https://doi.org/10.1029/134GM01},
url = {https://agupubs.onlinelibrary.wiley.com/doi/abs/10.1029/134GM01},
eprint = {https://agupubs.onlinelibrary.wiley.com/doi/pdf/10.1029/134GM01},
year = {2003}
}

@article{Arcodia_2023,
doi = {10.1088/2752-5295/aced60},
url = {https://doi.org/10.1088/2752-5295/aced60},
year = {2023},
month = {sep},
publisher = {IOP Publishing},
volume = {2},
number = {4},
pages = {045002},
author = {Arcodia, Marybeth C and Barnes, Elizabeth A and Mayer, Kirsten J and Lee, Jiwoo and Ordonez, Ana and Ahn, Min-Seop},
title = {Assessing decadal variability of subseasonal forecasts of opportunity using explainable AI},
journal = {Environmental Research: Climate}
}

@Article{Hall2017,
author={Hall, Richard J.
and Jones, Julie M.
and Hanna, Edward
and Scaife, Adam A.
and Erd{\'e}lyi, R{\'o}bert},
title={Drivers and potential predictability of summer time North Atlantic polar front jet variability},
journal={Climate Dynamics},
year={2017},
month={Jun},
day={01},
volume={48},
number={11},
pages={3869-3887},
issn={1432-0894},
doi={10.1007/s00382-016-3307-0},
url={https://doi.org/10.1007/s00382-016-3307-0}
}

@article {AVerifiedEstimationoftheElNioIndexNio34since1877,
      author = "Lucia Bunge and Allan J. Clarke",
      title = "A Verified Estimation of the El Niño Index Niño-3.4 since 1877",
      journal = "Journal of Climate",
      year = "2009",
      publisher = "American Meteorological Society",
      address = "Boston MA, USA",
      volume = "22",
      number = "14",
      doi = "10.1175/2009JCLI2724.1",
      pages=      "3979 - 3992",
      url = "https://journals.ametsoc.org/view/journals/clim/22/14/2009jcli2724.1.xml"
}

@article{smyle,
AUTHOR = {Yeager, S. G. and Rosenbloom, N. and Glanville, A. A. and Wu, X. and Simpson, I. and Li, H. and Molina, M. J. and Krumhardt, K. and Mogen, S. and Lindsay, K. and Lombardozzi, D. and Wieder, W. and Kim, W. M. and Richter, J. H. and Long, M. and Danabasoglu, G. and Bailey, D. and Holland, M. and Lovenduski, N. and Strand, W. G. and King, T.},
TITLE = {The Seasonal-to-Multiyear Large Ensemble (SMYLE) prediction system using the
Community Earth System Model version 2},
JOURNAL = {Geoscientific Model Development},
VOLUME = {15},
YEAR = {2022},
NUMBER = {16},
PAGES = {6451--6493},
URL = {https://gmd.copernicus.org/articles/15/6451/2022/},
DOI = {10.5194/gmd-15-6451-2022}
}

@misc{lang2024aifsecmwfsdatadriven,
	archiveprefix = {arXiv},
	author = {Simon Lang and Mihai Alexe and Matthew Chantry and Jesper Dramsch and Florian Pinault and Baudouin Raoult and Mariana C. A. Clare and Christian Lessig and Michael Maier-Gerber and Linus Magnusson and Zied Ben Bouall{\`e}gue and Ana Prieto Nemesio and Peter D. Dueben and Andrew Brown and Florian Pappenberger and Florence Rabier},
	eprint = {2406.01465},
	primaryclass = {physics.ao-ph},
	title = {AIFS -- ECMWF's data-driven forecasting system},
	url = {https://arxiv.org/abs/2406.01465},
	year = {2024}}

@article{Ferranti:2018aa,
	author = {Ferranti, Laura and Magnusson, Linus and Vitart, Fr{\'e}d{\'e}ric and Richardson, David S.},
	date = {2018/07/01},
	doi = {https://doi.org/10.1002/qj.3341},
	isbn = {0035-9009},
	journal = {Quarterly Journal of the Royal Meteorological Society},
	journal1 = {Quarterly Journal of the Royal Meteorological Society},
	journal2 = {Quarterly Journal of the Royal Meteorological Society},
	journal3 = {Q J R Meteorol Soc},
	month = {2025/03/17},
	n2 = {The potential of early warning for severe cold conditions is explored using the Subseasonal to Seasonal (S2S) Prediction research project data archive. We explore the use of a two-dimensional phase space based on the leading empirical orthogonal functions (EOFs) of mid-tropospheric flow computed over the Euro-Atlantic region in order to study the time evolution of flow patterns associated with high-impact temperature anomalies. We find that the phase space is an effective tool for monitoring predictions of regime transitions at medium and extended ranges. We show that a number of S2S systems have some skill in the prediction of cold spells over Europe, even beyond the medium range. In particular, the ECMWF (European Centre for Medium-Range Weather Forecasts) model represents well the observed preferential transition paths. We reveal that the impact of the Madden?Julian Oscillation (MJO) on the predictive skill of large-scale flow over Europe is asymmetric. The impact of the MJO on the Brier skill scores and reliability is significantly positive for predictions of the negative phase of the North Atlantic Oscillation (NAO): beyond week one, forecasts with the MJO in their initial state are significantly more reliable than forecasts with no MJO in their initial conditions. In contrast, the predictive skill for positive NAO shows little sensitivity to the MJO.},
	number = {715},
	pages = {1788--1802},
	publisher = {John Wiley \& Sons, Ltd},
	title = {How far in advance can we predict changes in large-scale flow leading to severe cold conditions over Europe?},
	url = {https://doi.org/10.1002/qj.3341},
	volume = {144},
	year = {2018},
	year1 = {2018}}

@article{Vitart:2022,
	author = {Fr{\'e}d{\'e}ric Vitart and Magdalena Alonso-Balmaseda and Laura Ferranti and Manuel Fuentes},
	chapter = {Earth System Science},
	doi = {10.21957/fv6k37c49h},
	issue = {173},
	month = {10},
	pages = {21-26},
	title = {The next extended-range configuration for IFS Cycle 48r1},
	url = {https://www.ecmwf.int/node/20521},
	year = {2022}}

@article{Pang:2024aa,
	author = {Pang, Da and Fang, Xianghui and Wang, Lei},
	date = {2024/12/18},
	doi = {10.1007/s00382-024-07515-5},
	id = {Pang2024},
	isbn = {1432-0894},
	journal = {Climate Dynamics},
	number = {1},
	pages = {47},
	title = {Feedback processes responsible for the deficiency of El Ni{\~n}o diversity in CESM2},
	url = {https://doi.org/10.1007/s00382-024-07515-5},
	volume = {63},
	year = {2024}}

@article{Wei:2021aa,
	author = {Wei, Ho-Hsuan and Subramanian, Aneesh C. and Karnauskas, Kristopher B. and DeMott, Charlotte A. and Mazloff, Matthew R. and Balmaseda, Magdalena A.},
	date = {2021/06/01},
	doi = {https://doi.org/10.1029/2020JC016967},
	isbn = {2169-9275},
	journal = {Journal of Geophysical Research: Oceans},
	journal1 = {Journal of Geophysical Research: Oceans},
	journal2 = {Journal of Geophysical Research: Oceans},
	journal3 = {J. Geophys. Res. Oceans},
	month = {2025/03/04},
	n2 = {Abstract Coupled processes and associated subsurface dynamics near the eastern edge of the Indo/western Pacific (WP) Warm Pool are important for air-sea interactions involved in tropical Pacific dynamics. We seek to shed light on the physical mechanisms governing air-sea interactions in the region and the impacts of their biases in models. In this study, we use the Ocean ReAnalysis System 5 (ORAS5) to identify mean-state biases in the National Center for Atmospheric Research Community Earth System Model version 2 (CESM2) with a particular focus on upper ocean properties and air-sea interaction processes. We show that the CESM2 has warm and fresh surface biases in the tropical Pacific Ocean, a barrier layer that is too thin in the WP, and an isothermal layer depth (ILD) that is too deep in the eastern Pacific (EP). These biases impact air-sea interaction processes involved in El Ni{\~n}o development. We compare the strong El Ni{\~n}o events in ORAS5 and CESM2 and show that biases in barrier layer thickness in the WP and in ILD in the EP are significant before the onset of the El Ni{\~n}o events. These biases then influence vertical mixing and entrainment processes, resulting in mixed layer cooling biases. Biases in the sea surface temperature seasonal cycle in the CESM2 also influence the development of the El Ni{\~n}o. We emphasize how the El Ni{\~n}o progression in models can be influenced by its sensitivity to the mean state biases in both subsurface ocean structure and seasonal cycle through local as well as the large-scale physical processes.},
	number = {6},
	pages = {e2020JC016967},
	publisher = {John Wiley \& Sons, Ltd},
	title = {Tropical Pacific Air-Sea Interaction Processes and Biases in CESM2 and Their Relation to El Ni{\~n}o Development},
	url = {https://doi.org/10.1029/2020JC016967},
	volume = {126},
	year = {2021},
	year1 = {2021}}

@article{Woelfle:2019aa,
	author = {Woelfle, M. D. and Bretherton, C. S. and Hannay, C. and Neale, R.},
	date = {2019/07/01},
	doi = {https://doi.org/10.1029/2019MS001647},
	journal = {Journal of Advances in Modeling Earth Systems},
	journal1 = {Journal of Advances in Modeling Earth Systems},
	journal2 = {Journal of Advances in Modeling Earth Systems},
	journal3 = {J. Adv. Model. Earth Syst.},
	month = {2025/03/04},
	n2 = {Abstract The structure of the east Pacific Intertropical Convergence Zone (ITCZ) as simulated in the Community Earth System Model version 2 (CESM2) is greatly improved as compared to its previous version, CESM version 1. Examination of intermediate model versions created as part of the development process for CESM2 shows the improvement in the ITCZ is well correlated with a reduction in the relative warmth of southeast Pacific sea surface temperatures (SSTs) as compared to the broader tropical mean. Cooling SST in this region enhances the zonal SST and surface pressure gradients and reduces the anomalously southward SST gradient present in boreal spring in early version of CESM2. The improvements in southeast Pacific SST are attributed to increases in low cloud cover and the associated shortwave cloud forcing over the southeast. Sensitivity tests using fixed SST simulations demonstrate the increase in cloud cover between two intermediate model versions, 119 and 125, to be driven by removal of the dependence of autoconversion and accretion rates on cloud water variance as well as the removal of a secondary condensation scheme. Both of these changes reduce drizzle rates in warm clouds increasing cloud lifetime and cloud fraction in the stratocumulus to trade cumulus transition region. The improvements in southeast Pacific shortwave cloud forcing and ITCZ climatology persist through subsequent changes to the cloud microphysics parameterizations. Despite improvements in the east Pacific ITCZ, the global mean ITCZ position and Pacific cold tongue bias strength do not exhibit a systematic improvement across the development simulations.},
	number = {7},
	pages = {1873--1893},
	publisher = {John Wiley \& Sons, Ltd},
	title = {Evolution of the Double-ITCZ Bias Through CESM2 Development},
	url = {https://doi.org/10.1029/2019MS001647},
	volume = {11},
	year = {2019},
	year1 = {2019}}

@misc{fernandez2025multiyeartodecadaltemperaturepredictionusing,
	archiveprefix = {arXiv},
	author = {M. A. Fernandez and Elizabeth A. Barnes},
	eprint = {2502.17583},
	primaryclass = {physics.ao-ph},
	title = {Multi-Year-to-Decadal Temperature Prediction using a Machine Learning Model-Analog Framework},
	url = {https://arxiv.org/abs/2502.17583},
	year = {2025}}

@article{Simpson:2023aa,
	address = {Boston MA, USA},
	author = {Simpson, Isla R. and Rosenbloom, Nan and Danabasoglu, Gokhan and Deser, Clara and Yeager, Stephen G. and McCluskey, Christina S. and Yamaguchi, Ryohei and Lamarque, Jean-Francois and Tilmes, Simone and Mills, Michael J. and Rodgers, Keith B.},
	date = {01 Sep. 2023},
	doi = {https://doi.org/10.1175/JCLI-D-22-0666.1},
	journal = {Journal of Climate},
	la = {English},
	number = {17},
	pages = {5687--5711},
	publisher = {American Meteorological Society},
	title = {The CESM2 Single-Forcing Large Ensemble and Comparison to CESM1: Implications for Experimental Design},
	url = {https://journals.ametsoc.org/view/journals/clim/36/17/JCLI-D-22-0666.1.xml},
	volume = {36},
	year = {2023}}

@article{Domeisen:2022aa,
	address = {Boston MA, USA},
	author = {Domeisen, Daniela I. V. and White, Christopher J. and Afargan-Gerstman, Hilla and Mu{\~n}oz, {\'A}ngel G. and Janiga, Matthew A. and Vitart, Fr{\'e}d{\'e}ric and Wulff, C. Ole and Antoine, Salom{\'e} and Ardilouze, Constantin and Batt{\'e}, Lauriane and Bloomfield, Hannah C. and Brayshaw, David J. and Camargo, Suzana J. and Charlton-P{\'e}rez, Andrew and Collins, Dan and Cowan, Tim and del Mar Chaves, Maria and Ferranti, Laura and G{\'o}mez, Rosario and Gonz{\'a}lez, Paula L. M. and Gonz{\'a}lez Romero, Carmen and Infanti, Johnna M. and Karozis, Stelios and Kim, Hera and Kolstad, Erik W. and LaJoie, Emerson and Lled{\'o}, Lloren{\c c} and Magnusson, Linus and Malguzzi, Piero and Manrique-Su{\~n}{\'e}n, Andrea and Mastrangelo, Daniele and Materia, Stefano and Medina, Hanoi and Palma, Llu{\'\i}s and Pineda, Luis E. and Sfetsos, Athanasios and Son, Seok-Woo and Soret, Albert and Strazzo, Sarah and Tian, Di},
	date = {01 Jun. 2022},
	doi = {https://doi.org/10.1175/BAMS-D-20-0221.1},
	journal = {Bulletin of the American Meteorological Society},
	la = {English},
	number = {6},
	pages = {E1473--E1501},
	publisher = {American Meteorological Society},
	title = {Advances in the Subseasonal Prediction of Extreme Events: Relevant Case Studies across the Globe},
	url = {https://journals.ametsoc.org/view/journals/bams/103/6/BAMS-D-20-0221.1.xml},
	volume = {103},
	year = {2022}}

@article{Zhao:2016aa,
	author = {Zhao, Zhizhen and Giannakis, Dimitrios},
	date = {2016/08/12},
	doi = {10.1088/0951-7715/29/9/2888},
	journal = {Nonlinearity},
	month = {08},
	title = {Analog Forecasting with Dynamics-Adapted Kernels},
	volume = {29},
	year = {2016}}

@article{Zhang:2013aa,
	address = {Boston MA, USA},
	author = {Zhang, Chidong},
	date = {01 Dec. 2013},
	doi = {https://doi.org/10.1175/BAMS-D-12-00026.1},
	journal = {Bulletin of the American Meteorological Society},
	la = {English},
	number = {12},
	pages = {1849--1870},
	publisher = {American Meteorological Society},
	title = {Madden--Julian Oscillation: Bridging Weather and Climate},
	url = {https://journals.ametsoc.org/view/journals/bams/94/12/bams-d-12-00026.1.xml},
	volume = {94},
	year = {2013}}

@article{Palmer:2004aa,
	address = {Boston MA, USA},
	author = {Palmer, T. N. and Alessandri, A. and Andersen, U. and Cantelaube, P. and Davey, M. and D{\'e}l{\'e}cluse, P. and D{\'e}qu{\'e}, M. and D{\'\i}ez, E. and Doblas-Reyes, F. J. and Feddersen, H. and Graham, R. and Gualdi, S. and Gu{\'e}r{\'e}my, J. -F. and Hagedorn, R. and Hoshen, M. and Keenlyside, N. and Latif, M. and Lazar, A. and Maisonnave, E. and Marletto, V. and Morse, A. P. and Orfila, B. and Rogel, P. and Terres, J. -M. and Thomson, M. C.},
	date = {01 Jun. 2004},
	doi = {https://doi.org/10.1175/BAMS-85-6-853},
	journal = {Bulletin of the American Meteorological Society},
	la = {English},
	number = {6},
	pages = {853--872},
	publisher = {American Meteorological Society},
	title = {DEVELOPMENT OF A EUROPEAN MULTIMODEL ENSEMBLE SYSTEM FOR SEASONAL-TO-INTERANNUAL PREDICTION (DEMETER)},
	url = {https://journals.ametsoc.org/view/journals/bams/85/6/bams-85-6-853.xml},
	volume = {85},
	year = {2004}}

@article{Krishnamurti:1999aa,
	author = {Krishnamurti, T. N. and Kishtawal, C. M. and LaRow, Timothy E. and Bachiochi, David R. and Zhang, Zhan and Williford, C. Eric and Gadgil, Sulochana and Surendran, Sajani},
	date = {1999/09/03},
	doi = {10.1126/science.285.5433.1548},
	journal = {Science},
	journal1 = {Science},
	journal2 = {Science},
	month = {2025/02/11},
	n2 = {A method for improving weather and climate forecast skill has been developed. It is called a superensemble, and it arose from a study of the statistical properties of a low-order spectral model. Multiple regression was used to determine coefficients from multimodel forecasts and observations. The coefficients were then used in the superensemble technique. The superensemble was shown to outperform all model forecasts for multiseasonal, medium-range weather and hurricane forecasts. In addition, the superensemble was shown to have higher skill than forecasts based solely on ensemble averaging.},
	number = {5433},
	pages = {1548--1550},
	publisher = {American Association for the Advancement of Science},
	title = {Improved Weather and Seasonal Climate Forecasts from Multimodel Superensemble},
	type = {doi: 10.1126/science.285.5433.1548},
	url = {https://doi.org/10.1126/science.285.5433.1548},
	volume = {285},
	year = {1999},
	year1 = {1999}}

@article{annurev:/content/journals/10.1146/annurev-earth-071719-055228,
	author = {AghaKouchak, Amir and Chiang, Felicia and Huning, Laurie S. and Love, Charlotte A. and Mallakpour, Iman and Mazdiyasni, Omid and Moftakhari, Hamed and Papalexiou, Simon Michael and Ragno, Elisa and Sadegh, Mojtaba},
	doi = {https://doi.org/10.1146/annurev-earth-071719-055228},
	issn = {1545-4495},
	journal = {Annual Review of Earth and Planetary Sciences},
	number = {Volume 48, 2020},
	pages = {519-548},
	publisher = {Annual Reviews},
	title = {Climate Extremes and Compound Hazards in a Warming World},
	type = {Journal Article},
	url = {https://www.annualreviews.org/content/journals/10.1146/annurev-earth-071719-055228},
	volume = {48},
	year = {2020}}

@article{Weisheimer:2014aa,
	author = {Weisheimer, A. and Palmer, Tim},
	date = {2014/07/06},
	doi = {10.1098/rsif.2013.1162},
	journal = {Journal of the Royal Society, Interface / the Royal Society},
	month = {07},
	pages = {20131162},
	title = {On the reliability of seasonal climate forecasts},
	volume = {11},
	year = {2014}}

@article{Leutbecher:2008aa,
	author = {Leutbecher, M. and Palmer, T. N.},
	date = {2008/03/20/},
	doi = {https://doi.org/10.1016/j.jcp.2007.02.014},
	isbn = {0021-9991},
	journal = {Journal of Computational Physics},
	journal1 = {Predicting weather, climate and extreme events},
	number = {7},
	pages = {3515--3539},
	title = {Ensemble forecasting},
	url = {https://www.sciencedirect.com/science/article/pii/S0021999107000812},
	volume = {227},
	year = {2008}}

@article{Hersbach:2020aa,
	author = {Hersbach, Hans and Bell, Bill and Berrisford, Paul and Hirahara, Shoji and Hor{\'a}nyi, Andr{\'a}s and Mu{\~n}oz-Sabater, Joaqu{\'\i}n and Nicolas, Julien and Peubey, Carole and Radu, Raluca and Schepers, Dinand and Simmons, Adrian and Soci, Cornel and Abdalla, Saleh and Abellan, Xavier and Balsamo, Gianpaolo and Bechtold, Peter and Biavati, Gionata and Bidlot, Jean and Bonavita, Massimo and De Chiara, Giovanna and Dahlgren, Per and Dee, Dick and Diamantakis, Michail and Dragani, Rossana and Flemming, Johannes and Forbes, Richard and Fuentes, Manuel and Geer, Alan and Haimberger, Leo and Healy, Sean and Hogan, Robin J. and H{\'o}lm, El{\'\i}as and Janiskov{\'a}, Marta and Keeley, Sarah and Laloyaux, Patrick and Lopez, Philippe and Lupu, Cristina and Radnoti, Gabor and de Rosnay, Patricia and Rozum, Iryna and Vamborg, Freja and Villaume, Sebastien and Th{\'e}paut, Jean-No{\"e}l},
	date = {2020/07/01},
	doi = {https://doi.org/10.1002/qj.3803},
	isbn = {0035-9009},
	journal = {Quarterly Journal of the Royal Meteorological Society},
	journal1 = {Quarterly Journal of the Royal Meteorological Society},
	journal2 = {Quarterly Journal of the Royal Meteorological Society},
	journal3 = {Q.J.R. Meteorol. Soc.},
	month = {2025/01/20},
	n2 = {Abstract Within the Copernicus Climate Change Service (C3S), ECMWF is producing the ERA5 reanalysis which, once completed, will embody a detailed record of the global atmosphere, land surface and ocean waves from 1950 onwards. This new reanalysis replaces the ERA-Interim reanalysis (spanning 1979 onwards) which was started in 2006. ERA5 is based on the Integrated Forecasting System (IFS) Cy41r2 which was operational in 2016. ERA5 thus benefits from a decade of developments in model physics, core dynamics and data assimilation. In addition to a significantly enhanced horizontal resolution of 31?km, compared to 80?km for ERA-Interim, ERA5 has hourly output throughout, and an uncertainty estimate from an ensemble (3-hourly at half the horizontal resolution). This paper describes the general set-up of ERA5, as well as a basic evaluation of characteristics and performance, with a focus on the dataset from 1979 onwards which is currently publicly available. Re-forecasts from ERA5 analyses show a gain of up to one day in skill with respect to ERA-Interim. Comparison with radiosonde and PILOT data prior to assimilation shows an improved fit for temperature, wind and humidity in the troposphere, but not the stratosphere. A comparison with independent buoy data shows a much improved fit for ocean wave height. The uncertainty estimate reflects the evolution of the observing systems used in ERA5. The enhanced temporal and spatial resolution allows for a detailed evolution of weather systems. For precipitation, global-mean correlation with monthly-mean GPCP data is increased from 67{\%} to 77{\%}. In general, low-frequency variability is found to be well represented and from 10?hPa downwards general patterns of anomalies in temperature match those from the ERA-Interim, MERRA-2 and JRA-55 reanalyses.},
	number = {730},
	pages = {1999--2049},
	publisher = {John Wiley \& Sons, Ltd},
	title = {The ERA5 global reanalysis},
	url = {https://doi.org/10.1002/qj.3803},
	volume = {146},
	year = {2020},
	year1 = {2020}}

@article{Romps_2022,
doi = {10.1088/1748-9326/ac8945},
url = {https://doi.org/10.1088/1748-9326/ac8945},
year = {2022},
month = {aug},
publisher = {IOP Publishing},
volume = {17},
number = {9},
pages = {094017},
author = {Romps, David M and Lu, Yi-Chuan},
title = {Chronically underestimated: a reassessment of US heat waves using the extended heat index},
journal = {Environmental Research Letters}
}

@article {SummerLandAtmosphereCouplingStrengthintheUnitedStatesComparisonamongObservationsReanalysisDataandNumericalModels,
      author = "Rui  Mei and Guiling  Wang",
      title = "Summer Land–Atmosphere Coupling Strength in the United States: Comparison among Observations, Reanalysis Data, and Numerical Models",
      journal = "Journal of Hydrometeorology",
      year = "2012",
      publisher = "American Meteorological Society",
      address = "Boston MA, USA",
      volume = "13",
      number = "3",
      doi = "10.1175/JHM-D-11-075.1",
      pages=      "1010 - 1022",
      url = "https://journals.ametsoc.org/view/journals/hydr/13/3/jhm-d-11-075_1.xml"
}

@article{Danabasoglu:2020aa,
	author = {Danabasoglu, G. and Lamarque, J. -F. and Bacmeister, J. and Bailey, D. A. and DuVivier, A. K. and Edwards, J. and Emmons, L. K. and Fasullo, J. and Garcia, R. and Gettelman, A. and Hannay, C. and Holland, M. M. and Large, W. G. and Lauritzen, P. H. and Lawrence, D. M. and Lenaerts, J. T. M. and Lindsay, K. and Lipscomb, W. H. and Mills, M. J. and Neale, R. and Oleson, K. W. and Otto-Bliesner, B. and Phillips, A. S. and Sacks, W. and Tilmes, S. and van Kampenhout, L. and Vertenstein, M. and Bertini, A. and Dennis, J. and Deser, C. and Fischer, C. and Fox-Kemper, B. and Kay, J. E. and Kinnison, D. and Kushner, P. J. and Larson, V. E. and Long, M. C. and Mickelson, S. and Moore, J. K. and Nienhouse, E. and Polvani, L. and Rasch, P. J. and Strand, W. G.},
	date = {2020/02/01},
	doi = {https://doi.org/10.1029/2019MS001916},
	journal = {Journal of Advances in Modeling Earth Systems},
	journal1 = {Journal of Advances in Modeling Earth Systems},
	journal2 = {Journal of Advances in Modeling Earth Systems},
	journal3 = {J. Adv. Model. Earth Syst.},
	month = {2025/01/17},
	n2 = {Abstract An overview of the Community Earth System Model Version 2 (CESM2) is provided, including a discussion of the challenges encountered during its development and how they were addressed. In addition, an evaluation of a pair of CESM2 long preindustrial control and historical ensemble simulations is presented. These simulations were performed using the nominal 1$\,^{\circ}$horizontal resolution configuration of the coupled model with both the ?low-top? (40 km, with limited chemistry) and ?high-top? (130 km, with comprehensive chemistry) versions of the atmospheric component. CESM2 contains many substantial science and infrastructure improvements and new capabilities since its previous major release, CESM1, resulting in improved historical simulations in comparison to CESM1 and available observations. These include major reductions in low-latitude precipitation and shortwave cloud forcing biases; better representation of the Madden-Julian Oscillation; better El Ni{\~n}o-Southern Oscillation-related teleconnections; and a global land carbon accumulation trend that agrees well with observationally based estimates. Most tropospheric and surface features of the low- and high-top simulations are very similar to each other, so these improvements are present in both configurations. CESM2 has an equilibrium climate sensitivity of 5.1?5.3 $\,^{\circ}$C, larger than in CESM1, primarily due to a combination of relatively small changes to cloud microphysics and boundary layer parameters. In contrast, CESM2's transient climate response of 1.9?2.0 $\,^{\circ}$C is comparable to that of CESM1. The model outputs from these and many other simulations are available to the research community, and they represent CESM2's contributions to the Coupled Model Intercomparison Project Phase 6.},
	number = {2},
	pages = {e2019MS001916},
	publisher = {John Wiley \& Sons, Ltd},
	title = {The Community Earth System Model Version 2 (CESM2)},
	url = {https://doi.org/10.1029/2019MS001916},
	volume = {12},
	year = {2020},
	year1 = {2020}}

@article{esd-13-1437-2022,
	author = {Mahmood, R. and Donat, M. G. and Ortega, P. and Doblas-Reyes, F. J. and Delgado-Torres, C. and Sams\'o, M. and Bretonni\`ere, P.-A.},
	doi = {10.5194/esd-13-1437-2022},
	journal = {Earth System Dynamics},
	number = {4},
	pages = {1437--1450},
	title = {Constraining low-frequency variability in climate projections to predict climate on decadal to multi-decadal timescales -- a poor man's initialized prediction system},
	url = {https://esd.copernicus.org/articles/13/1437/2022/},
	volume = {13},
	year = {2022}}

@article{Mullan:2006aa,
	author = {Mullan, A. Brett and Thompson, Craig S.},
	date = {2006/03/30},
	doi = {https://doi.org/10.1002/joc.1261},
	isbn = {0899-8418},
	journal = {International Journal of Climatology},
	journal1 = {International Journal of Climatology},
	journal2 = {International Journal of Climatology},
	journal3 = {Int. J. Climatol.},
	month = {2025/01/09},
	n2 = {Abstract An analogue forecast scheme is described for multifield prediction of monthly and seasonal New Zealand climate anomalies on the basis of the methodology of Livezey and Barnston (1988) for US seasonal temperatures. The method is applied to predicting terciles of temperature and precipitation for six regions of New Zealand. Empirical orthogonal function analysis is used to reduce sea surface temperature and sea-level pressure predictors down to a set of five independent indices, which incorporate variations due to El Ni{\~n}o-Southern Oscillation, Indian Ocean sea temperatures and a wave 3 pattern in the Southern Hemisphere westerlies. A full bootstrap cross-validation procedure is carried out, along with Monte Carlo tests, to assess the skill of the method on independent data and to determine the significance of the results. Significant skill is found for seasonal temperature forecasts for the summer and winter seasons; there is less success in predicting monthly temperatures or rainfall at either timescale. Considerable care is required to constrain the climate state vector, from which analogues are defined, and to constrain the search procedure itself, in order to produce results that are stable with respect to small parameter changes in the model. For the New Zealand region, 5 to 7 is found to be the optimum number of ?closest analogues?, and the inclusion of anti-analogues improves the predictions, at least in the seasonal case. Skill in predicting regional temperature and rainfall is shown to be related to a combination of skill in predicting sea-level pressure patterns and to how strongly these patterns project onto temperature and rainfall anomalies. Copyright ? 2005 Royal Meteorological Society.},
	number = {4},
	pages = {485--504},
	publisher = {John Wiley \& Sons, Ltd},
	title = {Analogue forecasting of New Zealand climate anomalies},
	url = {https://doi.org/10.1002/joc.1261},
	volume = {26},
	year = {2006},
	year1 = {2006}}

@article{Han:2023aa,
	author = {Han, Ji-Young and Kim, Sang-Wook and Park, Chang-Hyun and Son, Seok-Woo},
	date = {2023/08/19},
	doi = {10.1186/s40562-023-00292-9},
	id = {Han2023},
	isbn = {2196-4092},
	journal = {Geoscience Letters},
	number = {1},
	pages = {37},
	title = {Ensemble size versus bias correction effects in subseasonal-to-seasonal (S2S) forecasts},
	url = {https://doi.org/10.1186/s40562-023-00292-9},
	volume = {10},
	year = {2023}}

@misc{toride2024usingdeeplearningidentify,
	archiveprefix = {arXiv},
	author = {Kinya Toride and Matthew Newman and Andrew Hoell and Antonietta Capotondi and Jakob Schl{\"o}r and Dillon J. Amaya},
	eprint = {2404.15419},
	primaryclass = {physics.ao-ph},
	title = {Using Deep Learning to Identify Initial Error Sensitivity for Interpretable ENSO Forecasts},
	url = {https://arxiv.org/abs/2404.15419},
	year = {2024}}

@article{Rader:2023aa,
	author = {Rader, Jamin K. and Barnes, Elizabeth A.},
	date = {2023/12/16},
	doi = {https://doi.org/10.1029/2023GL104983},
	isbn = {0094-8276},
	journal = {Geophysical Research Letters},
	journal1 = {Geophysical Research Letters},
	journal2 = {Geophysical Research Letters},
	journal3 = {Geophys Res Lett},
	month = {2025/01/09},
	n2 = {Abstract Seasonal-to-decadal climate prediction is crucial for decision-making in a number of industries, but forecasts on these timescales have limited skill. Here, we develop a data-driven method for selecting optimal analogs for seasonal-to-decadal analog forecasting. Using an interpretable neural network, we learn a spatially-weighted mask that quantifies how important each grid point is for determining whether two climate states will evolve similarly. We show that analogs selected using this weighted mask provide more skillful forecasts than analogs that are selected using traditional spatially-uniform methods. This method is tested on two prediction problems using the Max Planck Institute for Meteorology Grand Ensemble: multi-year prediction of North Atlantic sea surface temperatures, and seasonal prediction of El Ni{\~n}o Southern Oscillation. This work demonstrates a methodical approach to selecting analogs that may be useful for improving seasonal-to-decadal forecasts and understanding their sources of skill.},
	number = {23},
	pages = {e2023GL104983},
	publisher = {John Wiley \& Sons, Ltd},
	title = {Optimizing Seasonal-To-Decadal Analog Forecasts With a Learned Spatially-Weighted Mask},
	url = {https://doi.org/10.1029/2023GL104983},
	volume = {50},
	year = {2023},
	year1 = {2023}}

@article{Wu:2023aa,
	author = {Wu, Yanling and Yan, Xiaoqin},
	doi = {10.3390/jmse11050980},
	isbn = {2077-1312},
	journal = {Journal of Marine Science and Engineering},
	number = {5},
	title = {Evaluating Changes in the Multiyear Predictability of the Pacific Decadal Oscillation Using Model Analogs since 1900},
	volume = {11},
	year = {2023}}

@article{VAN-DEN-DOOL:1994aa,
	author = {{Van den Dool}, H. M.},
	date = {1994/05/01},
	doi = {https://doi.org/10.1034/j.1600-0870.1994.t01-2-00006.x},
	isbn = {0280-6495},
	journal = {Tellus A},
	journal1 = {Tellus A},
	journal2 = {Tellus A},
	month = {2025/01/09},
	n2 = {ABSTRACT A three-way relationship is derived between the size of a library (M years) of historical atmospheric data, the distance between an arbitrarily picked state of the atmosphere and its nearest neighbor (or analogue), and the size of the spatial domain, as measured by the number of spatial degrees of freedom (N). It is found that it would take a library of order 1030?years to find 2 observed flows that match to within current observational error over a large area such as the Northern Hemisphere. Obviously, with only 10?100?years of data, the probability of finding natural analogous is very small, unless one is satisfied with analogy over small areas or in just 2 of 3 degrees of freedom as represented, for instance, by 2 or 3 leading empirical orthogonal modes. We further propose the notion that analogues can be constructed by combining a number of observed flow patterns. We have found at least one application where linearly constructed analogues are conclusively better at specifying US surface weather from concurrent 700?mb geopotential height than natural analogues are.},
	number = {3},
	pages = {314--324},
	publisher = {John Wiley \& Sons, Ltd},
	title = {Searching for analogues, how long must we wait?},
	url = {https://doi.org/10.1034/j.1600-0870.1994.t01-2-00006.x},
	volume = {46},
	year = {1994},
	year1 = {1994}}

@article{Ding:2018aa,
	address = {Boston MA, USA},
	author = {Ding, Hui and Newman, Matthew and Alexander, Michael A. and Wittenberg, Andrew T.},
	date = {15 Jul. 2018},
	doi = {https://doi.org/10.1175/JCLI-D-17-0661.1},
	journal = {Journal of Climate},
	la = {English},
	number = {14},
	pages = {5437--5459},
	publisher = {American Meteorological Society},
	title = {Skillful Climate Forecasts of the Tropical Indo-Pacific Ocean Using Model-Analogs},
	url = {https://journals.ametsoc.org/view/journals/clim/31/14/jcli-d-17-0661.1.xml},
	volume = {31},
	year = {2018}}

@article{Walsh:2021aa,
	author = {Walsh, John and Brettschneider, Brian and Kettle, Nathan and Thoman, Richard},
	date = {2021/01/01},
	doi = {10.4236/acs.2021.113028},
	journal = {Atmospheric and Climate Sciences},
	month = {01},
	pages = {469--485},
	title = {An Analog Method for Seasonal Forecasting in Northern High Latitudes},
	volume = {11},
	year = {2021}}

@article{Ding:2019aa,
	author = {Ding, Hui and Newman, Matthew and Alexander, Michael A. and Wittenberg, Andrew T.},
	date = {2019/02/16},
	doi = {https://doi.org/10.1029/2018GL080598},
	isbn = {0094-8276},
	journal = {Geophysical Research Letters},
	journal1 = {Geophysical Research Letters},
	journal2 = {Geophysical Research Letters},
	journal3 = {Geophys. Res. Lett.},
	month = {2025/01/09},
	n2 = {Abstract Retrospective tropical Indo-Pacific forecasts for 1961?2015 are made using 28 models from the fifth phase of the Coupled Model Intercomparison Project (CMIP5) plus four models from the North American Multi-Model Ensemble (NMME), using a model-analog technique. Forecast ensembles are extracted from preexisting model simulations, by finding those states that initially best match an observed anomaly and tracking their subsequent evolution, requiring no additional model integrations. Model-analog forecasts from the 10 ?best? CMIP5 models have skill for sea surface temperature and precipitation comparable to that of both the NMME model-analog forecast ensemble and (since 1982) traditional assimilation-initialized NMME hindcasts. The El Ni{\~n}o?Southern Oscillation (ENSO) forecast skill has no trend over the 55-year period, and its decadal variations appear largely random, although the skill does improve during epochs of increased ENSO activity. Including the CMIP5-projected effects of external radiative forcings improves the tropical sea surface temperature skill of the model-analog forecasts but not within the ENSO region.},
	number = {3},
	pages = {1721--1730},
	publisher = {John Wiley \& Sons, Ltd},
	title = {Diagnosing Secular Variations in Retrospective ENSO Seasonal Forecast Skill Using CMIP5 Model-Analogs},
	url = {https://doi.org/10.1029/2018GL080598},
	volume = {46},
	year = {2019},
	year1 = {2019}}

@article { TheNorthAmericanMultimodelEnsemblePhase1SeasonaltoInterannualPredictionPhase2towardDevelopingIntraseasonalPrediction,
      author = "Ben P. Kirtman and Dughong Min and Johnna M. Infanti and James L. Kinter and Daniel A. Paolino and Qin Zhang and Huug van den Dool and Suranjana Saha and Malaquias Pena Mendez and Emily Becker and Peitao Peng and Patrick Tripp and Jin Huang and David G. DeWitt and Michael K. Tippett and Anthony G. Barnston and Shuhua Li and Anthony Rosati and Siegfried D. Schubert and Michele Rienecker and Max Suarez and Zhao E. Li and Jelena Marshak and Young-Kwon Lim and Joseph Tribbia and Kathleen Pegion and William J. Merryfield and Bertrand Denis and Eric F. Wood",
      title = "The North American Multimodel Ensemble: Phase-1 Seasonal-to-Interannual Prediction; Phase-2 toward Developing Intraseasonal Prediction",
      journal = "Bulletin of the American Meteorological Society",
      year = "2014",
      publisher = "American Meteorological Society",
      address = "Boston MA, USA",
      volume = "95",
      number = "4",
      doi = "10.1175/BAMS-D-12-00050.1",
      pages=      "585 - 601",
      url = "https://journals.ametsoc.org/view/journals/bams/95/4/bams-d-12-00050.1.xml"
}

@article{Lou:2023aa,
	author = {Lou, Jiale and Newman, Matthew and Hoell, Andrew},
	date = {2023/07/14},
	doi = {10.1038/s41612-023-00417-z},
	id = {Lou2023},
	isbn = {2397-3722},
	journal = {npj Climate and Atmospheric Science},
	number = {1},
	pages = {89},
	title = {Multi-decadal variation of ENSO forecast skill since the late 1800s},
	url = {https://doi.org/10.1038/s41612-023-00417-z},
	volume = {6},
	year = {2023}}

@article{Lorenz:1969aa,
	address = {Boston MA, USA},
	author = {Lorenz, Edward N.},
	date = {01 Jul. 1969},
	doi = {https://doi.org/10.1175/1520-0469(1969)26<636:APARBN>2.0.CO;2},
	journal = {Journal of Atmospheric Sciences},
	la = {English},
	number = {4},
	pages = {636--646},
	publisher = {American Meteorological Society},
	title = {Atmospheric Predictability as Revealed by Naturally Occurring Analogues},
	url = {https://journals.ametsoc.org/view/journals/atsc/26/4/1520-0469_1969_26_636_aparbn_2_0_co_2.xml},
	volume = {26},
	year = {1969}}

@article{Pegion:2019aa,
	address = {Boston MA, USA},
	author = {Pegion, Kathy and Kirtman, Ben P. and Becker, Emily and Collins, Dan C. and LaJoie, Emerson and Burgman, Robert and Bell, Ray and DelSole, Timothy and Min, Dughong and Zhu, Yuejian and Li, Wei and Sinsky, Eric and Guan, Hong and Gottschalck, Jon and Metzger, E. Joseph and Barton, Neil P and Achuthavarier, Deepthi and Marshak, Jelena and Koster, Randal D. and Lin, Hai and Gagnon, Normand and Bell, Michael and Tippett, Michael K. and Robertson, Andrew W. and Sun, Shan and Benjamin, Stanley G. and Green, Benjamin W. and Bleck, Rainer and Kim, Hyemi},
	date = {01 Oct. 2019},
	doi = {https://doi.org/10.1175/BAMS-D-18-0270.1},
	journal = {Bulletin of the American Meteorological Society},
	la = {English},
	number = {10},
	pages = {2043--2060},
	publisher = {American Meteorological Society},
	title = {The Subseasonal Experiment (SubX): A Multimodel Subseasonal Prediction Experiment},
	url = {https://journals.ametsoc.org/view/journals/bams/100/10/bams-d-18-0270.1.xml},
	volume = {100},
	year = {2019}}

@article{Breeden:2022aa,
	address = {Boston MA, USA},
	author = {Breeden, Melissa L. and Albers, John R. and Butler, Amy H. and Newman, Matthew},
	date = {01 Oct. 2022},
	doi = {https://doi.org/10.1175/MWR-D-22-0062.1},
	journal = {Monthly Weather Review},
	la = {English},
	number = {10},
	pages = {2617--2628},
	publisher = {American Meteorological Society},
	title = {The Spring Minimum in Subseasonal 2-m Temperature Forecast Skill over North America},
	url = {https://journals.ametsoc.org/view/journals/mwre/150/10/MWR-D-22-0062.1.xml},
	volume = {150},
	year = {2022}}

@article{Merryfield:2020aa,
	address = {Boston MA, USA},
	author = {Merryfield, William J. and Baehr, Johanna and Batt{\'e}, Lauriane and Becker, Emily J. and Butler, Amy H. and Coelho, Caio A. S. and Danabasoglu, Gokhan and Dirmeyer, Paul A. and Doblas-Reyes, Francisco J. and Domeisen, Daniela I. V. and Ferranti, Laura and Ilynia, Tatiana and Kumar, Arun and M{\"u}ller, Wolfgang A. and Rixen, Michel and Robertson, Andrew W. and Smith, Doug M. and Takaya, Yuhei and Tuma, Matthias and Vitart, Frederic and White, Christopher J. and Alvarez, Mariano S. and Ardilouze, Constantin and Attard, Hannah and Baggett, Cory and Balmaseda, Magdalena A. and Beraki, Asmerom F. and Bhattacharjee, Partha S. and Bilbao, Roberto and de Andrade, Felipe M. and DeFlorio, Michael J. and D{\'\i}az, Leandro B. and Ehsan, Muhammad Azhar and Fragkoulidis, Georgios and Gonzalez, Alex O. and Grainger, Sam and Green, Benjamin W. and Hell, Momme C. and Infanti, Johnna M. and Isensee, Katharina and Kataoka, Takahito and Kirtman, Ben P. and Klingaman, Nicholas P. and Lee, June-Yi and Mayer, Kirsten and McKay, Roseanna and Mecking, Jennifer V. and Miller, Douglas E. and Neddermann, Nele and Justin Ng, Ching Ho and Oss{\'o}, Albert and Pankatz, Klaus and Peatman, Simon and Pegion, Kathy and Perlwitz, Judith and Recalde-Coronel, G. Cristina and Reintges, Annika and Renkl, Christoph and Solaraju-Murali, Balakrishnan and Spring, Aaron and Stan, Cristiana and Sun, Y. Qiang and Tozer, Carly R. and Vigaud, Nicolas and Woolnough, Steven and Yeager, Stephen},
	date = {01 Jun. 2020},
	doi = {https://doi.org/10.1175/BAMS-D-19-0037.1},
	journal = {Bulletin of the American Meteorological Society},
	la = {English},
	number = {6},
	pages = {E869--E896},
	publisher = {American Meteorological Society},
	title = {Current and Emerging Developments in Subseasonal to Decadal Prediction},
	url = {https://journals.ametsoc.org/view/journals/bams/101/6/bamsD190037.xml},
	volume = {101},
	year = {2020}}

@article{Albers:2019aa,
	author = {Albers, John R. and Newman, Matthew},
	date = {2019/11/16},
	doi = {https://doi.org/10.1029/2019GL085270},
	isbn = {0094-8276},
	journal = {Geophysical Research Letters},
	journal1 = {Geophysical Research Letters},
	journal2 = {Geophysical Research Letters},
	journal3 = {Geophys. Res. Lett.},
	month = {2025/01/09},
	n2 = {Abstract The current generation of subseasonal operational model forecasts has, on average, low skill for leads beyond 3 weeks. This is likely a fundamental property of the climate system, due to the relative high amplitude of unpredictable weather variability compared to potentially predictable, but generally weaker, climate signals. Thus, for subseasonal forecasts to be useful, their high versus low skill events should be identified at time of forecast. We show that a linear inverse model (LIM), an empirical-dynamical model constructed from covariability statistics of wintertime (December?March) weekly averaged observational analyses, can be used to identify, a priori, the expected extratropical subseasonal surface and midtropospheric forecast skill. The LIM's predicted signal-to-noise ratio identifies the subset (10{\%}?30{\%}) of Weeks 3?6 forecasts?of the LIM and two operational models from the National Centers for Environmental Prediction and the European Centre for Medium-Range Weather Forecasts?with relatively higher skill versus the much larger remainder of forecasts whose skill cannot be distinguished from random chance.},
	number = {21},
	pages = {12527--12536},
	publisher = {John Wiley \& Sons, Ltd},
	title = {A Priori Identification of Skillful Extratropical Subseasonal Forecasts},
	url = {https://doi.org/10.1029/2019GL085270},
	volume = {46},
	year = {2019},
	year1 = {2019}}

@article{Mayer:2021aa,
	author = {Mayer, Kirsten J. and Barnes, Elizabeth A.},
	date = {2021/05/28},
	doi = {https://doi.org/10.1029/2020GL092092},
	isbn = {0094-8276},
	journal = {Geophysical Research Letters},
	journal1 = {Geophysical Research Letters},
	journal2 = {Geophysical Research Letters},
	journal3 = {Geophys Res Lett},
	month = {2025/01/09},
	n2 = {Abstract Midlatitude prediction on subseasonal timescales is difficult due to the chaotic nature of the atmosphere and often requires the identification of favorable atmospheric conditions that may lead to enhanced skill (?forecasts of opportunity?). Here, we demonstrate that an artificial neural network (ANN) can identify such opportunities for tropical-extratropical circulation teleconnections within the North Atlantic (40$\,^{\circ}$N, 325$\,^{\circ}$E) at a lead of 22 days using the network's confidence in a given prediction. Furthermore, layer-wise relevance propagation (LRP), an ANN explainability technique, pinpoints the relevant tropical features the ANN uses to make accurate predictions. We find that LRP identifies tropical hot spots that correspond to known favorable regions for midlatitude teleconnections and reveals a potential new pattern for prediction in the North Atlantic on subseasonal timescales.},
	number = {10},
	pages = {e2020GL092092},
	publisher = {John Wiley \& Sons, Ltd},
	title = {Subseasonal Forecasts of Opportunity Identified by an Explainable Neural Network},
	url = {https://doi.org/10.1029/2020GL092092},
	volume = {48},
	year = {2021},
	year1 = {2021}}

@article{Mariotti:2020aa,
	address = {Boston MA, USA},
	author = {Mariotti, Annarita and Baggett, Cory and Barnes, Elizabeth A. and Becker, Emily and Butler, Amy and Collins, Dan C. and Dirmeyer, Paul A. and Ferranti, Laura and Johnson, Nathaniel C. and Jones, Jeanine and Kirtman, Ben P. and Lang, Andrea L. and Molod, Andrea and Newman, Matthew and Robertson, Andrew W. and Schubert, Siegfried and Waliser, Duane E. and Albers, John},
	date = {01 May. 2020},
	doi = {https://doi.org/10.1175/BAMS-D-18-0326.1},
	journal = {Bulletin of the American Meteorological Society},
	la = {English},
	number = {5},
	pages = {E608--E625},
	publisher = {American Meteorological Society},
	title = {Windows of Opportunity for Skillful Forecasts Subseasonal to Seasonal and Beyond},
	url = {https://journals.ametsoc.org/view/journals/bams/101/5/bams-d-18-0326.1.xml},
	volume = {101},
	year = {2020}}

@misc{ling2024fengwuw2sdeeplearningmodel,
	archiveprefix = {arXiv},
	author = {Fenghua Ling and Kang Chen and Jiye Wu and Tao Han and Jing-Jia Luo and Wanli Ouyang and Lei Bai},
	eprint = {2411.10191},
	primaryclass = {cs.LG},
	title = {FengWu-W2S: A deep learning model for seamless weather-to-subseasonal forecast of global atmosphere},
	url = {https://arxiv.org/abs/2411.10191},
	year = {2024}}

@article{Peng:2023aa,
	author = {Peng, Yihao and Liu, Xiaolei and Su, Jingzhi and Liu, Xinli and Zhang, Yixu},
	date = {2023/09/01/},
	doi = {https://doi.org/10.1016/j.aosl.2023.100357},
	isbn = {1674-2834},
	journal = {Atmospheric and Oceanic Science Letters},
	journal1 = {Special Issue: Climate Variability, Climate Prediction and Climate Extremes - to commemorate the 20th Anniversary of Nansen-Zhu International Research Centre},
	number = {5},
	pages = {100357},
	title = {Skill improvement of the yearly updated reforecasts in ECMWF S2S prediction from 2016 to 2022},
	url = {https://www.sciencedirect.com/science/article/pii/S1674283423000351},
	volume = {16},
	year = {2023}}

@article{Vitart:2018aa,
	author = {Vitart, Fr{\'e}d{\'e}ric and Robertson, Andrew W.},
	date = {2018/03/12},
	doi = {10.1038/s41612-018-0013-0},
	id = {Vitart2018},
	isbn = {2397-3722},
	journal = {npj Climate and Atmospheric Science},
	number = {1},
	pages = {3},
	title = {The sub-seasonal to seasonal prediction project (S2S) and the prediction of extreme events},
	url = {https://doi.org/10.1038/s41612-018-0013-0},
	volume = {1},
	year = {2018}}

@article{Robertson:2018ab,
	author = {Robertson, Andrew W. and Camargo, Suzana J. and Sobel, Adam and Vitart, Frederic and Wang, Shuguang},
	date = {2018/02/21},
	doi = {10.1038/s41612-017-0009-1},
	id = {Robertson2018},
	isbn = {2397-3722},
	journal = {npj Climate and Atmospheric Science},
	number = {1},
	pages = {20178},
	title = {Summary of workshop on sub-seasonal to seasonal predictability of extreme weather and climate},
	url = {https://doi.org/10.1038/s41612-017-0009-1},
	volume = {1},
	year = {2018}}

@article{Chen:2024aa,
	author = {Chen, Lei and Zhong, Xiaohui and Li, Hao and Wu, Jie and Lu, Bo and Chen, Deliang and Xie, Shang-Ping and Wu, Libo and Chao, Qingchen and Lin, Chensen and Hu, Zixin and Qi, Yuan},
	date = {2024/07/30},
	doi = {10.1038/s41467-024-50714-1},
	id = {Chen2024},
	isbn = {2041-1723},
	journal = {Nature Communications},
	number = {1},
	pages = {6425},
	title = {A machine learning model that outperforms conventional global subseasonal forecast models},
	url = {https://doi.org/10.1038/s41467-024-50714-1},
	volume = {15},
	year = {2024}}

@article{White:2017aa,
	author = {White, Christopher J. and Carlsen, Henrik and Robertson, Andrew W. and Klein, Richard J.T. and Lazo, Jeffrey K. and Kumar, Arun and Vitart, Frederic and Coughlan de Perez, Erin and Ray, Andrea J. and Murray, Virginia and Bharwani, Sukaina and MacLeod, Dave and James, Rachel and Fleming, Lora and Morse, Andrew P. and Eggen, Bernd and Graham, Richard and Kjellstr{\"o}m, Erik and Becker, Emily and Pegion, Kathleen V. and Holbrook, Neil J. and McEvoy, Darryn and Depledge, Michael and Perkins-Kirkpatrick, Sarah and Brown, Timothy J. and Street, Roger and Jones, Lindsey and Remenyi, Tomas A. and Hodgson-Johnston, Indi and Buontempo, Carlo and Lamb, Rob and Meinke, Holger and Arheimer, Berit and Zebiak, Stephen E.},
	doi = {https://doi.org/10.1002/met.1654},
	eprint = {https://rmets.onlinelibrary.wiley.com/doi/pdf/10.1002/met.1654},
	journal = {Meteorological Applications},
	number = {3},
	pages = {315-325},
	title = {Potential applications of subseasonal-to-seasonal (S2S) predictions},
	url = {https://rmets.onlinelibrary.wiley.com/doi/abs/10.1002/met.1654},
	volume = {24},
	year = {2017}}
\clearpage





\renewcommand{\thesection}{S\arabic{section}}
\renewcommand{\thesubsection}{S\arabic{section}.\arabic{subsection}}
\renewcommand{\thesubsubsection}{S\arabic{section}.\arabic{subsection}.\arabic{subsubsection}}
\setcounter{section}{0}
\renewcommand{\thefigure}{S\arabic{figure}}
\setcounter{figure}{0}
\renewcommand{\thetable}{S\arabic{table}}
\setcounter{table}{0}
\title[]{Supplemental Information for \change{AI-Informed Model Analogs for S2S Prediction}{AI-Informed Model-Analogs for Understanding Subseasonal-to-Seasonal Jet Stream and North American Temperature Predictability}}
\maketitle
\doublespacing
\section{Week 3-4 Windows Southern California Predictions}
\label{sec:CA}
\subsection{Daily Data}
We employ a 7-day sliding window to smooth daily CESM2-LE and ERA5 data, using a backward moving average for input data and a forward moving average for target data. All smoothed daily data is regridded via bilinear interpolation to 2.5° x 2.5° resolution. We use a coarser resolution for the daily data to reduce the memory load, as our analog library of daily data climate maps is $\sim10\times$ larger than the library of monthly data. This data is similarly converted to anomalies about the seasonal cycle and then to standard deviations at each grid point.
For the daily CESM2-LE data, we subtract the linear trend from each calendar day at each grid point. We include a shorter timespan of data from each member (1850-1949) than the monthly data, as each daily-data year contains more than $30\times$ the amount of samples. The analog library is composed of fields from the first 5 members, while fields from the next 4 members (with a 2/1/1 training/validation/testing split) make up the SOIs (see Table \ref{tab:daily_ensemble_members} for member details). For daily ERA5 data, we use dates between 1942-2023, fitting and subtracting a third-order polynomial at each grid point and each calendar day to define detrended anomalies.

\subsection{Week 3-4 Results}
\begin{figure}[ht]
    \centering
    \includegraphics[width=\textwidth]{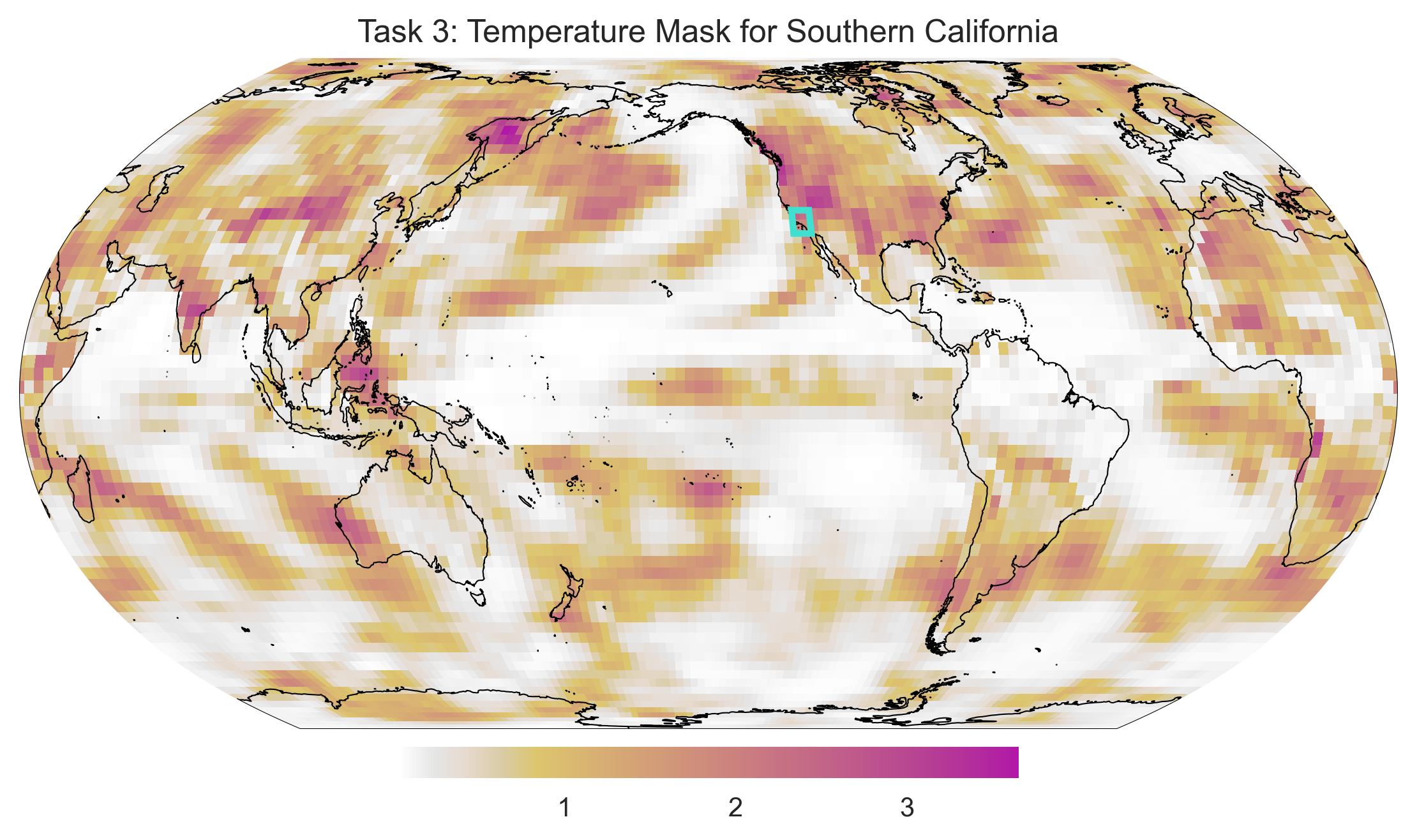}
    \caption{The learned mask for Task \#3, Southern California summer temperature classification. The cyan box outlines the target region.}
    \label{fig:california_mask}
\end{figure}

We also assess the short-range S2S skill of the AI-based model-analog approach by classifying Week 3-4 Southern California ($32^\circ-37^\circ \mathrm{N},\ 116^\circ-121^\circ \mathrm{W}$) summer temperatures (Task \#1). The three target classes (cold, neutral, and warm) are formed as in Task \#2, where target temperatures are split into terciles, ensuring all classes are equally sized. We predict each 2-week period from the third week of June through the third week of September, using the learned weights in Figure \ref{fig:california_mask}. This mask exhibits weights that are distributed globally, yet unevenly, with noticeably increased weight around the western U.S. as well as in the North Pacific. The more diffuse weighting pattern, as compared to Tasks \#1 and \#2, likely reflects the noisier synoptic patterns present in the daily data used to train this mask.

\begin{figure}[ht!]
    \centering
    \setlength{\fboxrule}{0.1pt} 
    \setlength{\fboxsep}{1pt}    
    \fbox{
        \begin{minipage}{0.97\textwidth} 
            \centering
            \begin{subfigure}{0.49\textwidth}
                \centering
                \includegraphics[width=\textwidth]{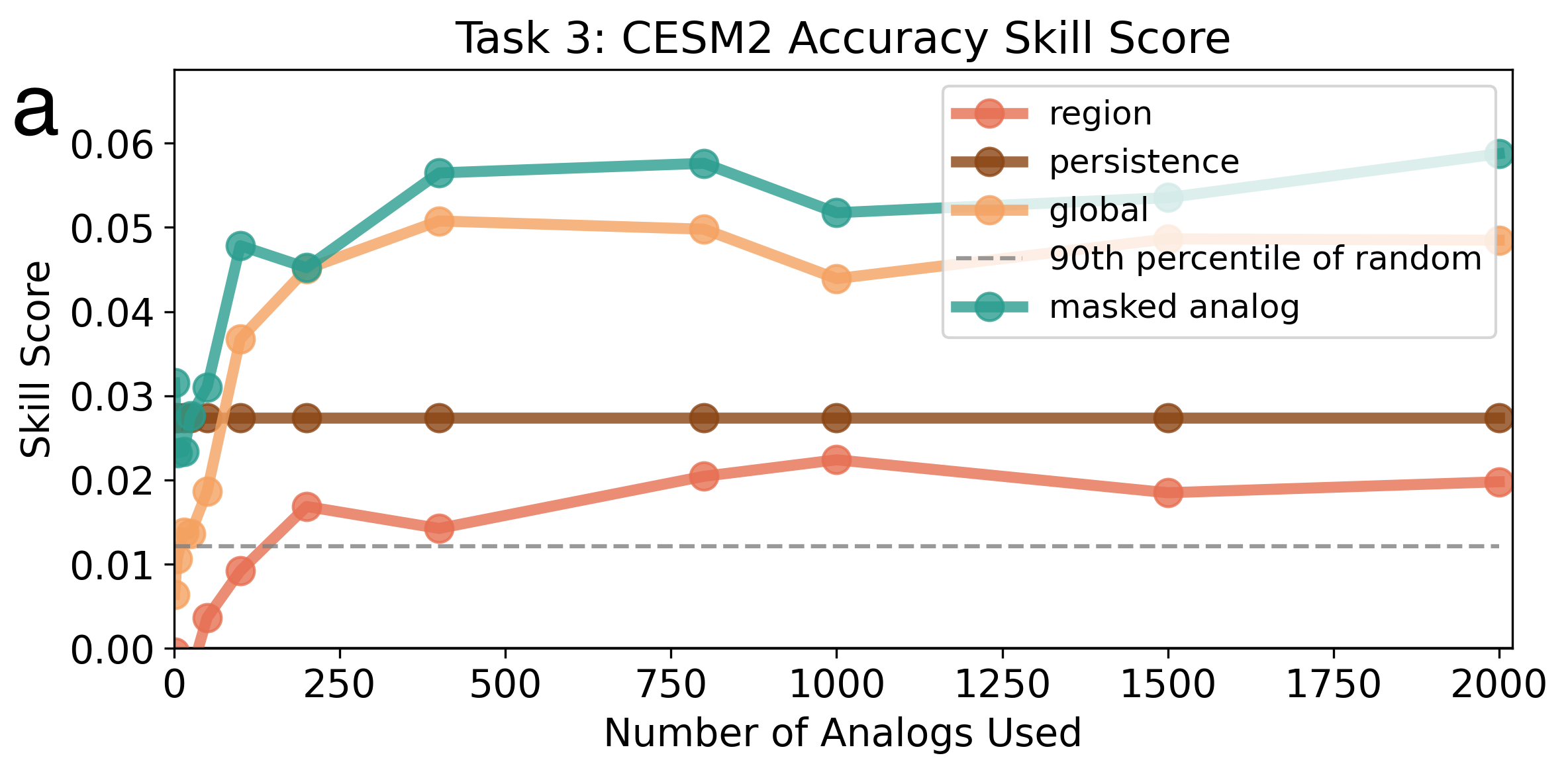}
            \end{subfigure}
            \hfill
            \begin{subfigure}{0.49\textwidth}
                \centering
                \includegraphics[width=\textwidth]{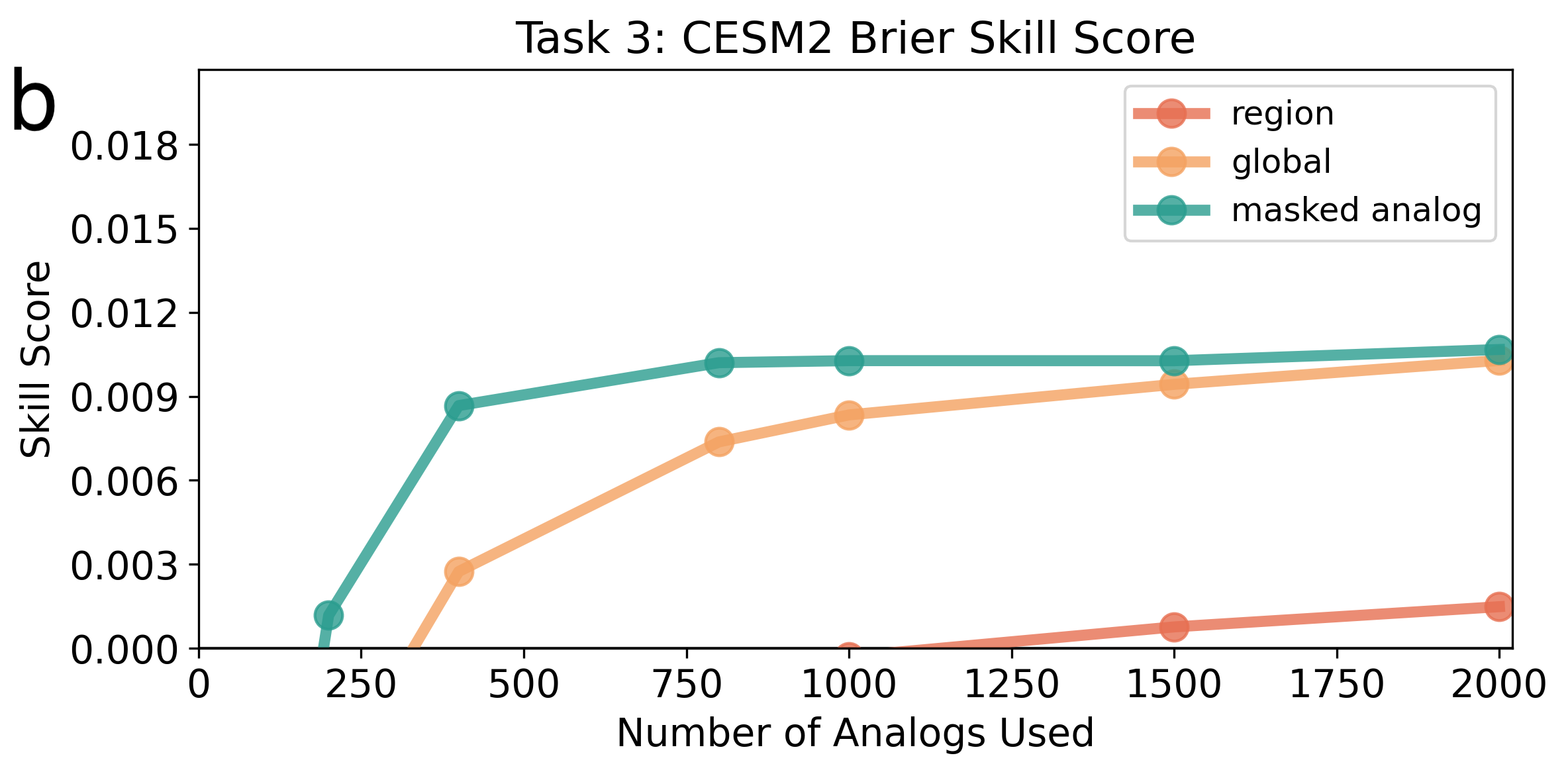}
            \end{subfigure}

            \vspace{2mm} 

            \begin{subfigure}{0.49\textwidth}
                \centering
                \includegraphics[width=\textwidth]{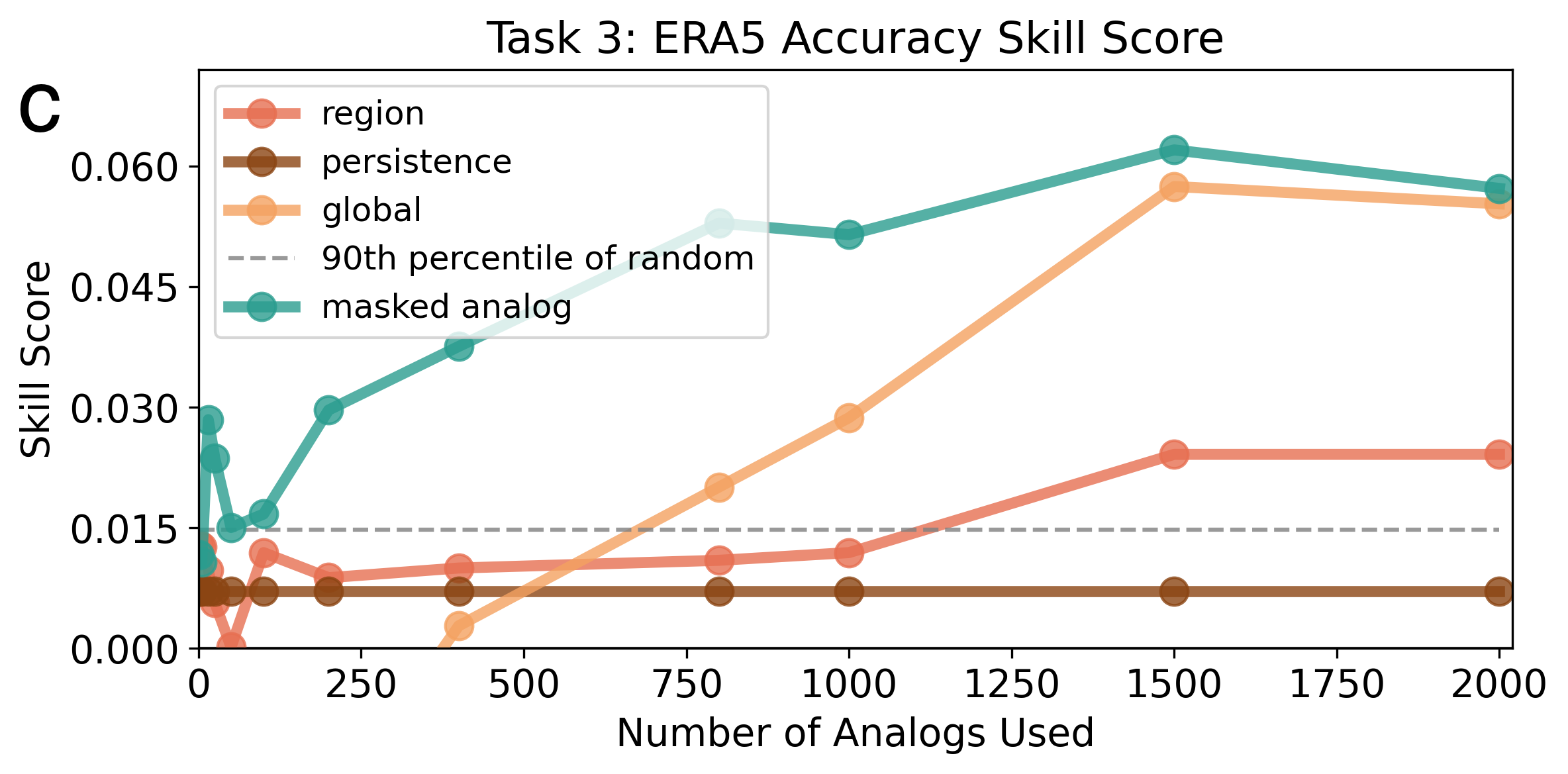}
            \end{subfigure}
            \hfill
            \begin{subfigure}{0.49\textwidth}
                \centering
                \includegraphics[width=\textwidth]{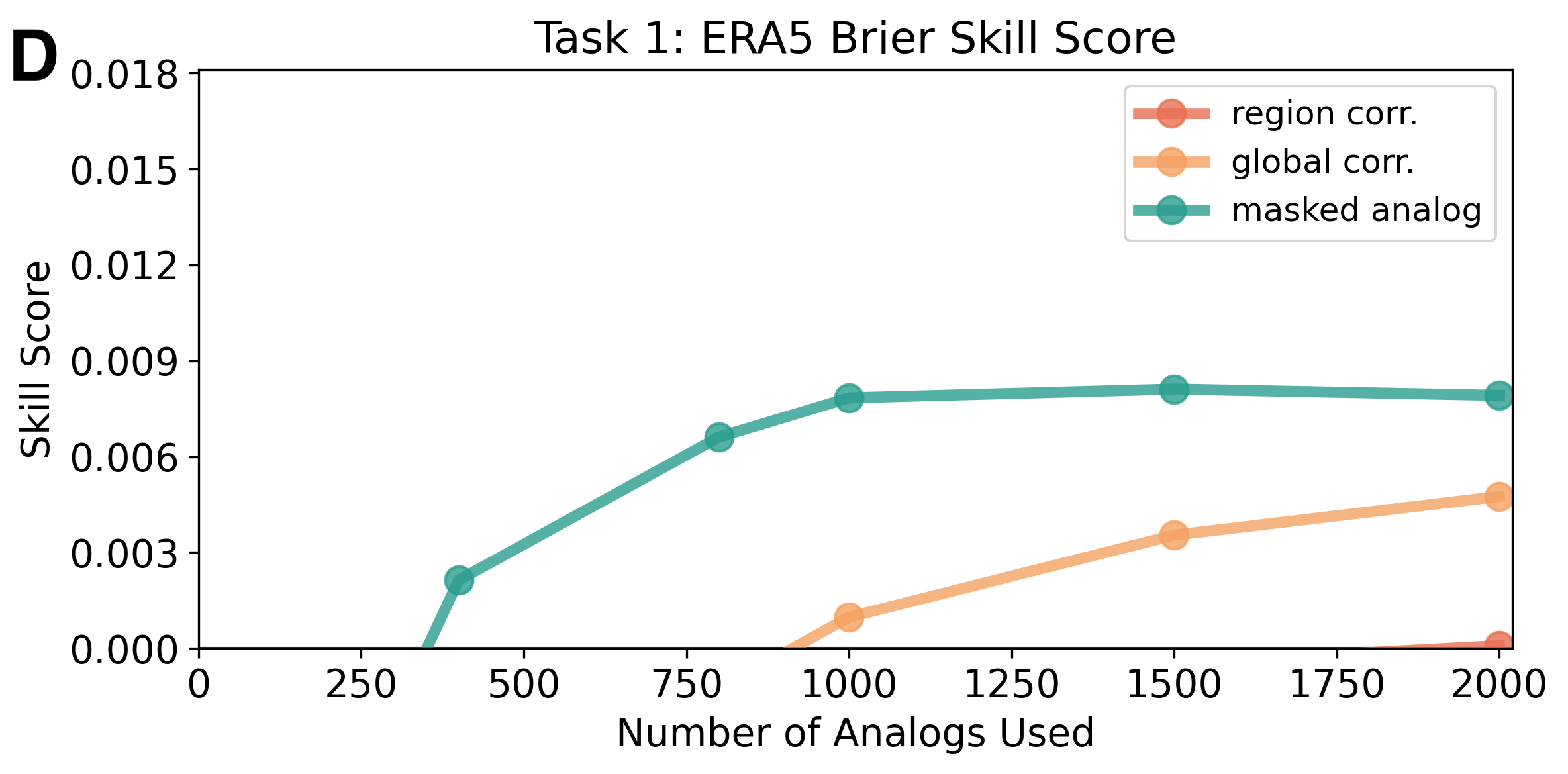}
            \end{subfigure}
        \end{minipage}
    }
    \caption{Skill scores for a) CESM2-LE accuracy, b) CESM2-LE BS, c) ERA5 accuracy, and d) ERA5 BS for Week 3-4 Southern California temperature classification.}
    \label{fig:CA_skill}
\end{figure}

 With the learned mask, MAE and BS skill scores exceed other baselines, with increases in MAE skill of 16\% and 18\% and CRPS skill of 4\% and 71\% when testing on CESM2-LE and ERA5 data, respectively (Figure \ref{fig:CA_skill}).  With the CESM2-LE test set, the highest skill is reached at 2000 analogs, while for ERA5, the skill score peaks at 1500 analogs.

\begin{figure}
    \centering
    \includegraphics[width=\textwidth]{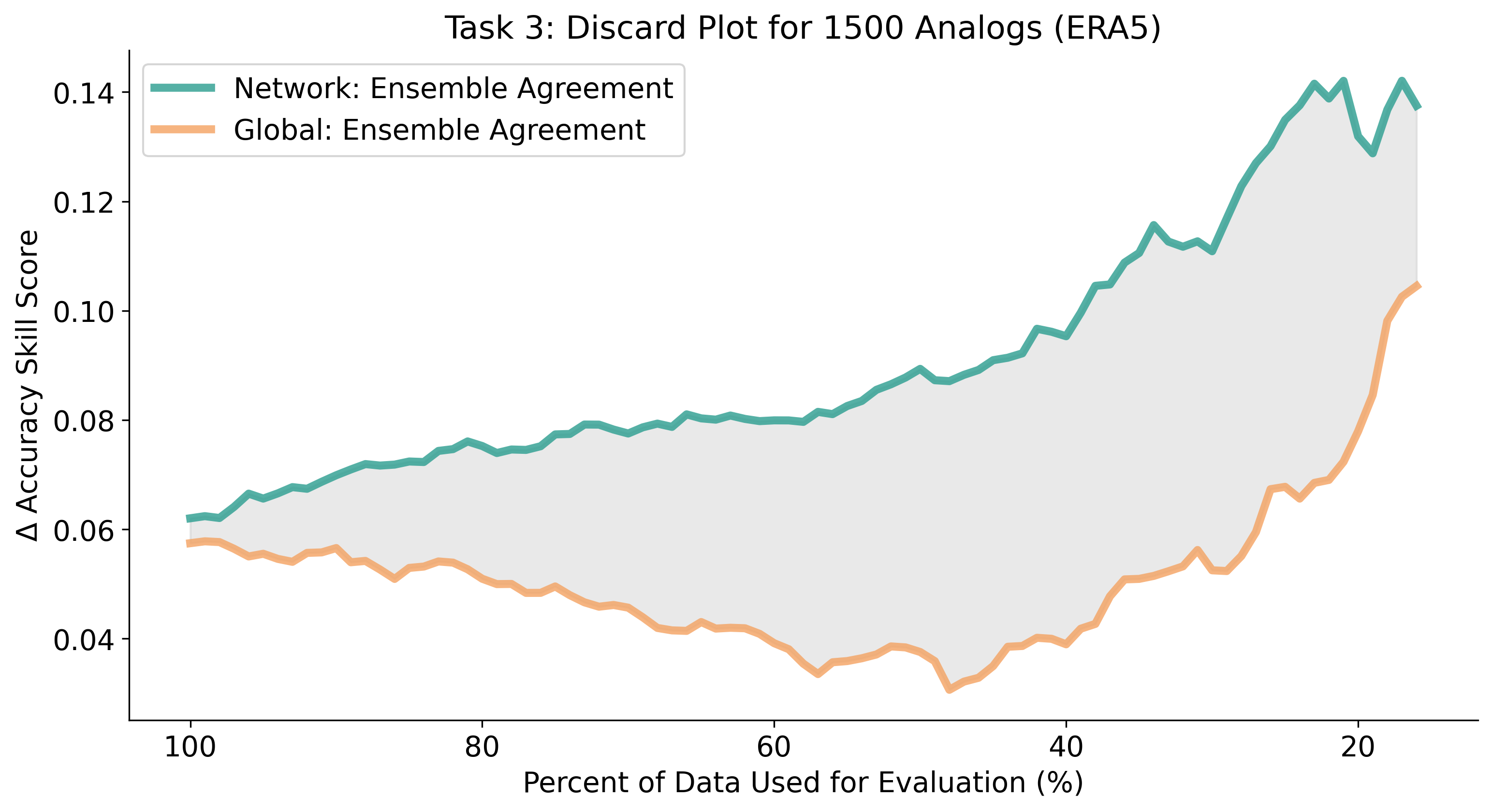}
    \caption{Discard plot based on ensemble agreement for Week 3-4 Southern California temperature classification using 1500 analogs, testing on ERA5 data. Data with the lowest ensemble agreement is progressively discarded, with the x-axis showing the percentage of data remaining.}
    \label{fig:CA_discard}
\end{figure}

\begin{figure}[ht]
    \centering
    \includegraphics[width=.9\textwidth]{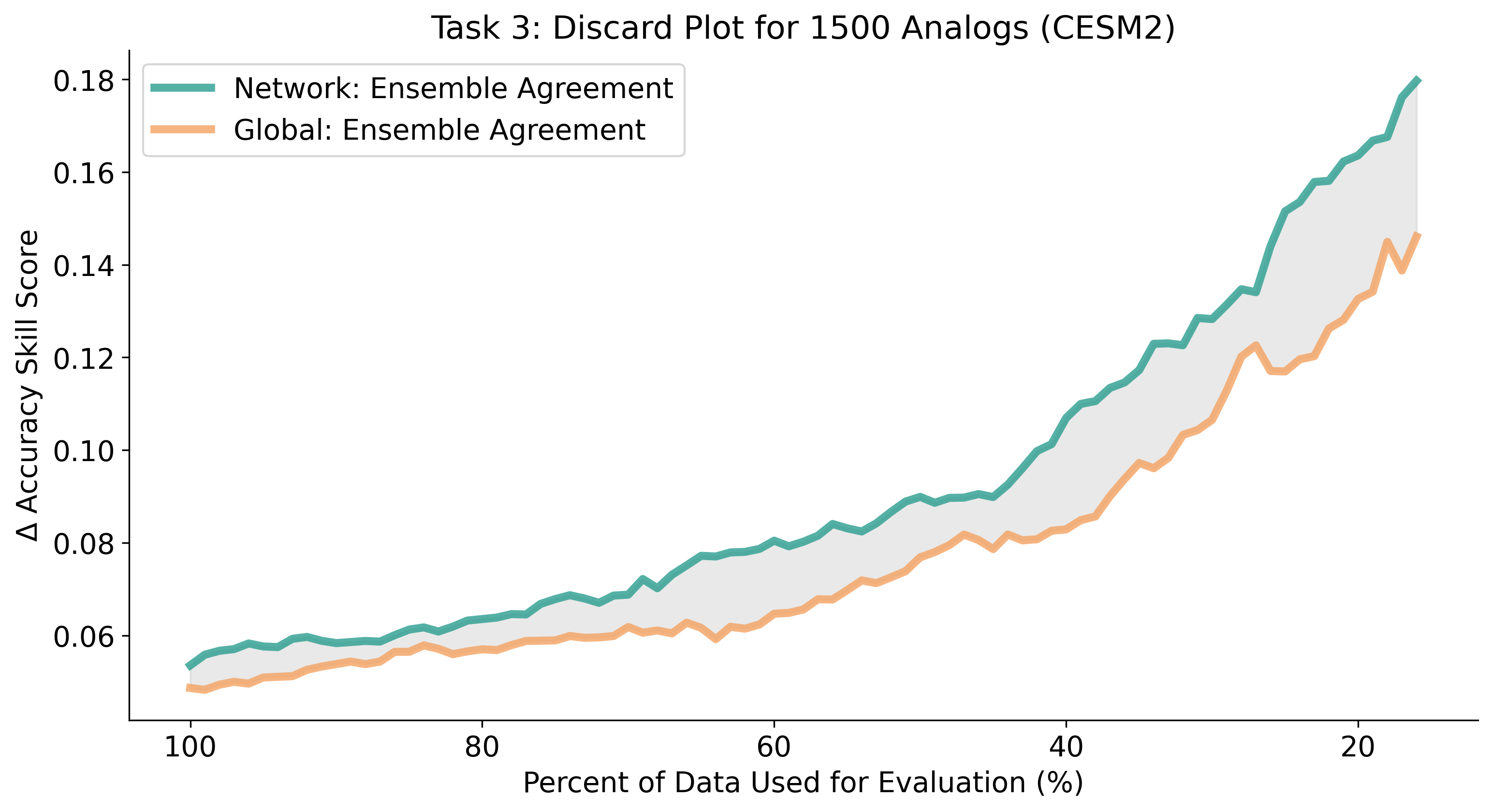}
    \caption{Discard plot for Week 3-4 Southern California temperature classification using 1500 analogs, testing on CESM2-LE data. Data with the lowest ensemble agreement is progressively discarded, with the x-axis showing the percentage of data remaining. Unlike when testing on reanalysis data, global ensemble agreement serves as a reasonable metric for forecast uncertainty. However, the learned-mask ensemble agreement still showcases a greater decrease in error for states with higher ensemble agreement.}
    \label{fig:CA_cesm2_discard}
\end{figure}

We diagnose whether the analog ensembles can offer insights into windows of opportunity via discard plots. Figure \ref{fig:CA_discard} uses ERA5 data and a 1500-analog ensemble to show the change in accuracy skill from a climatological forecast as samples with lower ensemble agreement are discarded (for CESM2 data see Figure \ref{fig:CA_cesm2_discard}). Here, ensemble agreement is computed as the fraction of ensemble members that agree on the majority prediction. We see that over all samples the mask offers just over a $4\%$ improvement in accuracy relative to climatology, but this improvement grows essentially monotonically to over $9\%$ for the $\sim25\%$ of samples with the highest ensemble agreement. This is not the case with a global mask's ensemble, which does not exhibit as precipitous of an increase in accuracy skill score and actually \textit{decreases} in accuracy until the $\sim50\%$ cutoff mark. 

\section{Neural Network}
\label{sec:supp_nn}
\subsection{Hyperparameters}
We use the following parameters for the neural network:
\begin{table}[ht!]
    \centering
    \begin{tabular}{ll}
        \toprule
        \textbf{Parameter} & \textbf{Value} \\
        \midrule
        Optimizer & Adam \\
        Learning rate & .0001 \\
        Batch size & 32 \\
        Loss function & Mean squared error \\
        Validation batch size & 1000 \\
        Early stopping patience & 30 \\
        Early stopping minimum delta & 0.0001 \\
        \bottomrule
    \end{tabular}
    \caption{Hyperparameters used in the model training.}
    \label{tab:hyperparameters}
\end{table}

\subsection[Constrained Inverse L2 Regularization]{Constrained Inverse $L_2$ Regularization}

\label{sec:L2}
To increase mask sparsity, we implement constrained inverse $L_2$ regularization. We do so to compare how post-hoc thresholding and learned sparsity compare in terms of model performance. We add the following term to the loss function:
\begin{equation}
  \frac{\lambda_2}{\sqrt{\sum_{i=1}^n w_i^2}}
\end{equation}
where $\lambda_2$ is the regularization strength, $w_i$ is the weight of the $i$th grid point, and $n$ is the total number of grid points in the weighted mask.

\begin{figure}[ht]
    \centering
    \includegraphics[width=\textwidth]{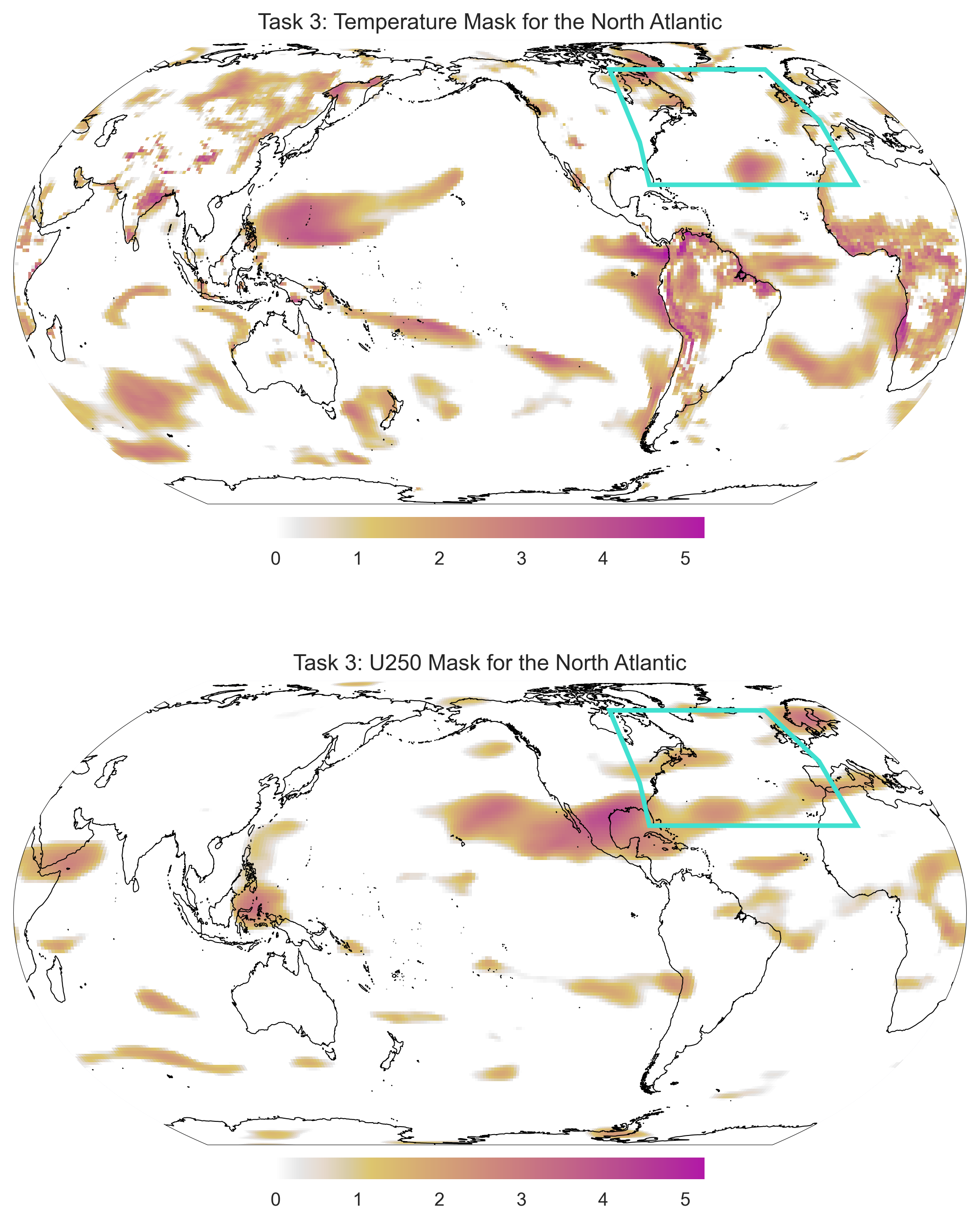}
    \caption{Using inverse $L_2$ regularization with $\lambda_2 = 100$ for the North Atlantic (Task \#2) results in a sparser map.}
    \label{fig:NAO_L2}
\end{figure}

This term is restricted during training such that $\sum_{i=1}^n w_i = n$. With this constraint, the regularization term is maximized (high loss) when $\forall i, \, w_i = 1$ and minimized when, for some $j$, $w_{j} = n$ and $w_{i \neq j} = 0$. Thus, this term promotes having more disparate weight values, with some weights of very high values and some weights of very low values. We generate the mask in Figure \ref{fig:NAO_L2} for the North Atlantic (Task \#3), using this regularization term with $\lambda_2 = 100$.
\section{CESM2-LE Members}

For monthly CESM2-LE data, we use the ensemble members listed in Table \ref{tab:monthly_ensemble_members}, while for daily CESM2-LE data, we use the ensemble members listed in Table \ref{tab:daily_ensemble_members}. The first four numbers of the member names correspond to the chosen model start dates (varying initial climate conditions), while the last three numbers indicate the realization (small perturbations to initial conditions). While we use a mixture of ocean initializations, these differences have been found to not impact S2S prediction five centuries beyond their initialization date \parencite{Arcodia_2023}.
\begin{table}[ht!]
  \centering
  \renewcommand{\arraystretch}{1.5}  
  \begin{tabular}{lp{0.65\textwidth}}  
    \toprule
    \textbf{Monthly Member Type} & \textbf{Members} \\
    \midrule
    Analog Library Members & 1301.020, 1301.019, 1301.018, 1301.017, 1301.016, 1301.015, 1301.014, 1301.013, 1301.012, 1301.011, 1281.020, 1281.019, 1281.018, 1281.017, 1281.016, 1281.015, 1281.014, 1281.013, 1281.012 \\
    SOI Train Members & 1301.010, 1301.009, 1301.008, 1301.007, 1301.006, 1301.005, 1301.004, 1301.003, 1301.002, 1301.001 \\
    SOI Validation Members & 1281.010, 1281.009 \\
    SOI Test Members & 1281.001, 1281.002 \\
    \bottomrule
  \end{tabular}
  \caption{List of monthly ensemble members used for the analog library, training SOIs, validation SOIs, and testing SOIs.}
  \label{tab:monthly_ensemble_members}
\end{table}

\begin{table}[ht!]
  \centering
  \renewcommand{\arraystretch}{1.5}  
  \begin{tabular}{lp{0.65\textwidth}}  
    \toprule
    \textbf{Daily Member Type} & \textbf{Members} \\
    \midrule
    Analog Library Members & 1231.012, 1251.013, 1251.014, 1281.015, 1281.016 \\
    SOI Train Members & 1301.017, 1301.018 \\
    SOI Validation Members & 1281.017 \\
    SOI Test Members & 1251.019 \\
    \bottomrule
  \end{tabular}
  \caption{List of daily ensemble members used for the analog library, training SOIs, validation SOIs, and testing SOIs.}
  \label{tab:daily_ensemble_members}
\end{table}

\section{Model Baslines}
\label{sec:model_baselines}
\add{While our work focuses on highlighting AI analogs utility for improving S2S model-analog forecasting and for understanding sources of predictability, here we offer a limited comparison to operational models to masked analogs Task \#1 as a reference point. We compute the MAE skill of 3 models from the North American Multi-Model Ensemble (NMME): NCEP-CFSv2 (NCEP), ECCC-CanESM5 (CAN), and ECCC-GEM5.2-NEMO (GEM) \protect{\parencite{TheNorthAmericanMultimodelEnsemblePhase1SeasonaltoInterannualPredictionPhase2towardDevelopingIntraseasonalPrediction}. We use the ensemble mean over 25 NCEP ensemble members' predictions from 1982-2010 and 21 CAN and GEM ensemble members' predictions from 1990-2023. We regrid all predictions to match the ERA5 grid, detrend, and convert to anomalies as described in Section \ref{sec:data}. These models are reported at .5- and 1.5-month leads, both of which we include both in Table \ref{tab:model_skill_scores}. We find that analog forecasting achieves lower skill than NCEP, but higher than CAN and GEM (Table \ref{tab:model_skill_scores}). Moreover, for the 15\% of most extreme forecasts, AI-based model-analog forecasting again achieves similar skill to the NMME models. AI-based model-analogs are outperformed by one model (CAN) but outperform the other two (Table \ref{tab:model_skill_scores}). Although a limited comparison is done here, these results suggest that skill for AI-based model-analog forecasting is in line with operational S2S models.}}

\begin{table}[ht!]
  \centering
  \renewcommand{\arraystretch}{1.5}  
  \begin{tabular}{l c c c}  
    \toprule
    \textbf{Model} & \textbf{Lead} & \textbf{Total Skill} & \textbf{Skill on Most Extreme Months} \\
    \midrule
    \textbf{Analogs} & \textbf{1 Month}   & \textbf{.068} & \textbf{0.24} \\
    NCEP & 0.5 Months & 0.14 & 0.23 \\
    NCEP & 1.5 Months & 0.085 & 0.19 \\
    CAN & 0.5 Months & $< 0$ & 0.30 \\
    CAN & 1 Month    & $< 0$ & 0.29 \\
    GEM & 0.5 Months & $< 0$ & 0.23 \\
    GEM & 1.5 Months & 0.015 & 0.0015 \\
    \bottomrule
  \end{tabular}
  \caption{Skill scores for the AI-based analog approach as compared to operational models (where a higher score indicates a more skillful forecast relative to climatology). We find that analog forecasting (bolded) has performance in line with NMME models for Task \#1. It outperforms two models and is outperformed by one, both for total skill and skill on the most extreme months.}
  \label{tab:model_skill_scores}
\end{table}

\section{Figures}

\begin{figure}[ht]
  \centering
  \includegraphics[width=.9\textwidth]{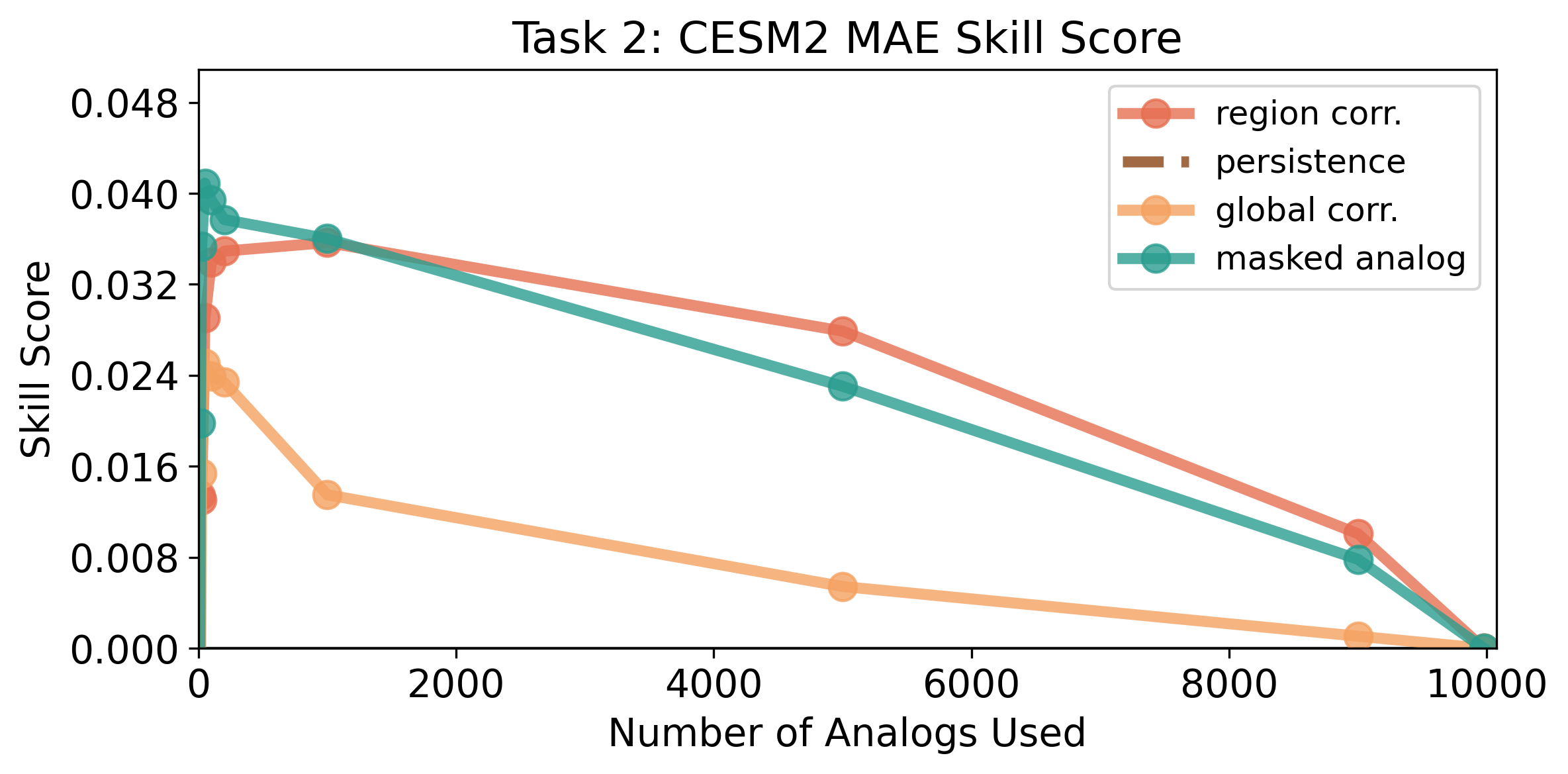}
  \caption{With regression problems, as the number of analogs approaches the total library size, the prediction becomes more and more similar to a climatological prediction. In this example the library size is $\sim10000$, where the skill is 0.}
  \label{fig:regression_to_mean}
\end{figure}
\begin{figure}[ht]
    \centering
    \includegraphics[width=.9\textwidth]{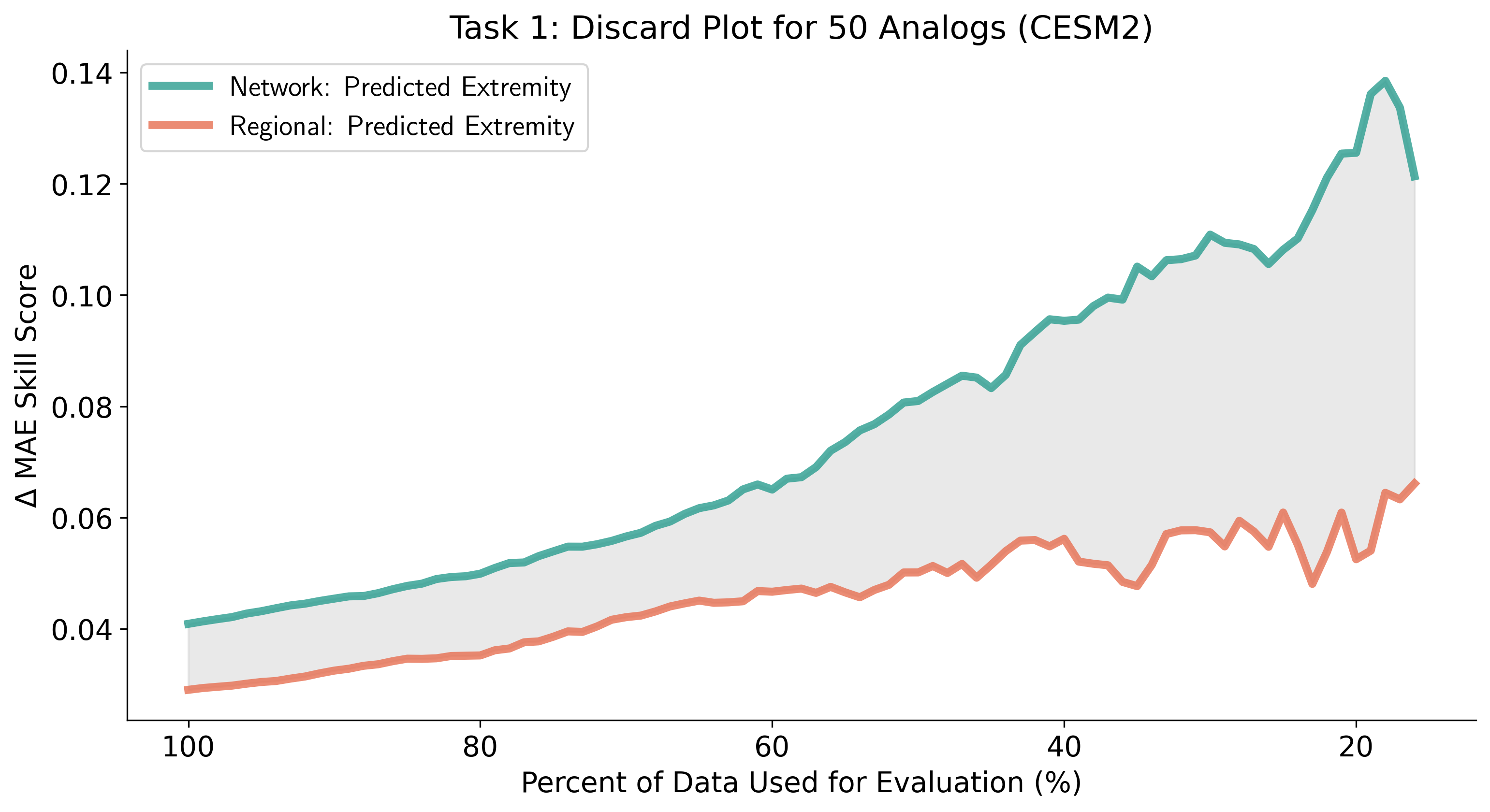}
    \caption{A discard plot for an ensemble of 50 analogs for midwestern U.S. summer temperature
    regression, testing on CESM2-LE data. Data with the lowest extremity is progressively discarded, with the x-axis showing the percentage of data remaining. While the regional mask performs relatively better on extreme predictions for CESM2-LE compared to ERA5 data, there is still a much larger improvement for extreme predictions with the learned mask.}
    \label{fig:midwest_cesm2_discard}
\end{figure}
\begin{figure}[ht]
  \centering
  \includegraphics[width=.85\textwidth]{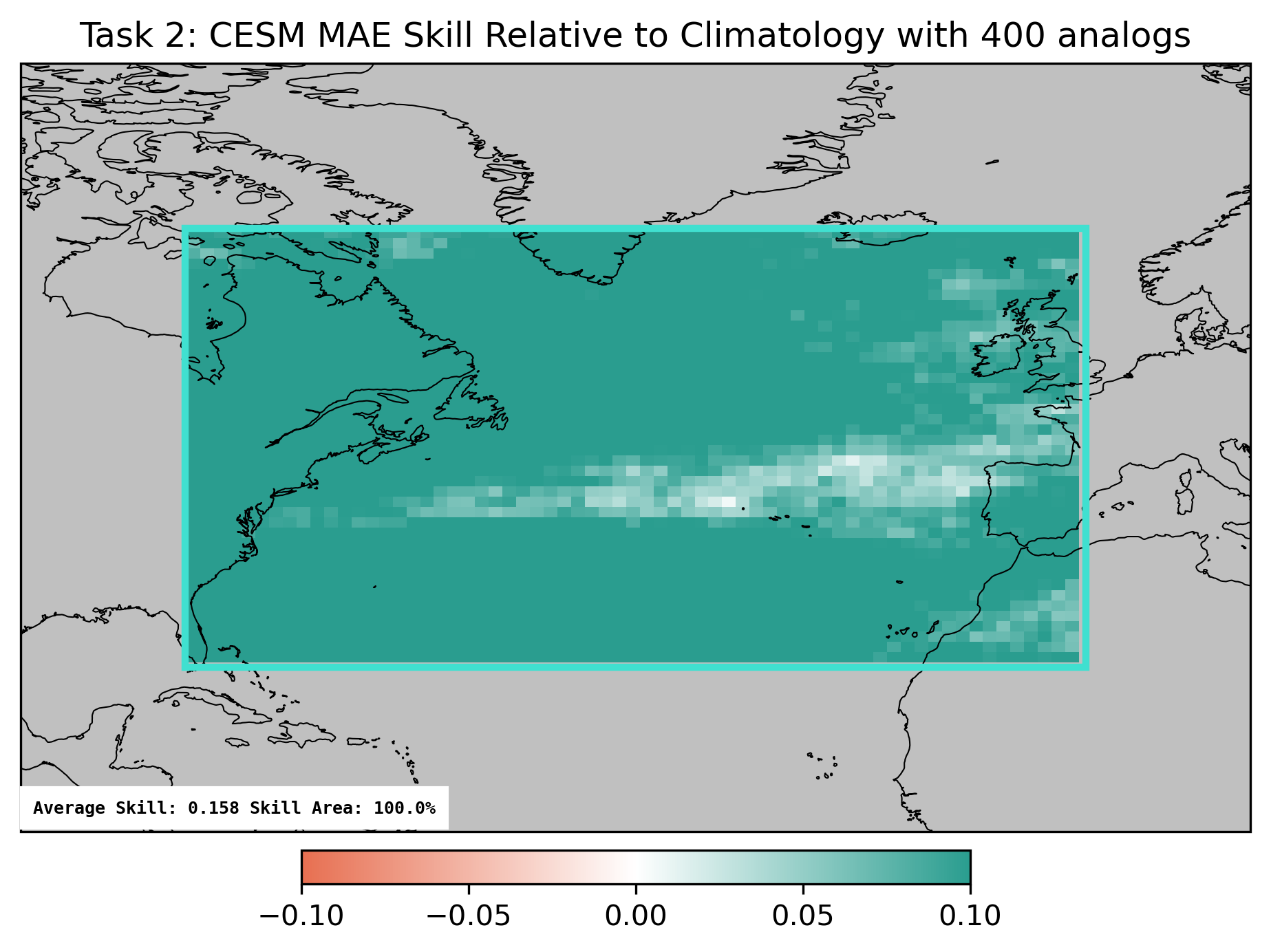}
  \caption{In Task \#2 a field of values rather than the average value across the region of interest is predicted. This map shows the average skill across SOIs when using a 400 analog ensemble on ERA5 data. At almost every grid point, the learned mask outperforms the climatological prediction.}
  \label{fig:NA_field_skill}
\end{figure}

\begin{figure}[ht!]
  \centering
  \setlength{\fboxrule}{0.1pt} 
  \setlength{\fboxsep}{1pt}    
  \fbox{
      \begin{minipage}{0.97\textwidth} 
          \centering
          \begin{subfigure}{0.49\textwidth}
              \centering
              \includegraphics[width=\textwidth]{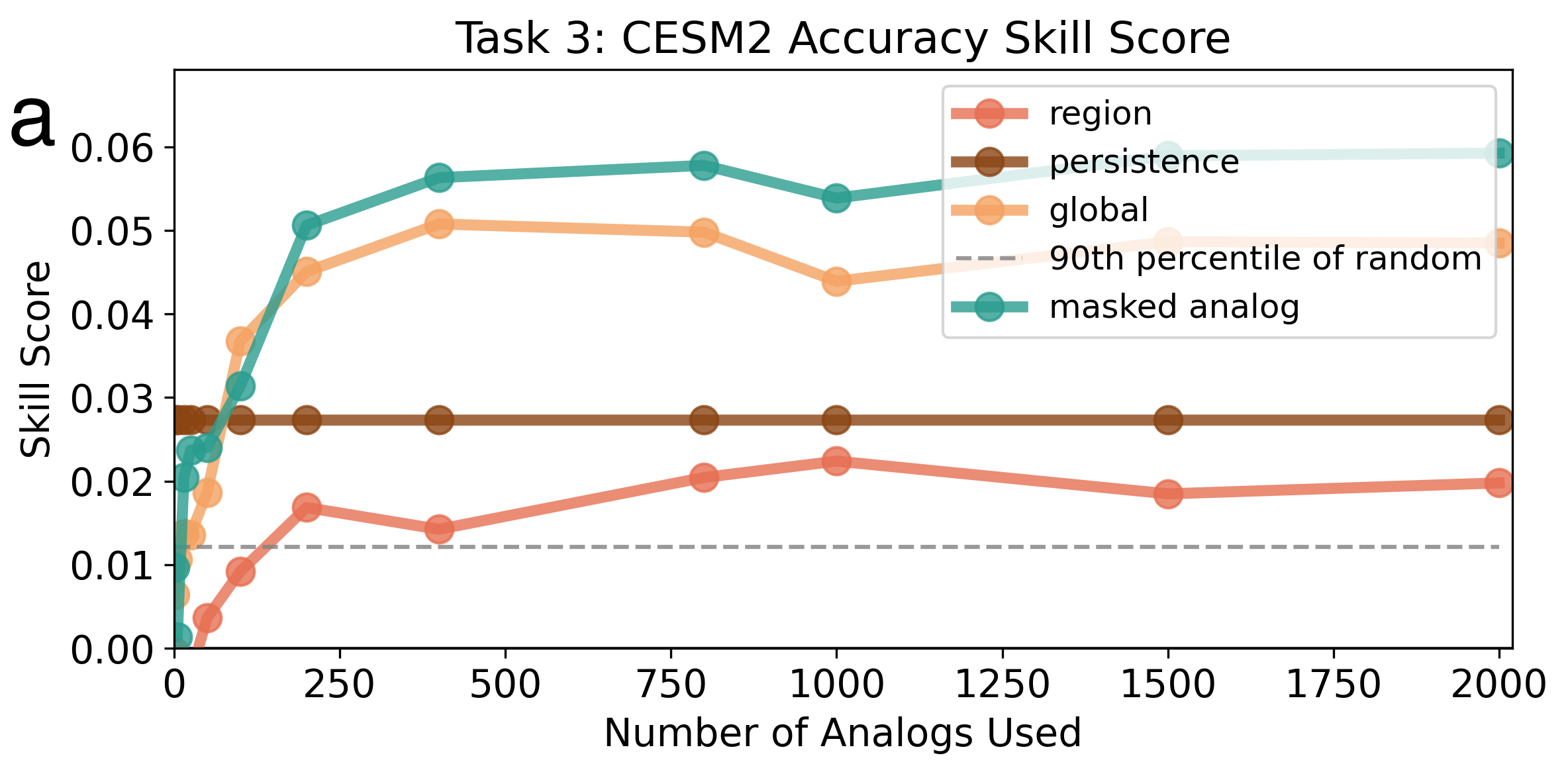}
          \end{subfigure}
          \hfill
          \begin{subfigure}{0.49\textwidth}
              \centering
              \includegraphics[width=\textwidth]{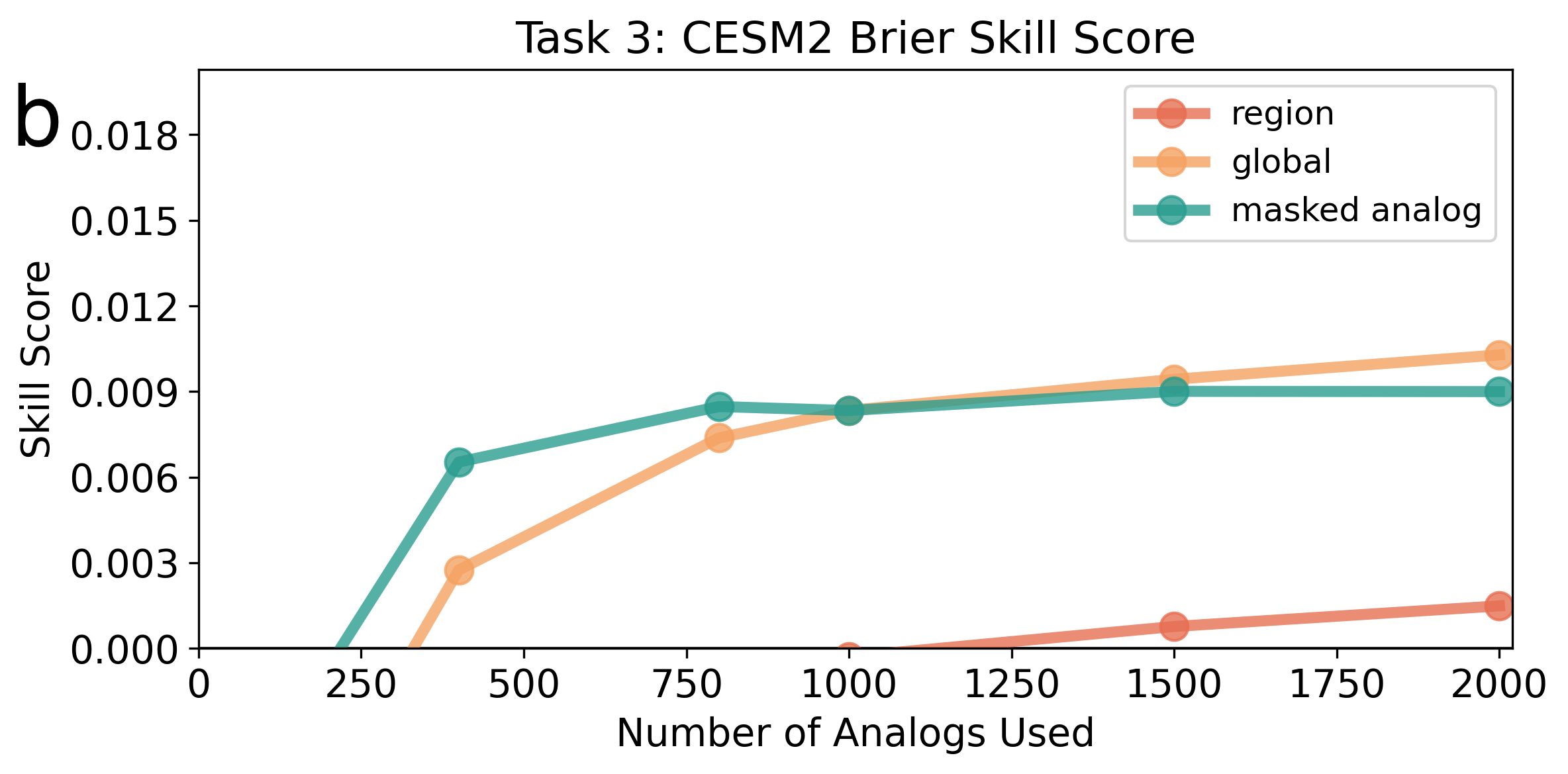}
          \end{subfigure}

          \vspace{2mm} 

          \begin{subfigure}{0.49\textwidth}
              \centering
              \includegraphics[width=\textwidth]{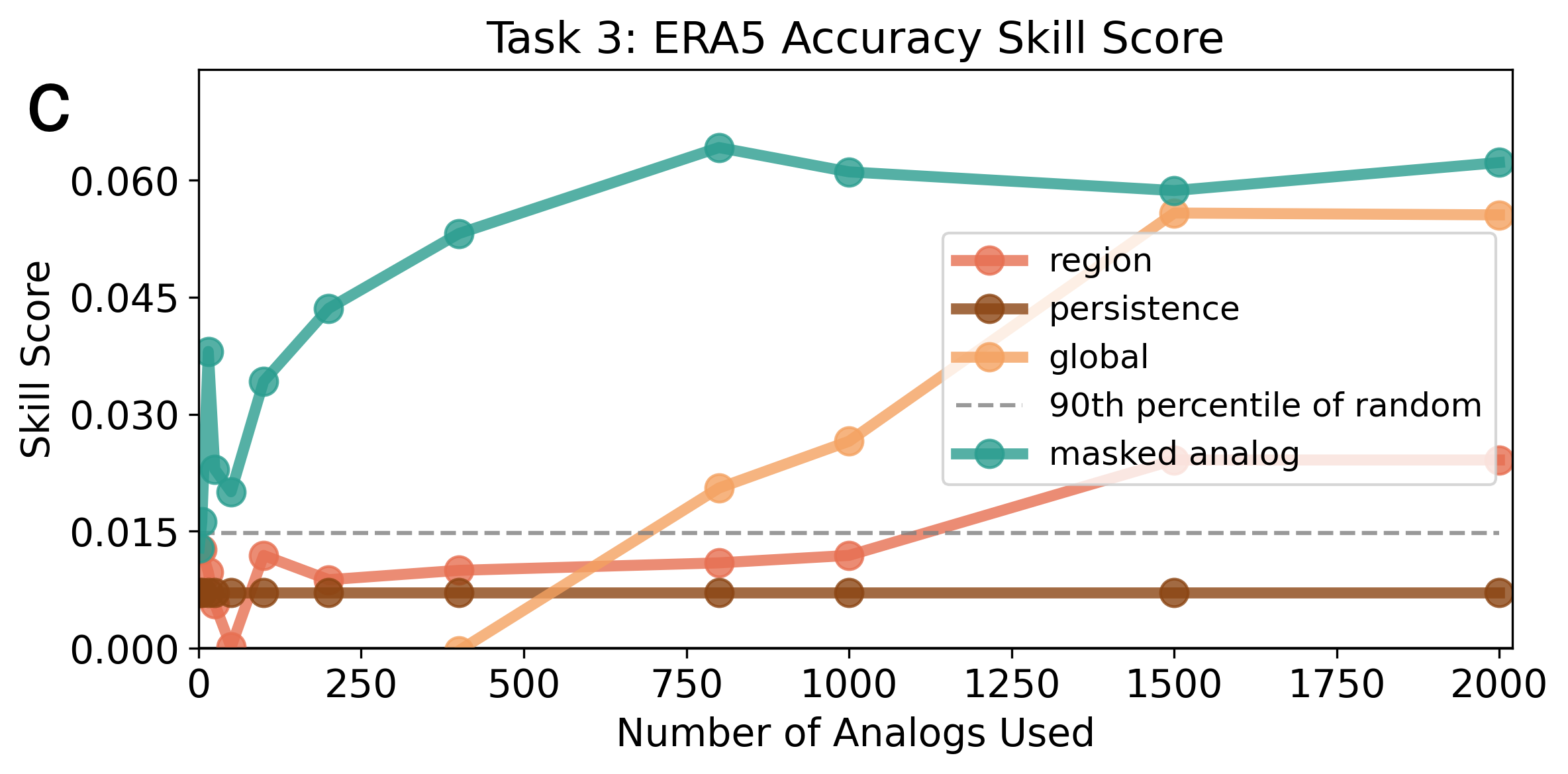}
          \end{subfigure}
          \hfill
          \begin{subfigure}{0.49\textwidth}
              \centering
              \includegraphics[width=\textwidth]{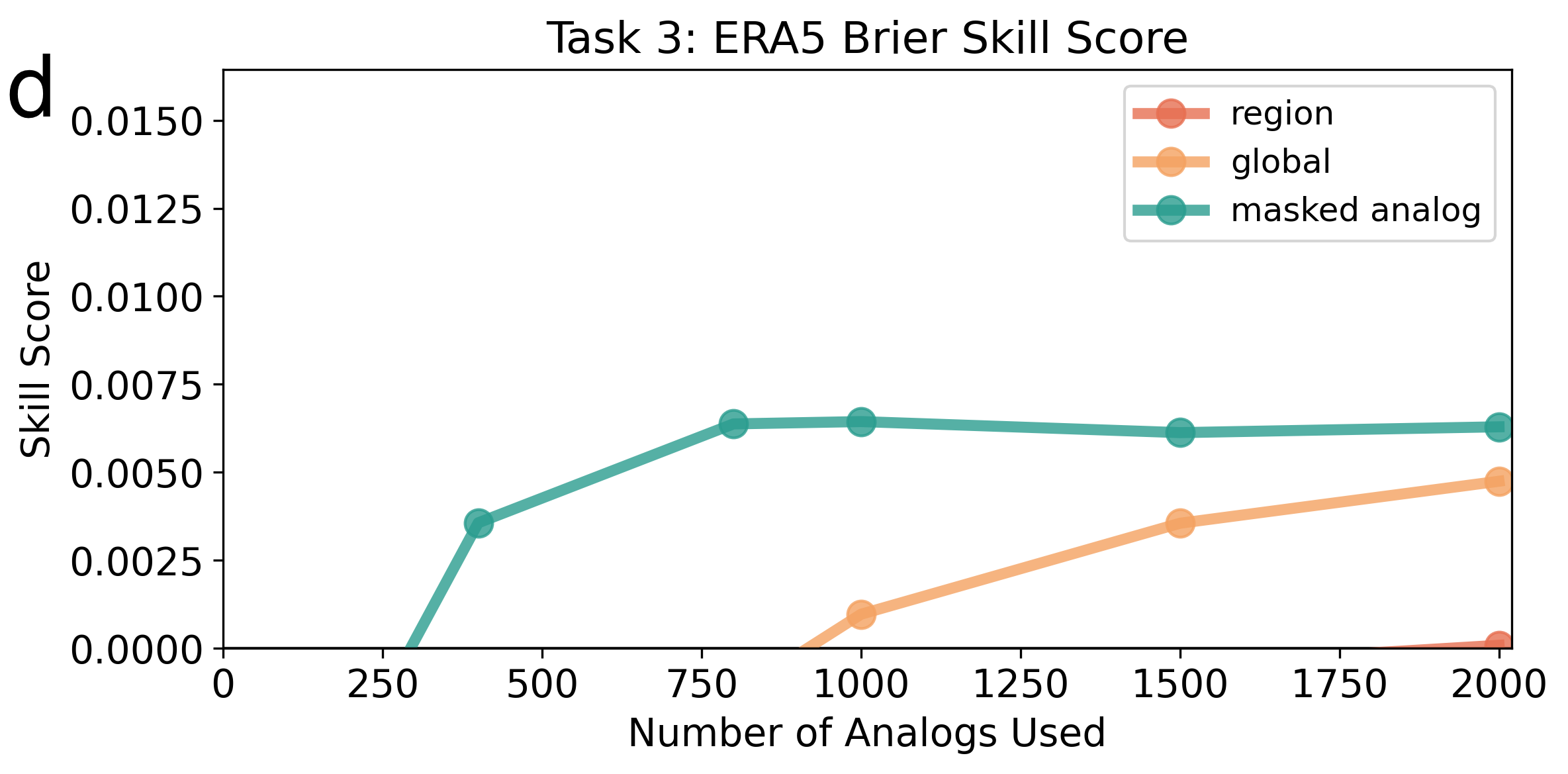}
          \end{subfigure}
      \end{minipage}
  }
  \caption{Skill scores using a 90th-percentile-thresholded mask for a) CESM2-LE accuracy, b) CESM2-LE BS, c) ERA5 accuracy, and d) ERA5 BS for Week 3-4 Southern California temperature classification. There is little difference in skill between the 90th-percentile-thresholded mask and the learned mask, although there is a slight decrease in Brier Skill Score (BS) for the 90th-percentile-thresholded mask.}
  \label{fig:CA_90th_percentile_mask}
\end{figure}

\begin{figure}[ht]
    \centering
    \includegraphics[width=\textwidth]{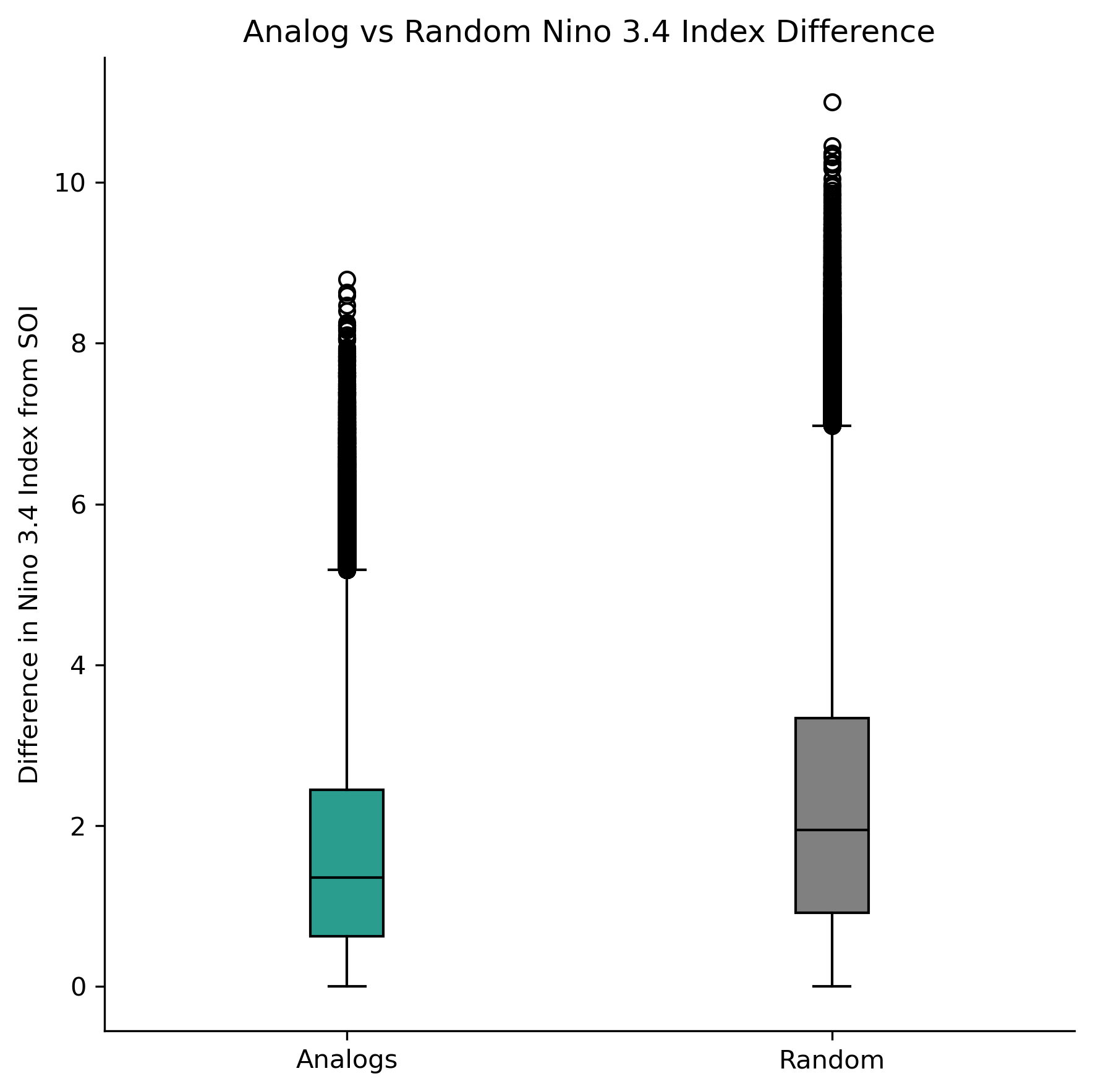}
    \caption{Difference in average Niño-3.4 index between the 400 best selected analogs and a random selection of 400. The analogs had a more similar mean Niño-3.4 index compared to the random selection.}
    \label{fig:analog_vs_random_nino34}
\end{figure}

\begin{figure}[ht]
    \centering
    \includegraphics[width=\textwidth]{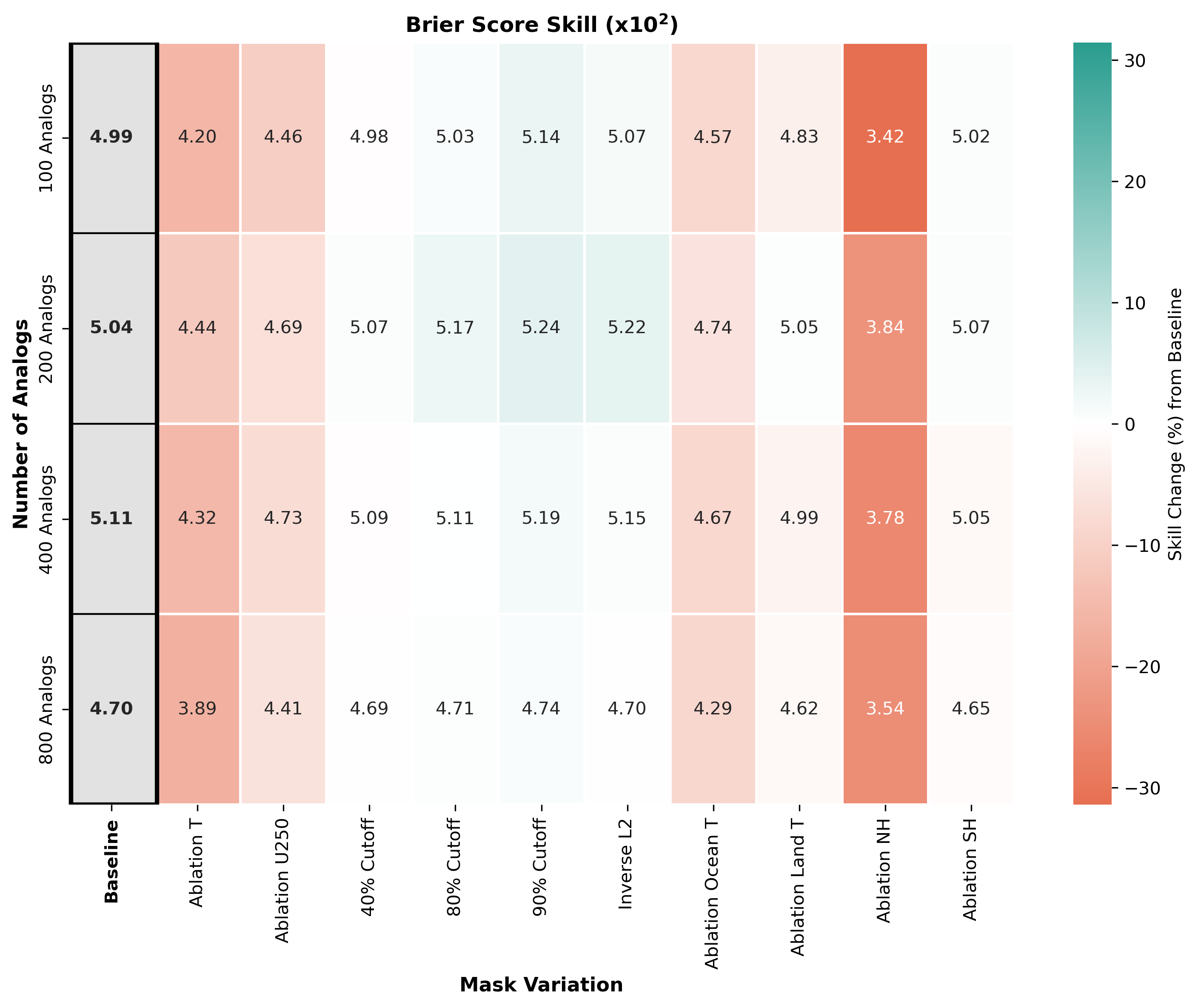}
    \caption{Expanded table, showing changes in BS skill with different ablation methods, for 100-800 analogs}
    \label{fig:bs_expanded_skill}
\end{figure}

\end{document}